\pdfoutput=1

\documentclass[aps,prd,preprint,groupaddress,amssymb,amsmath,nofootinbib]{revtex4}

\usepackage{color}
\usepackage{graphicx}
\usepackage{epstopdf}
\usepackage{hyperref} 
\usepackage{epsfig,bbm,amsmath,amssymb}
\usepackage{multirow}

\def\vev#1{\langle #1 \rangle}
\def\be{\begin{equation}}
\def\ee{\end{equation}}
\def\bea{\begin{eqnarray}}
\def\eea{\end{eqnarray}}

\def\bali{\begin{align}}
\def\ealign{\end{align}}

\def\slash#1{\setbox0=\hbox{$#1$}#1\hskip-\wd0\dimen0=5pt\advance
       \dimen0 by-\ht0\advance\dimen0 by\dp0\lower0.5\dimen0\hbox
         to\wd0{\hss\sl/\/\hss}}

\begin{document}

\preprint{}


\title{Trilepton Signatures of Light Charged and CP-odd Higgs Bosons in Top Quark Decays}

\author{Radovan Derm\' \i\v sek}
\email[]{dermisek@indiana.edu}
\affiliation{Physics Department, Indiana University, Bloomington, IN 47405, USA}

\author{Enrico Lunghi}
\email[]{elunghi@indiana.edu}
\affiliation{Physics Department, Indiana University, Bloomington, IN 47405, USA}

\author{Aditi Raval}
\email[]{adiraval@indiana.edu}
\affiliation{Physics Department, Indiana University, Bloomington, IN 47405, USA}


\date{December 19, 2012}

\begin{abstract}

In singlet extensions of the two Higgs doublet model, a light CP-odd Higgs boson ($A$) can significantly modify decay modes of the charged Higgs, without necessarily affecting decay modes of the standard model-like Higgs boson. These effects can be sizable even if the doublet component of the light CP-odd Higgs is small, so that constraints from Upsilon decays and direct CP-odd Higgs production do not apply. We study a scenario in which the charged Higgs is produced in top quark decays, $t \to H^\pm b$, with dominant $H^\pm \to W^\pm A$. We focus on the CP-odd Higgs mass range below the $b\bar b$ threshold. We summarize all current experimental constraints and find that, as a result of $H^\pm \to W^\pm A$ dominance, this scenario is not constrained in a large region of $\tan\beta$, $m_{H^\pm}$, and CP-odd Higgs doublet component. We discuss search strategies based on $A \to \tau^+ \tau^-$, with both taus decaying leptonically, and on the subleading decay mode $A \to \mu^+\mu^-$. We show that a search for $t \to H^+ b, \; H^+ \to W^+ A, \; A \to \mu^+ \mu^- $ in 20 fb$^{-1}$ of 8 TeV LHC data can constrain most of the currently allowed parameter space. Existing trilepton searches are not sensitive to this signal due to the adopted isolation criteria.

\end{abstract}


\maketitle

\section{Introduction}

The exciting discovery of a new particle at the LHC~\cite{:2012gk, :2012gu} raises many questions and creates new challenges. Although current data indicate that the particle has properties close to what is expected from the Higgs boson in the standard model (SM), much larger  data sets will be needed to determine whether it really is the Higgs boson that completes the standard model  or a particle that points to a new physics. A significant deviation in  production cross sections  or branching ratios in various channels from those predicted by the SM can be observed at any point. In addition, this particle could be accompanied by more Higgs  states.

In  well motivated theories beyond the SM the Higgs sector is typically more complicated. Models with two Higgs doublets, for example the minimal supersymmetric model (MSSM), contain five Higgs bosons in the spectrum:  light and heavy CP-even Higgses, $h$ and $H$, the CP-odd Higgs, $A$, and a pair of charged Higgs bosons, $H^\pm$. Singlet extensions of the two Higgs doublet models, for example the next-to-minimal supersymmetric model (NMSSM),  contain three CP-even Higgs bosons, 
two CP-odd Higgs bosons, 
and a pair of charged Higgs bosons,
and there are many simple models with even more complicated Higgs sectors. 
  It is usually the case that there is one Higgs boson with properties (couplings to $W$ and $Z$ bosons) of the SM Higgs, we refer to it as the SM-like Higgs. However,  such a Higgs does not necessarily decay in the same way as the SM Higgs, and a significant model dependence of decay modes applies to other Higgses as well. 

It is well known that the presence of a light CP-odd Higgs boson can modify the decay modes of heavier Higgses. 
We study a scenario with the charged Higgs lighter than the top quark, so that  $t \to H^\pm b$ decay mode is present, and a light CP-odd Higgs
so that $H^\pm \to W^\pm A$ decay mode is open and can dominate.  This scenario generically appears and is phenomenologically viable in singlet extensions of the two Higgs doublet model with or without supersymmetry~\cite{Dermisek:2008id, Dermisek:2008sd, Dermisek:2008uu}. 
In order for our study to be applicable to a variety of models we only assume couplings of Higgs bosons originating from doublets to fermions, gauge bosons, and other Higgses according to the two Higgs doublet model Type II (supersymmetric models are of this type). The two CP-odd Higgses originating from doublets and the singlet can mix, and thus the light CP-odd Higgs is allowed to have an arbitrary doublet fraction. 
 We do not assume any specific mass relations in  Higgs spectrum,  and we do not include any kind of corrections originating from additional particles in specific models, such as  SUSY corrections. We summarize experimental constraints from direct searches for $H^\pm \to \tau \nu$, $H^\pm \to c s$, and $H^\pm \to W^\pm A$ followed by 
$A \to \tau^+ \tau^- $, and show that, as a result of $H^\pm \to W^\pm A$ dominance, this scenario is not constrained in a large range of $\tan\beta$, $m_{H^\pm}$, and the doublet fraction of the CP-odd Higgs. We do not impose any constraint from flavor changing rare decays (e.g. $B\to X_s \gamma$, $B_s\to \mu^+\mu^-$, etc.) as they may strongly depend on other sectors of a complete model.

A light CP-odd Higgs can simultaneously modify decay modes of other Higgses. This has been extensively studied  in the context of possible modifications of decay modes of the SM-like Higgs boson ($h \to AA$ can be the dominant decay mode), see for instance Refs.~\cite{Dermisek:2005ar,  Dermisek:2005gg,  Dermisek:2007yt,  Chang:2008cw}, motivated by considerations of naturalness of electroweak symmetry breaking in supersymmetric models~\cite{NEWSB}. Current data do not favor non-standard dominant decay modes of the SM-like Higgs. 
However, we show that $H^\pm \to W^\pm A$ can be the dominant decay mode of the charged Higgs boson even if the CP-odd Higgs has only a tiny doublet component so that it is not constrained by direct searches, Upsilon decays, and it would not significantly affect the decay modes of the SM-like CP-even Higgs boson. As an example, we discuss the viability of this scenario in the NMSSM.

The existence of a light CP-odd Higgs boson can be motivated by approximate global symmetries and thus it occurs in a variety of models. Examples include the NMSSM with R-symmetry or Peccei-Quinn symmetry~\cite{Dobrescu:2000yn, Dermisek:2006wr},  models with new strong dynamics~\cite{Gripaios:2009pe}, little Higgs models~\cite{Bellazzini:2009xt, Bellazzini:2009kw}, and models with hidden sectors~\cite{Mardon:2009gw}. 

In this paper we discuss search strategies for the charged Higgs boson appearing in top quark decays, $t \to H^\pm b$,  $H^\pm \to W^\pm A$ with either $A\to \mu^+\mu^-$ or $A \to \tau^+ \tau^-$ (with both $\tau$s decaying into leptons). This is especially advantageous if the CP-odd Higgs is below the $b \bar b $ threshold, that is the scenario we investigate, but is also applicable to heavier CP-odd Higgses. The outline of the strategy and some preliminary results were given in Ref.~\cite{Dermisek:2010af}.\footnote{A complementary search strategy for $t \to H^\pm b$,  $H^\pm \to W^\pm A$ with the CP-odd Higgs boson above the $b \bar b $ threshold was recently discussed in Ref.~\cite{Rathsman:2012dp}.}
 To suppress background, especially from semileptonic bottom and charm decays, we require an additional lepton ($e$ or $\mu$) from one of the two $W$s. In addition to the case when $A \to \mu^+ \mu^- $, in which the mass of the CP-odd Higgs can be reconstructed, there is a complementary signal in which the $e$ and $\mu$ leptons originate from $A \to \tau^+ \tau^-$. The latter can also be observed especially for trilepton combinations with two same sign and same flavor leptons, {\it e.g.}  $\mu^+ \mu^+ e^-$, which have very low background. We show that these searches in  20 fb$^{-1}$ of data from the 8 TeV run of the LHC can constrain a large region of the parameter space to which other searches are not sensitive. In particular, we show that existing trilepton searches are not sensitive to this signal due to the adopted isolation criteria. We also discuss the reach of the LHC at 14 TeV center-of-mass energy. Additional improvements of the analysis are possible with b-tagging that we do not require here.

The decay mode $H^\pm \to W^\pm A$ that we are studying in singlet extensions of the 2HDM, has been investigated in the   MSSM~\cite{Moretti:1994ds, Djouadi:1995gv}\footnote{In the MSSM $H^\pm \to W^\pm A$   is typically not the dominant decay mode since in most of the parameter space it is not kinematically open as a two body decay. It can be the dominant decay mode of the charged Higgs only  for a sufficiently  light CP-odd Higgs~\cite{Akeroyd:2002hh},~\cite{Dermisek:2008id}. However, such a scenario is no longer phenomenologically viable in the MSSM.}, in a  variety of  scenarios with two or more Higgs doublets~\cite{Borzumati:1998xr, Akeroyd:2012yg},  and scenarios beyond the MSSM that can be parameterized by certain higher dimensional operators~\cite{Bae:2010cd, Carena:2010cs}. A light charge Higgs scenario  was also discussed in connection with the $2.8 \sigma$ deviation from lepton universality in $W$ decays measured at LEP~\cite{:2004qh}, to which a charged Higgs with mass close to the mass of the $W$ boson can contribute~\cite{Park:2006gk, Dermisek:2008dq}, and  $H^\pm \to W^\pm A$ decay mode can provide a viable option to avoid LEP limits~\cite{Dermisek:2008dq}.

This paper is organized as follows. In Section~\ref{sec:model} we discuss properties of the charged and CP-odd Higgs bosons in singlet extensions of the two Higgs doublet model Type II (as an example we discuss this scenario in the NMSSM), and describe the region of the parameter space  in which $H^\pm \to W^\pm A$  dominates. We discuss constraints from direct production of the CP-odd Higgs, Upsilon decays and also from the requirement that the decay modes of the SM-like Higgs boson are not significantly modified. 
In Section~\ref{sec:br} we present numerical study of $t \to H^\pm b$,  $H^\pm \to W^\pm A$, explore the parameter space of the model, and summarize experimental constraints from direct searches for $H^\pm \to \tau \nu$, $H^\pm \to c s$, and $H^\pm \to W^\pm A$ followed by 
$A \to \tau^+ \tau^- $.
The LHC study of search strategies based on leptonic decays of $\tau$s from $A \to \tau^+ \tau^- $, and a subleading decay mode of the CP-odd Higgs, $A \to \mu^+ \mu^- $, is presented in Section~\ref{sec:lhc}. We also discuss relevant background in detail, and discuss the reach of the LHC at 8 TeV and 14 TeV in this Section. 
 We conclude and discuss possible improvements of the analysis in Section~\ref{sec:conclusions}.
Formulas for decay widths are summarized in the Appendix.

\section{Light Charged Higgs boson with dominant $H^\pm \to W^\pm A$ }
\label{sec:model}

Couplings of the charged Higgs to quarks and leptons can be written as:
\begin{equation}
{\cal L} \supset \frac{g_2}{\sqrt{2} M_W}  \left[ \bar u_i V_{ij} \left( m_{d_j} X  P_R + m_{u_i} Y P_L  \right) d_j  \; + \;   \bar \nu_i   m_{e_i} Z P_R  e_i \right]  H^+  + h.c. 
\end{equation}
where $i,j$ are flavor indices and $V$ is the CKM matrix. In general, coefficients $X$, $Y$, $Z$ can be arbitrary complex numbers. However, in the two Higgs doublet model (2HDM)
 of Type~II and supersymmetric models (that we will focus on), $X=Y=1/Z = \tan \beta$, where $\tan \beta$ is the ratio of vacuum expectation values of the two Higgs doublets.
 In these models, if the charged Higgs is lighter than the top quark, the dominant decay modes of the charged Higgs are: $H^\pm \to cs$ for small $\tan \beta$ and $H^\pm \to \tau \nu$ for large $\tan \beta$. 
 
 In presence of a light CP-odd Higgs boson $A$, another decay mode, $H^\pm \to W^\pm A$, can easily dominate. This is because the strength of the $W^\pm-H^\pm-A$ interaction is given by the gauge coupling $g_2$ of $SU(2)$ which is much larger than Yukawa couplings of the charm and strange quarks and of the tau lepton that control the decays modes to fermions. In models with  more than one CP-odd Higgs boson, {\it e.g.} in singlet extensions of the two Higgs doublet model, the  $W^\pm-H^\pm-A$ coupling is proportional to the doublet component of the corresponding mass eigenstate:
 \begin{equation}
A =  \cos \theta_A \, A_{2HDM} + \sin \theta_A \, A_S.
\label{eq:a1_composition_2HDM}
\end{equation}

The partial width of the Charged Higgs boson decaying into tau lepton and neutrino is given by (neglecting the mass of the tau)
\begin{equation}
\Gamma (H^\pm \to \tau \nu) \simeq \frac{G_F}{\sqrt{2}} \frac{M_{H^\pm}}{4\pi} m^2_\tau \tan^2 \beta,
\label{eq:H_to_tau_nu}
\end{equation}
while the partial width of the Charged Higgs boson decaying into $W$ boson and the CP-odd Higgs, when kinematically possible,  is given by (neglecting the mass of the CP-odd Higgs):
\begin{equation}
\Gamma (H^\pm \to W^\pm A) \simeq \frac{G_F}{\sqrt{2}} \frac{M^3_{H^\pm}}{8 \pi} \left(1 - \frac{M_W^2}{M^2_{H^\pm}}\right)^3 \cos^2 \theta_A,
\label{eq:H_to_W_A}
\end{equation}
where $\cos \theta_A$ is the doublet component of the CP-odd Higgs mass eigenstate (at the amplitude level). Thus the $H^\pm \to W^\pm A$ decay mode dominates for:
\begin{equation}
 \cos^2 \theta_A > \frac{2 m^2_\tau}{M^2_{H^\pm} \left(1 - \frac{M_W^2}{M^2_{H^\pm}}\right)^3} \tan^2 \beta.
\end{equation}
This equation receives corrections proportional to $m_{A,\tau}^2/m^2_{H^\pm}$ that can be derived from the exact formulae presented in the appendix.

For medium or large $\tan \beta$ the $H^\pm \to W^\pm A$ mode is sizable only if the CP-odd Higgs has a significant doublet component; however as $\tan \beta $ approaches 1, $H^\pm \to W^\pm A$ dominates even if the CP-odd Higgs has only  ${\cal O} (1\%)$ doublet component (i.e. $\cos^2 \theta_A \simeq 0.01$). For  $\tan \beta \simeq 1$ also the $H^\pm \to c s$ decay becomes relevant; this, however, does not change our conclusions. The minimum of the doublet component of the CP-odd Higgs ($\cos^2\theta_A$) required for $H^\pm \to W^\pm A$ to dominate is plotted in Fig.~\ref{fig:HWA} as a function of $M_{H^\pm}$ and $\tan \beta$ (see Sec.~\ref{sec:br} for a detailed discussion).

In the scenario that we consider in the following sections, the couplings of Higgs doublets to gauge bosons and fermions are identical to those in the Two Higgs Doublet Model Type II, and the lightest CP-odd Higgs mass eigenstate can have an arbitrary doublet component. We do not consider corrections to Higgs couplings and masses that occur in specific models, e.g. in supersymmetric models. By minimizing the model dependence of our study, all the results we obtain can be easily reinterpreted in a variety of models. Nevertheless, in order to further motivate this scenario we demonstrate, in the following subsection, that it is also viable in supersymmetric models in which the Higgs sector is more constrained.

\subsection{Light charged and CP-odd Higgs bosons in the NMSSM}

The scenario we discussed in the previous section occurs and is phenomenologically viable in the simplest extension of the MSSM, the next-to-minimal supersymmetric model (NMSSM), which adds only one singlet chiral superfield, $\widehat{S}$. Its particle content differs from the MSSM by the addition of one CP-even and one CP-odd state in the neutral Higgs sector (assuming CP conservation), and one additional neutralino.  
Apart from the usual quark and lepton Yukawa couplings, the scale invariant superpotential
\begin{align}
\lambda \; \widehat{S} \widehat{H}_u \widehat{H}_d + \frac{\kappa}{3} \ \widehat{S}^3\; ,
\end{align}
depends on two dimensionless couplings $\lambda$ and $\kappa$ beyond the MSSM. The associated soft SUSY breaking trilinear  terms are
\begin{align}
\lambda \; A_{\lambda} S H_u H_d + \frac{\kappa}{3} A_\kappa S^3 
\end{align}
(hatted  letters denote superfields, while unhatted correspond to their scalar components).
The Higgs sector of the NMSSM is described by these trilinear and soft trilinear couplings, together with two further input parameters, $\tan \beta = v_u/v_d$ and $ \mu_\mathrm{eff} = \lambda s$, where $v_u\equiv \vev {H_u}$, $v_d\equiv \vev{H_d}$ and $s\equiv \vev S$.

The scenario that we consider hinges on several properties of the lightest CP-odd Higgs. First of all, we need a CP-odd Higgs lighter than $\sim 10$ GeV so that it dominantly decays into $\tau^+ \tau^-$. This can be achieved by approximate global symmetries; for instance, in the $U(1)_R$ symmetry limit of the NMSSM, $A_\kappa, A_\lambda \to 0$, the CP-odd Higgs is exactly massless (for more details and formulas, see e.g. Ref.~\cite{Dermisek:2006wr}). The second important property is the strength of its couplings to other Higgses, gauge bosons and fermions. This is controlled by the doublet component of the lightest CP-odd Higgs mass eigenstate, $\cos^2 \theta_A$: 
\begin{equation}
A =  \cos \theta_A \, A_{MSSM} + \sin \theta_A \, A_S.
\label{eq:a1_composition}
\end{equation}
For example, in the limit $A_\kappa, A_\lambda \to 0$ we find $\cos\theta_A\simeq (v/s)\sin2\beta$.

The discussion of  decay modes of the charged Higgs is basically identical to our general discussion, namely formulas for partial decay widths of the charged Higgs given in Eq.~(\ref{eq:H_to_tau_nu}) and Eq.~(\ref{eq:H_to_W_A}) are only slightly modified by SUSY corrections. 
However, a light CP-odd Higgs can also significantly modify the decay modes of the SM-like Higgs which is not favored by current data. In what follows we show that 
in the NMSSM the decay mode $H^\pm \to W^\pm A$ can  dominate  while the decay modes of the SM-like Higgs are not significantly modified.

The partial width of the lightest doublet-like CP-even Higgs boson decaying into $b \bar b$  is given by (neglecting phase space suppression):
\begin{equation}
\Gamma (h \to b \bar b) \simeq \frac{3 g_2^2 }{32 \pi M_W^2} \left( \frac{\cos \alpha}{\cos \beta} \right)^2M_{h} \; m^2_b ,
\end{equation}
while its partial width into two CP-odd Higgs bosons, when kinematically allowed, is given by (neglecting phase space suppression)~\cite{Dermisek:2006wr}:
\begin{equation}
\Gamma (h \to AA) \simeq \frac{M_W^2}{32 \pi g_2^2 M_{h}} \left[ \frac{g_1^2 + g_2^2}{2} \cos 2 \beta \sin (\beta + \alpha) \cos^2 \theta_A  + f (\lambda^2,\lambda \kappa, \kappa^2, \lambda A_\lambda, \kappa A_\kappa) \right]^2.
\label{eq:h_to_aa}
\end{equation}
The first term corresponds to the two Higgs doublet contribution and is the only term that survives in the $\cos^2 \theta_A = 1$ limit. The second piece represents terms originating from couplings of the singlet. Their exact form  can be found in Ref.~\cite{Dermisek:2006wr}. In the limit $\alpha \to \beta - \pi/2$ the lightest doublet-like CP-even Higgs boson has couplings to gauge bosons and fermion as the SM Higgs boson. In this case 
$\cos \alpha / \cos \beta \to \tan \beta$, $\sin (\beta + \alpha) \to - \cos 2 \beta$, and considering, for the time being, only the first term in Eq.~(\ref{eq:h_to_aa}) we find: 
\begin{equation}
\frac{\Gamma (h \to b \bar b)}{\Gamma (h \to AA)} \simeq \left( \frac{2 \sqrt{3} g_2^2 }{g_1^2 + g_2^2} \frac{M_{h} m_b}{M_W^2}  \frac{\tan \beta}{\cos^2 2 \beta} \frac{1}{\cos^2 \theta_A}\right)^2.
\end{equation}
For any $1 <\tan \beta < 50$ we find $\tan \beta / \cos^2 2 \beta \gtrsim 5$ with the minimum at $\tan \beta \simeq 3$. Thus, considering only the first term in Eq.~(\ref{eq:h_to_aa}), the $h \to AA$ mode dominates only if the CP-odd Higgs has a significant doublet component,  $\cos^2 \theta_A \gtrsim 0.5$.

The second piece in Eq.~(\ref{eq:h_to_aa}) depends strongly on the choice of parameters: it can be very small, or it can easily make $h \to AA$ to dominate for any $\tan \beta$ even for very singlet-like CP-odd Higgs boson.  This was investigated in detail in Ref.~\cite{Dermisek:2006wr} where the conditions under which $h \to AA$ dominates were studied. It was found that a non-zero soft trilinear couplings, especially $A_\lambda$, was crucial. For example, in the limit $A_\lambda, A_\kappa \to 0$, the branching ratio $h \to AA$ is at most of order 10\% while $\cos\theta_A\simeq (v/s)\sin2\beta$. This is an example of the parameter space in which the doublet component of the CP-odd Higgs can be large enough for  $H^\pm \to W^\pm A$ to dominate while $B(h \to AA)$ is small. Indeed, in the numerical study, scenarios with negligible $B(h \to AA)$ and $\cos^2 \theta_A \simeq 0.1$ appeared, see Fig. 12 in ~\cite{Dermisek:2006wr}. Similar scenarios also showed up in Ref.~\cite{Dermisek:2008uu} which studied non-standard decay modes of the charged and heavy CP-even Higgs bosons.\footnote{Refs.~\cite{Dermisek:2006wr} and \cite{Dermisek:2008uu} focused on a light SM-like Higgs boson below LEP limits that dominantly decays to $AA$ motivated by arguments of fine-tuning in the electroweak symmetry breaking; thus the superpartner masses were typically fixed to small values.
In light of the Higgs boson discovery and direct constraints on SUSY spectrum, these choices are no longer phenomenologically viable. However, rising the superpartner masses above limits and increasing stop masses and mixing that would make the SM like Higgs boson sufficiently heavy does not change our discussion since these parameters do not enter in branching ratios of the CP-even or charged Higgs bosons.    
In addition, some scenarios in Ref.~\cite{Dermisek:2008uu} had negligible $B(h \to AA)$, dominant $H^\pm \to W^\pm A$, and a SM-like Higgs in $\sim 125$ GeV mass range.}

 \subsection{Constraints from Upsilon decays and direct searches for the CP-odd Higgs}
 \label{sec:upsilon}
Constraints on a light CP-odd Higgs scenario come also from B factories and LHC data. These two sets of experimental results provide the strongest limits and thus we will not discuss other experimental constraints.

At B factories a light CP-odd Higgs can be produced  in Upsilon decays, $\Upsilon \to A \gamma$, with sizable rates in the NMSSM~\cite{Dermisek:2006py}.
At present, the strongest constraints come from BaBar~\cite{Aubert:2009cp, Aubert:2009cka} that sets limits  on $B (\Upsilon \to A \gamma)$ with $A \to \tau^+ \tau^-$ at the level of $10^{-5}$  and $A \to \mu^+ \mu^-$  at the level of $10^{-6}$ (these are the strongest limits, the exact exclusion limits depend on the mass of the CP-odd Higgs boson and are typically weaker as $m_A$  increases). 

In addition, a CP-odd Higgs can be directly produced at the LHC in the gluon fusion channel~\cite{Dermisek:2009fd}. This search strategy has several advantages. A possible discovery does not rely on searches in Higgs decays, for which the branching ratios are highly model dependent. Moreover, this mode does not suffer from phase space limitations of searches in  Upsilon decays (for $m_A$ approaching the mass of the Upsilon).  In spite of the large production cross section, picking up the signal on huge background is a serious problem. The dominant decay mode, $a \to  \tau^+ \tau^-$ seems to be hopeless; however searching for the subleading decay mode, $A \to \mu^+ \mu^-$,  is actually very promising~\cite{Dermisek:2009fd}.

Recently, searches for a light CP-odd Higgs boson were performed at the LHC using data sample corresponding to an integrated luminosity of $39.3 \; {\rm pb^{-1}}$ at ATLAS~\cite{Atlas_lightA} and $1.3 {\rm fb^{-1}}$ at CMS~\cite{Chatrchyan:2012am} collected at $\sqrt{s} = 7$ TeV. Stronger limits come from CMS which sets upper limits on the production cross section times decay branching ratio $\sigma (pp \to A)  \times  {\rm BR}(A \to \mu^+\mu^-)$ in the range of 1.5--7.5 pb~\cite{Chatrchyan:2012am}. 

Both searches, $\Upsilon \to A \gamma $ with $A \to \tau^+ \tau^-, \; \mu^+\mu^-$, and $gg \to A \to \mu^+ \mu^-$, are mainly sensitive to the coupling of the CP-odd Higgs to down type fermions which is given by $\tan \beta \cos \theta_A$. They nicely complement each other. Babar limits are stronger for smaller CP-odd Higgs masses, however for $m_A \gtrsim 7$ GeV the CMS limits take over. As a rough summary of the limits, for $m_A$ up to $\sim 9 $ GeV  and for any $1< \tan \beta < 50$ the product $\tan \beta \cos \theta_A$ is constrained to be less than $\sim 0.5$~\cite{Dermisek:2010mg,Chatrchyan:2012am}.

In summary, the scenario with dominant $H^\pm \to W^\pm A$ and small $B(h \to AA)$  that we consider in this paper occurs  and is well motivated for small to medium $\tan \beta$ and small doublet fraction of the CP-odd Higgs. Predictions from this region of the parameter space for the processes discussed above  comfortably satisfy (and are typically far below) the experimental limits which can be roughly translated into $\tan \beta \cos \theta_A \lesssim 0.5$.

\section{Charged and CP-odd Higgs bosons in top quark decays}
\label{sec:br}
\begin{figure}
\begin{center}
\includegraphics[width=0.33 \linewidth]{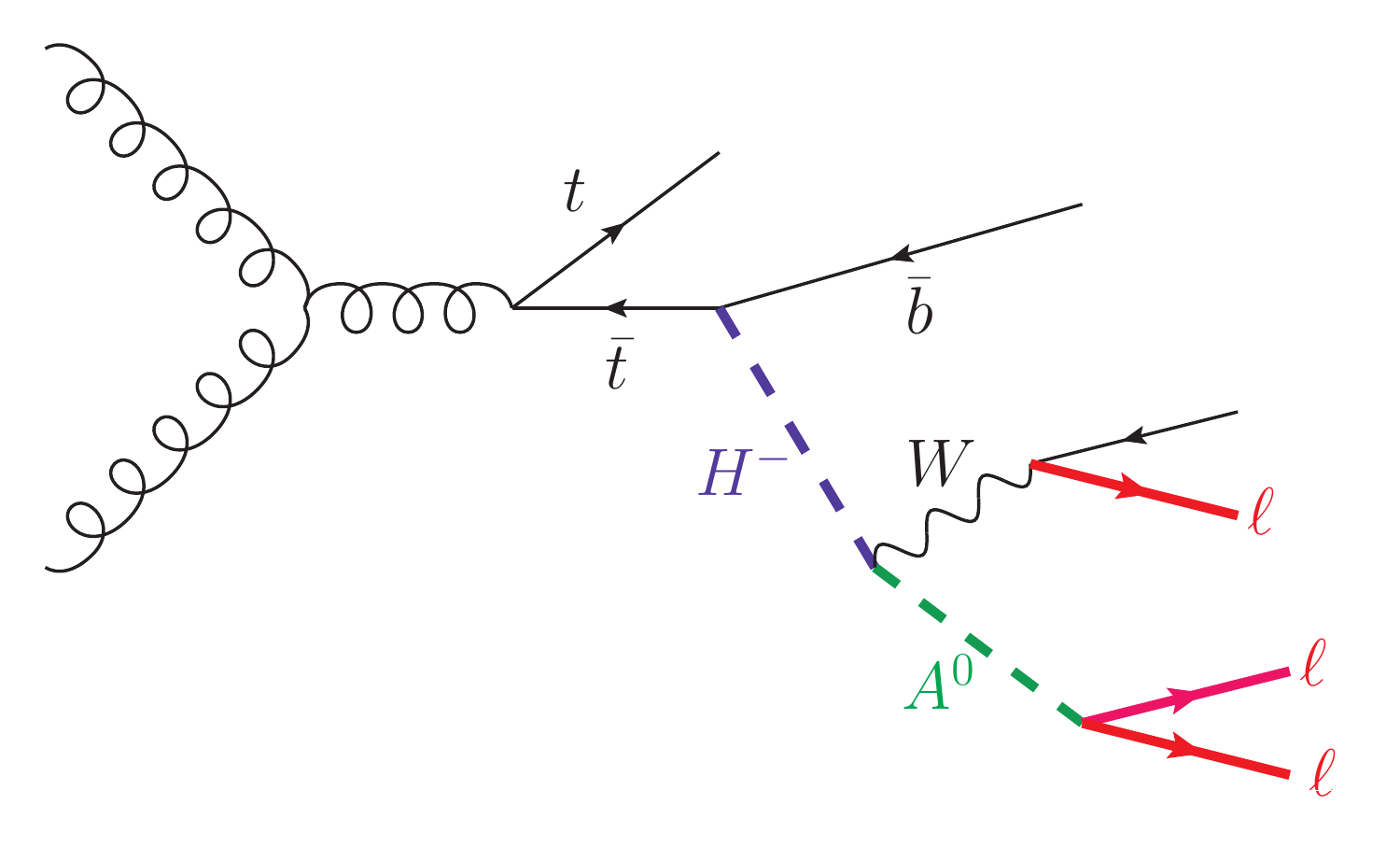}
\includegraphics[width=0.33 \linewidth]{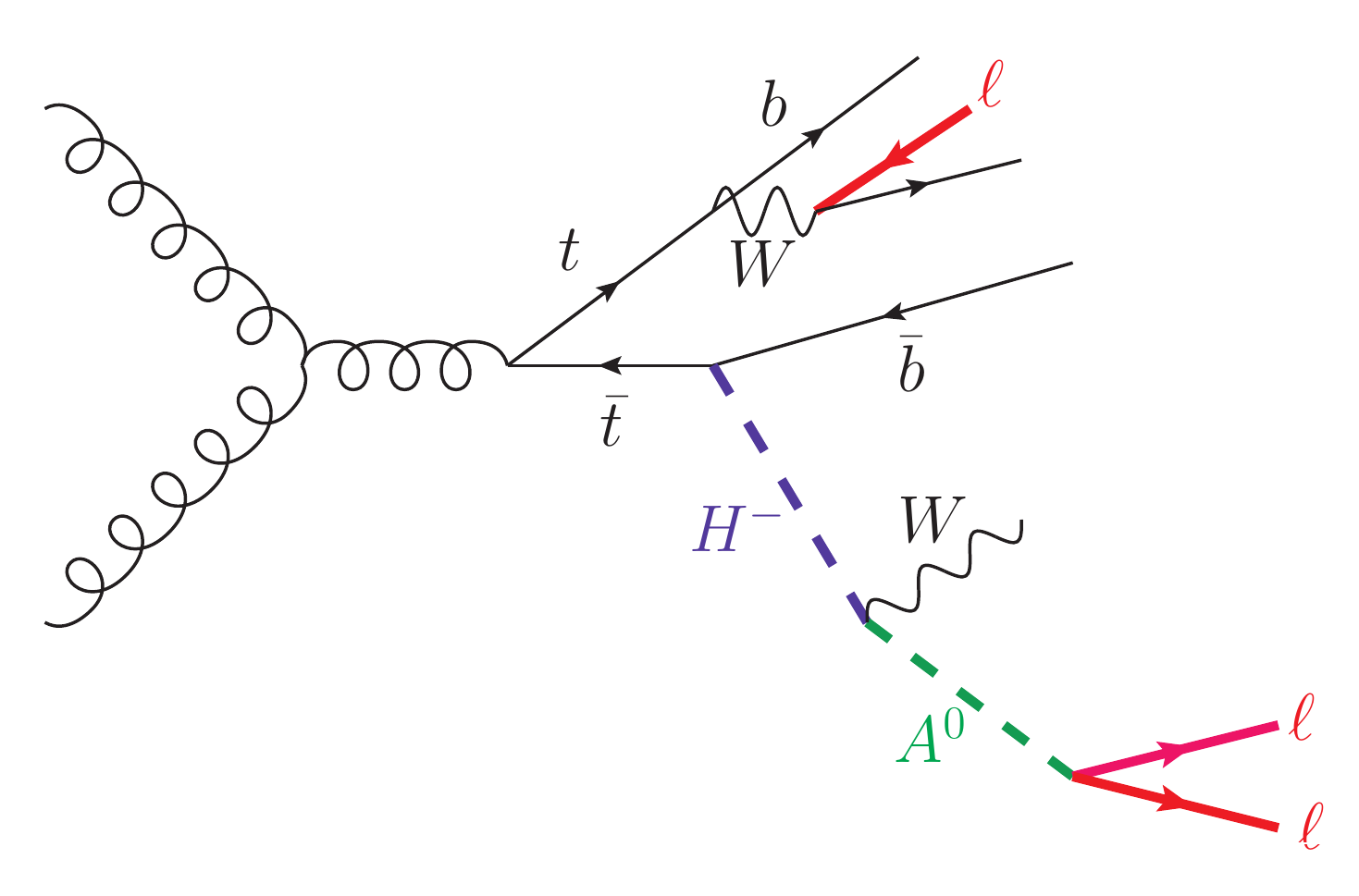}
\includegraphics[width=0.3 \linewidth]{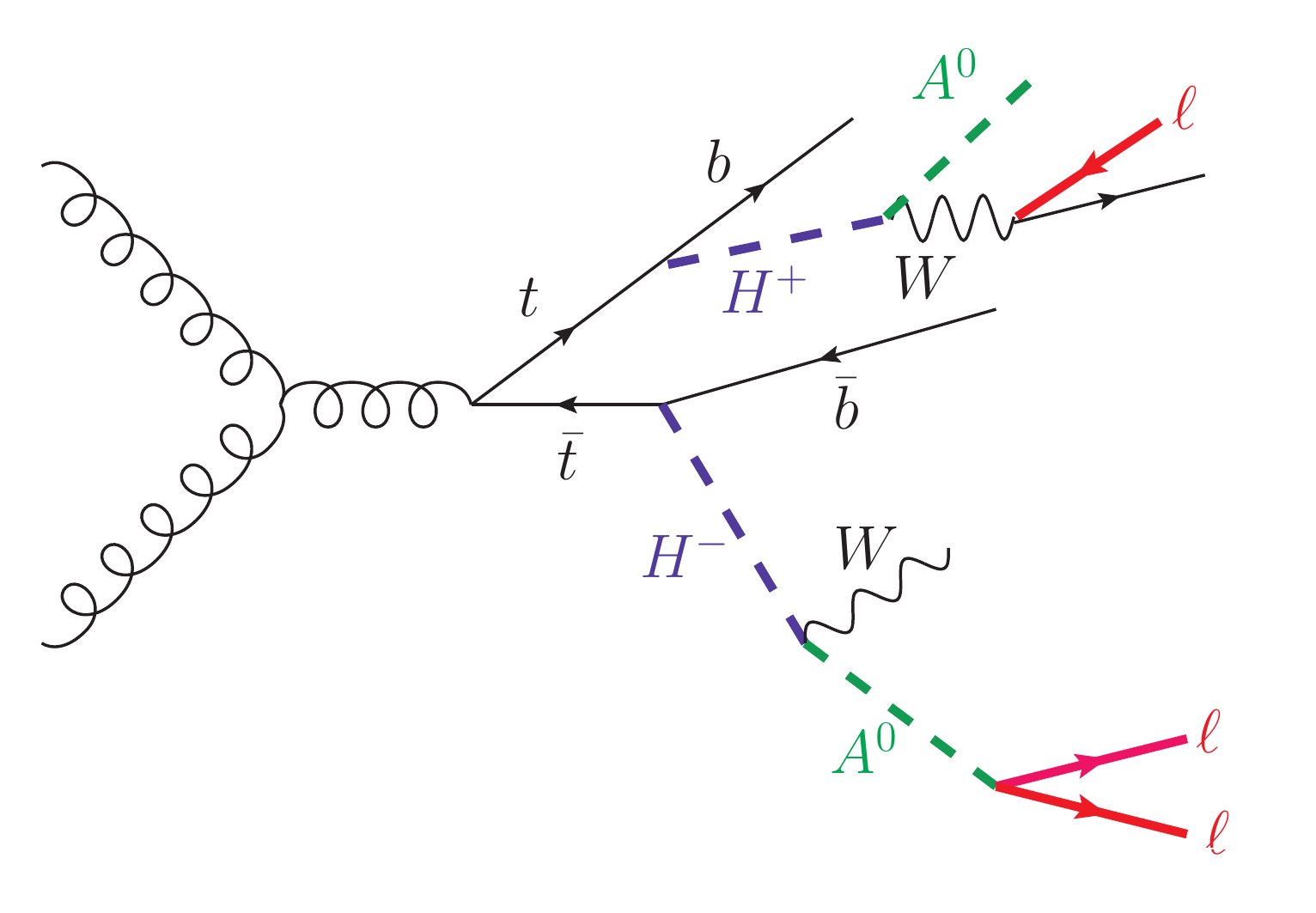}
\end{center}
\vskip -0.7cm
\caption{\it Feynman diagrams for charged Higgs production in top quark decays with three leptons in the final state.  \label{fig:process}}
\end{figure}

The main thrust of this paper is the investigation of the process
\bea
p p \to t \bar t \to t \; (\bar b   H^-) \to t \;(\bar b   W^- A^0) \to 
 t \; (\bar b \;  W^- \ell^+ \ell^-) 
\label{signal}
\eea
with one additional lepton coming from the $W^-$ (first diagram in Fig.~\ref{fig:process}) or from the $t$. In the latter case, the lepton originates from the $W^+$ produced in either $t\to W^+ b$ (second diagram in Fig.~\ref{fig:process}) or $t\to H^+ b \to W^+ A^0 b $ (third diagram in Fig.~\ref{fig:process}). In this section we identify regions of the  $(m_{H^\pm},\tan\beta,\cos^2\theta_A)$  parameter space that are already excluded by present experimental data and the ones in which we expect a sizable signal. 

Direct searches for the process in Eq.~(\ref{signal}) can be expressed in terms of constraints on the following combinations of branching ratios:
\begin{align}
{\cal B}_{\rm \tau\tau} &= X  \times  {\cal B} (t \to b H^\pm) \times {\cal B} (H^\pm \to W^{(*)} A) \times {\cal B} (A \to \tau\tau) \; ,\\
{\cal B}_{\rm \mu\mu} &=   X\times {\cal B} (t \to b H^\pm) \times {\cal B} (H^\pm \to W^{(*)} A) \times {\cal B} (A \to \mu\mu) \; .
\end{align}
where the factor $X \sim 1$ takes into account the different possible production mechanisms of the $W$--boson that produces the third lepton. It is given by~\footnote{In reference to Fig.~\ref{fig:process}, the third lepton can be produced by the $W$ coming from the charged Higgs decay (first term in the numerator), from a standard $t\to b W$ decay of the upper top (second term) or from the $t\to H^+ b \to W^+ A^0 b $ chain (third term).}
\begin{align}
X &= \frac{1 + {\cal B}(t\to bW) + {\cal B}(t\to bH^\pm) \times {\cal B}(H^\pm \to W^{(*)} A)}{2} \; ,
\label{X}
\end{align}
which reduces to $X=1$ under the assumption ${\cal B} (H^\pm \to W^{(*)} A) =1$. The explicit expressions for the top and charged Higgs widths that we use in this study are collected in the appendix.

\begin{figure}
\begin{center}
\includegraphics[width=0.45 \linewidth]{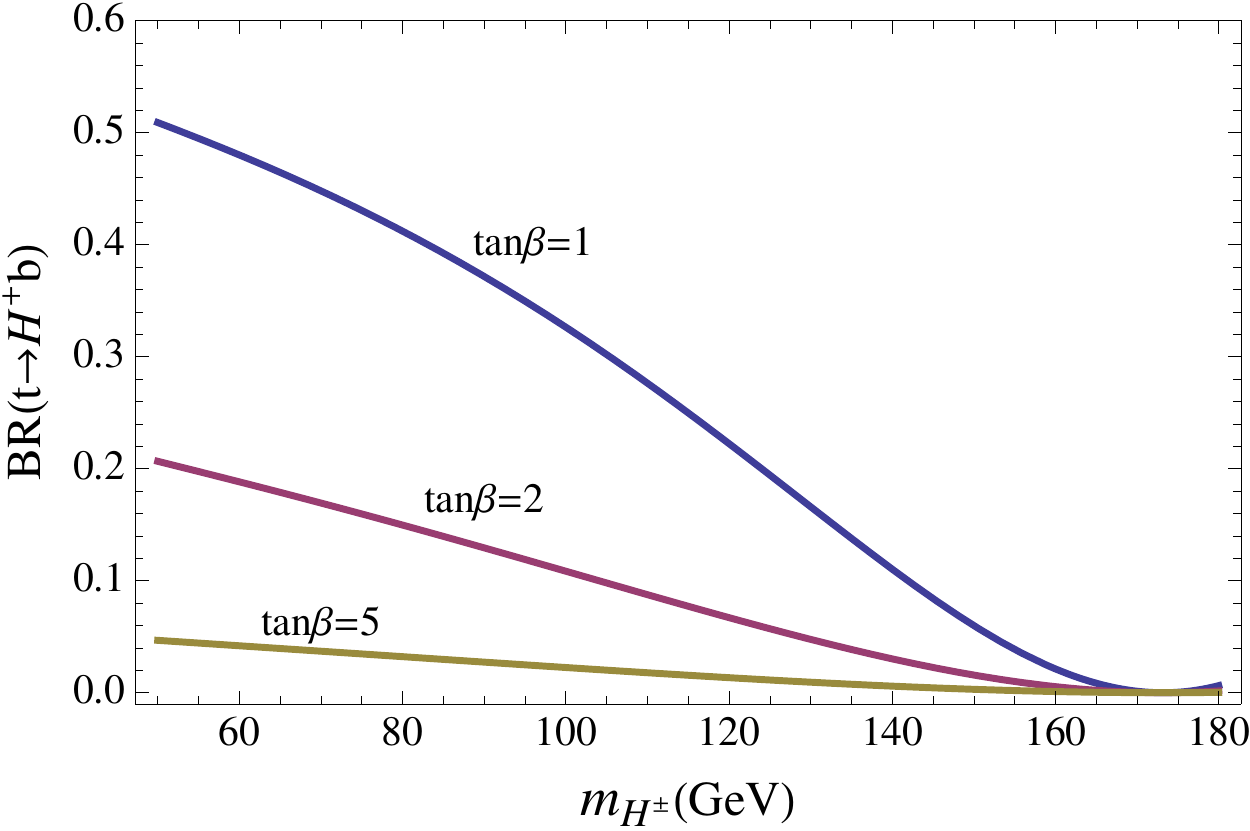}
\includegraphics[width=0.45 \linewidth]{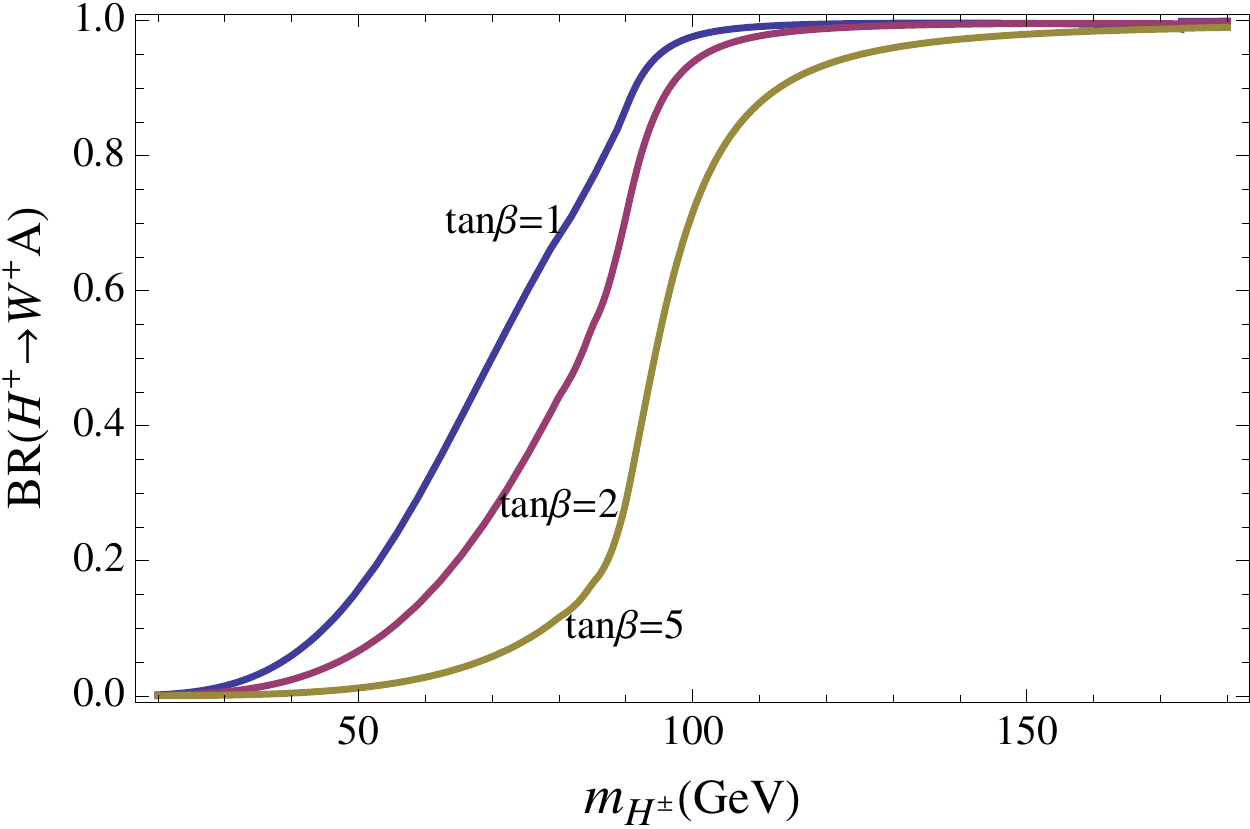}
\includegraphics[width=0.45 \linewidth]{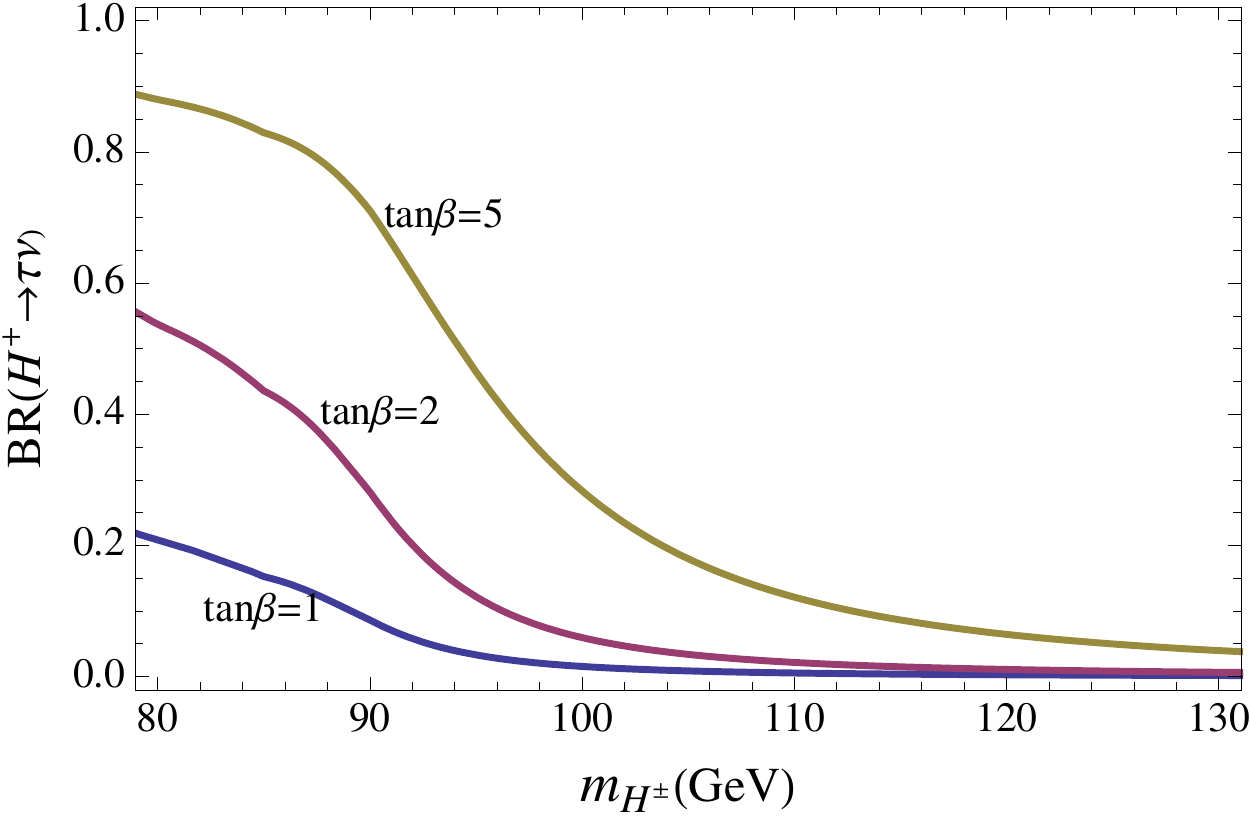}
\includegraphics[width=0.45 \linewidth]{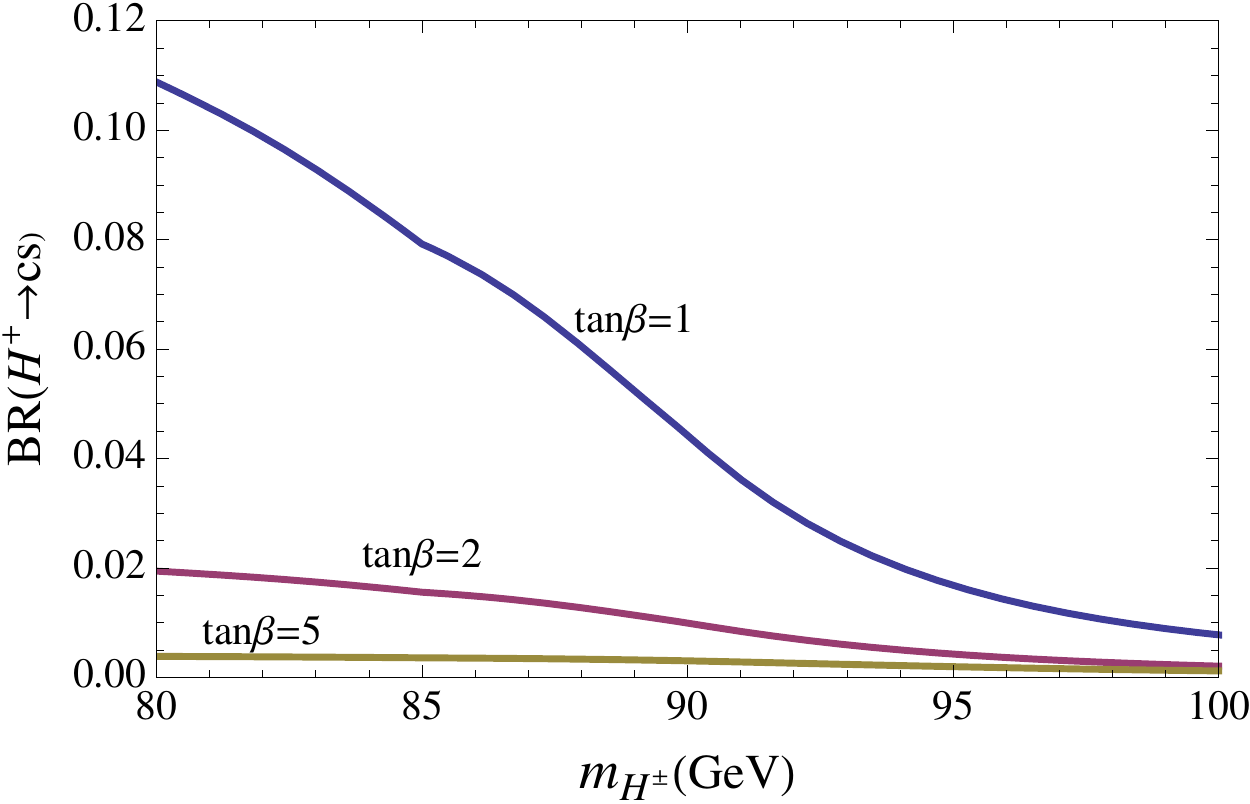}
\end{center}
\vskip -0.7cm
\caption{\it Branching ratios for $t\to H^+ b$ and various $H^\pm$ modes as functions of $m_{H^\pm}$ and $\tan\beta$ for fixed $m_A = 8\; {\rm GeV}$ and $\cos^2\theta_A=1$.\label{fig:BR}}
\end{figure}

\begin{figure}
\begin{center}
\includegraphics[width=0.45 \linewidth]{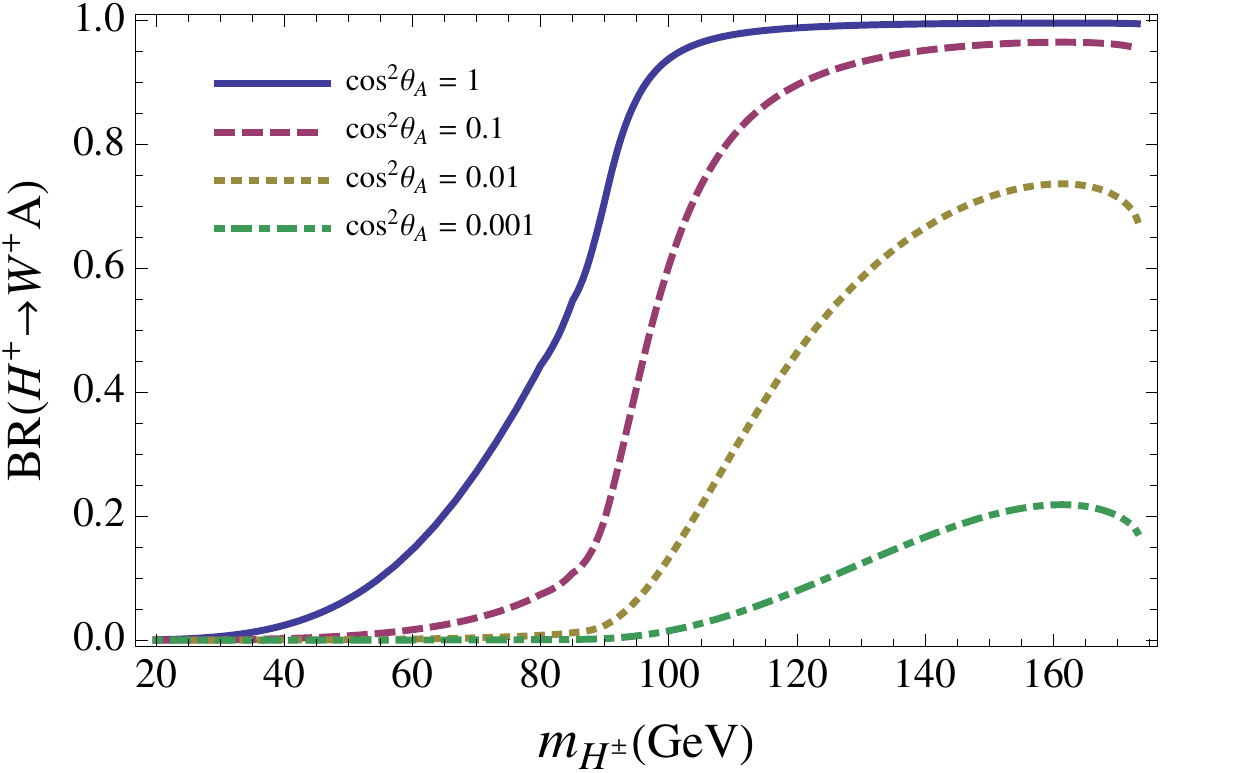}
\includegraphics[width=0.45 \linewidth]{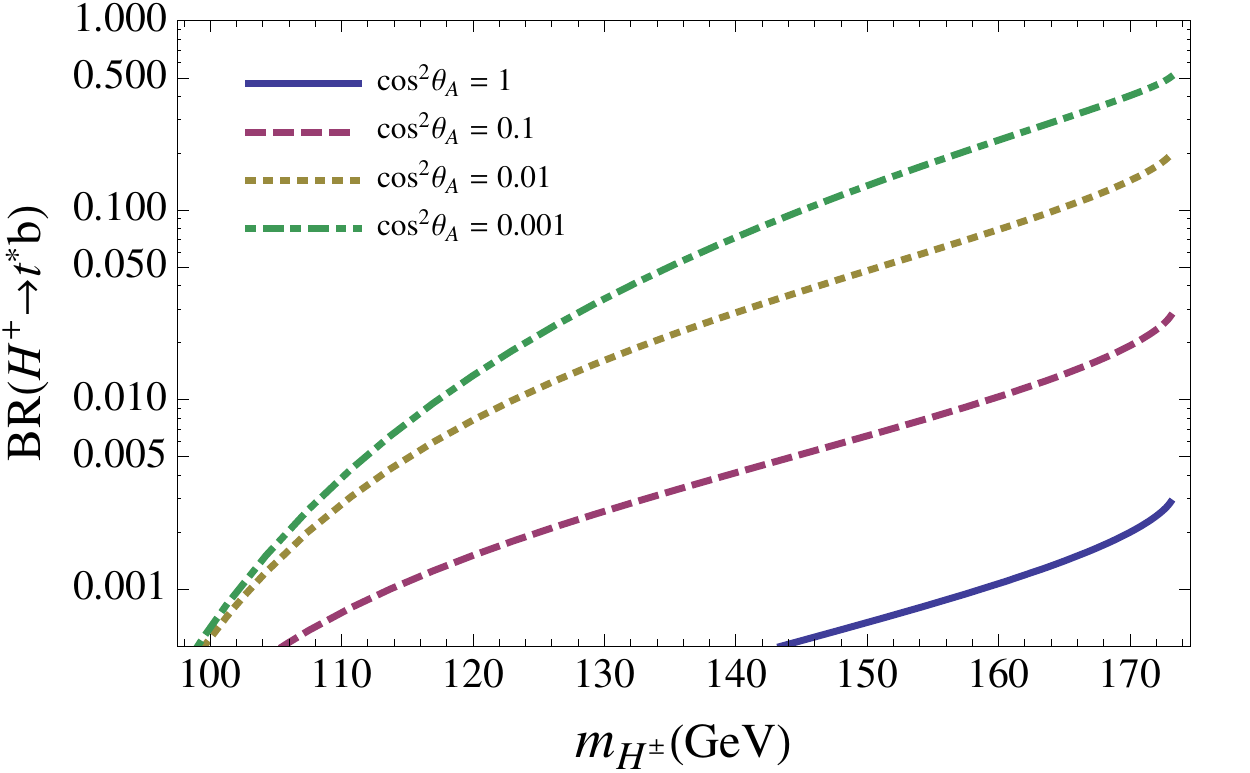}
\includegraphics[width=0.45 \linewidth]{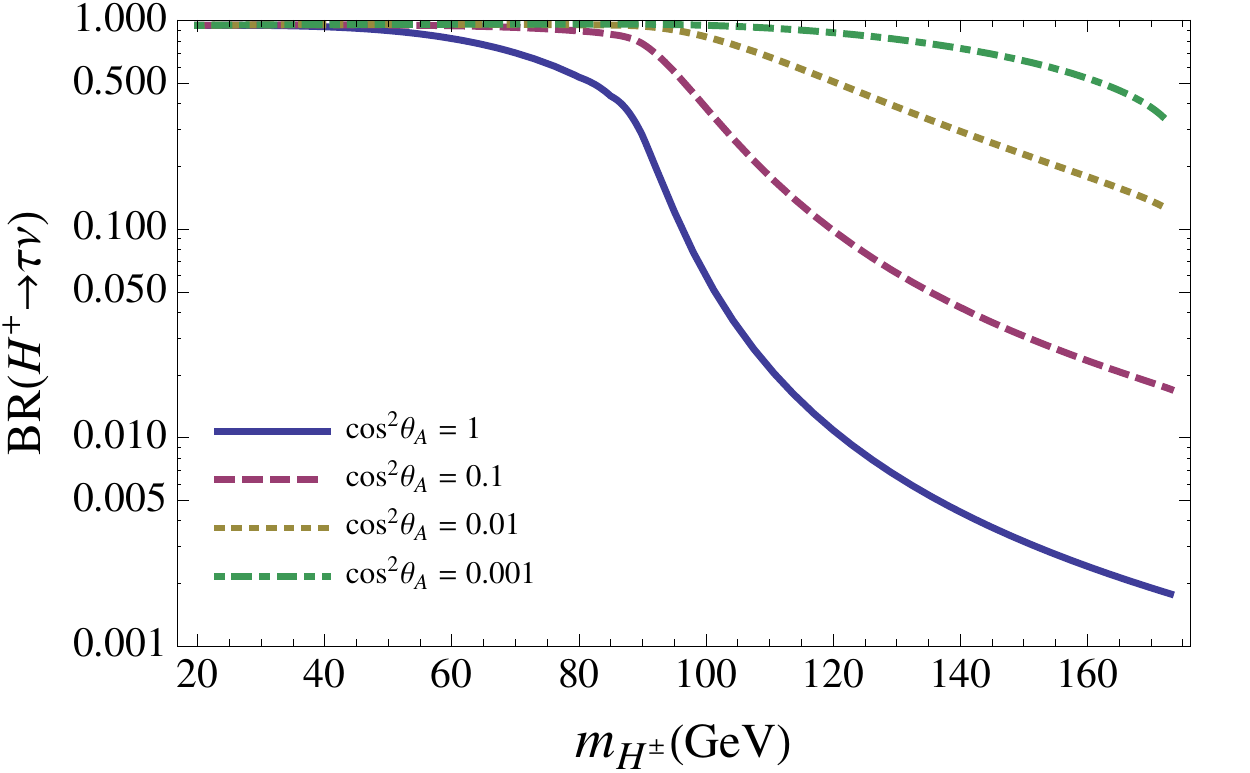}
\includegraphics[width=0.45 \linewidth]{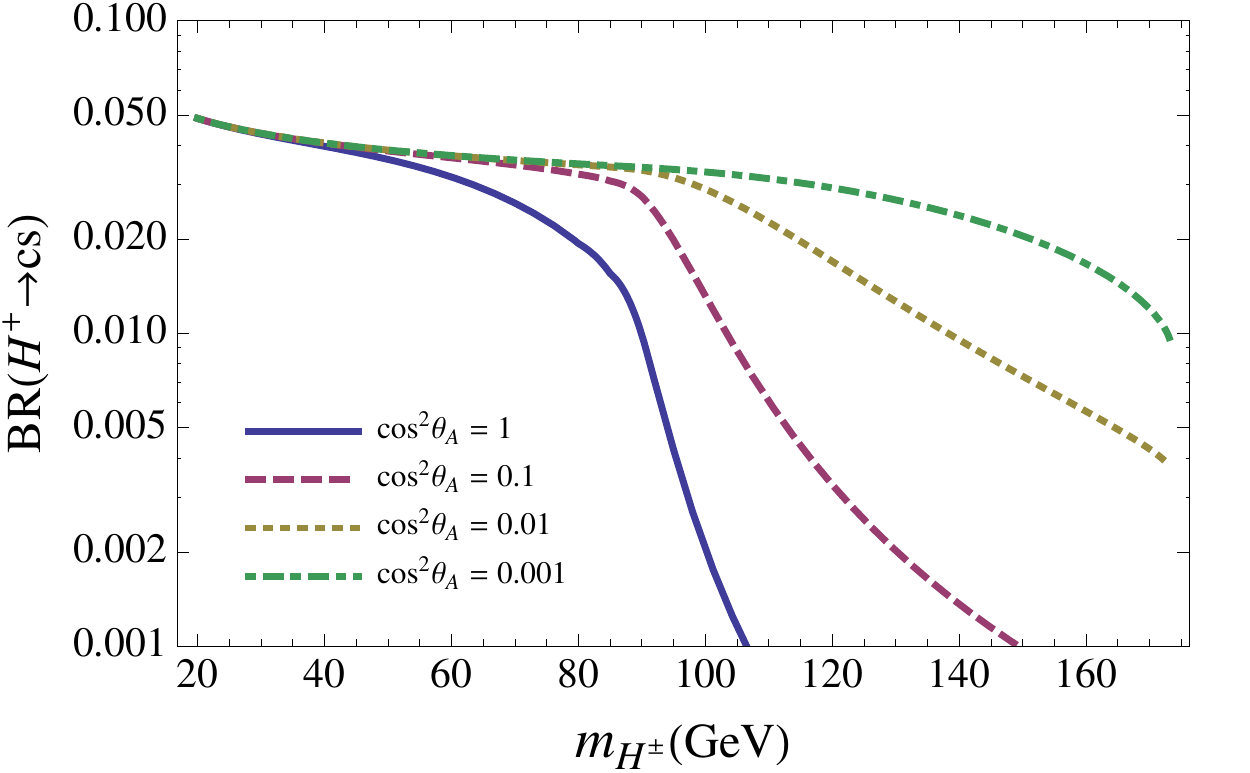}
\end{center}
\vskip -0.7cm
\caption{\it Branching ratios of various $H^\pm$ modes as functions of $\cos^2\theta_A$ and $m_{H^\pm}$ for fixed $m_A = 8\; {\rm GeV}$ and $\tan\beta=2$.\label{fig:BRxi}}
\end{figure}

We begin our analysis by showing in Figs.~\ref{fig:BR} and \ref{fig:BRxi} the relevant top and charged Higgs branching ratios as functions of $m_{H^\pm}$ for fixed $m_A = 8 \; {\rm GeV}$ and various values of $\tan\beta$ and $\cos^2 \theta_A$. Note that the mass of the CP-odd Higgs affects the kinematic threshold in the $H^\pm \to W^\pm A$ decay; this channel can still be dominant below the $WA$ threshold due to the off-shell $H^\pm \to W^* A$ decay. The doublet fraction of the CP-odd Higgs impacts directly the $H^\pm \to W^{(*)} A$ partial width, $\Gamma (H^\pm \to W^{(*)} A) \propto \cos^2\theta_A$, and indirectly all other channels ($\tau\nu$ and $\bar cs$) through contributions to the total charged Higgs width (in the $\cos^2\theta_A=0$ limit we have ${\cal B} (H^\pm \to \tau \nu) + {\cal B} (H^\pm \to \bar c s)  \simeq 1$).

\begin{figure}
\begin{center}
\includegraphics[width=0.49 \linewidth]{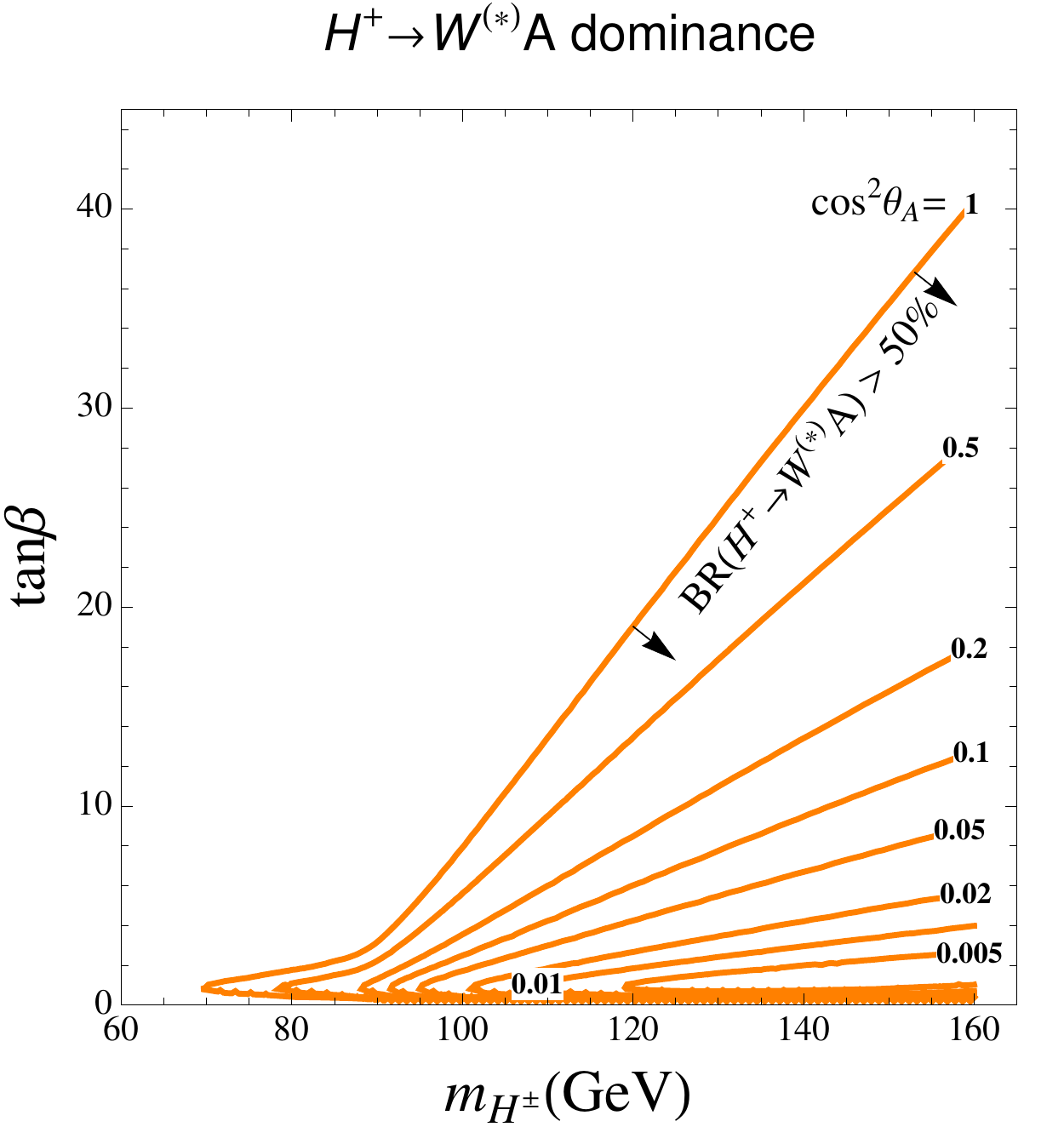}
\includegraphics[width=0.49 \linewidth]{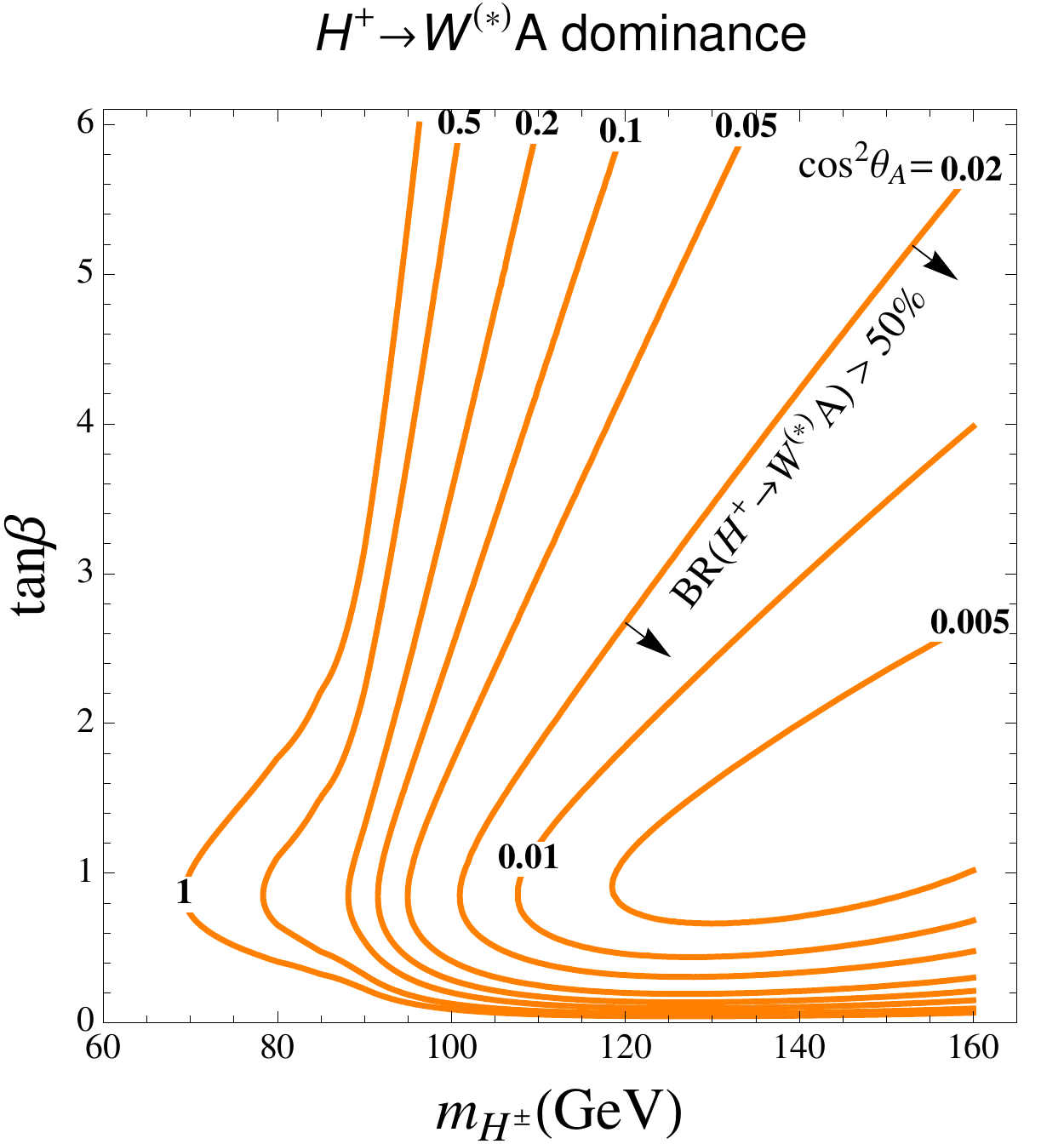}
\end{center}
\vskip -0.7cm
\caption{\it The contours correspond to a fixed value of $\cos^2 \theta_A$ (that appears in the label) and are obtained imposing the ${\rm BR}(H^\pm \to W^{(*)} A) = 50\%$ constraint. The region of the $[m_{H^\pm},\tan\beta]$ plane to the right of a given contour has ${\rm BR}(H^\pm \to W^{(*)} A) \geq 50\%$ for any value of $\cos^2 \theta_A$ than the indicated one. The plot on the right zooms on to the low $\tan\beta$ region.
\label{fig:HWA}}
\end{figure}

The most crucial ingredient in our analysis is the $H^\pm \to W^{(*)} A$ branching ratio. Each contour in Fig.~\ref{fig:HWA} corresponds to a different $\cos^2\theta_A$ value (that label the contour itself) and the region to the right of each curve corresponds to ${\cal B} (H^\pm \to W^{(*)} A) \geq 50\%$. The panel on the right zooms on low--$\tan\beta$: for $\tan\beta \lesssim 6$ we find large regions of $H^\pm \to W^{(*)} A$ dominance even for a very small mixing angle.

\begin{figure}
\begin{center}
\includegraphics[width=1 \linewidth]{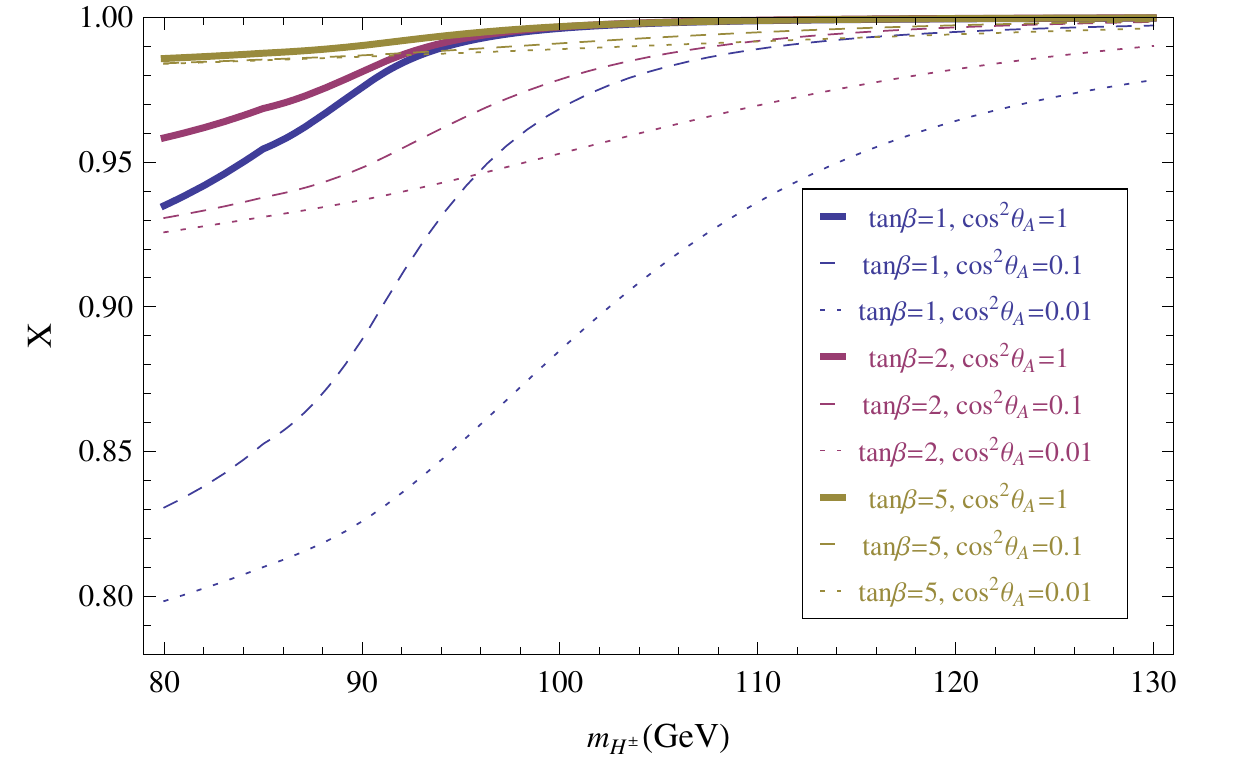}
\end{center}
\vskip -0.7cm
\caption{\it Ratio $X$ defined in Eq.~(\ref{X}) for various values of $\tan\beta$ and $\cos^2\theta_A$. 
\label{fig:X}}
\end{figure}

The final ingredient of our discussion is the ratio $X$ defined in Eq.~(\ref{X}). The deviation of this ratio from 1 is controlled by the $H^\pm \to W^{(*)} A$ branching ratio, and therefore by the value of $\cos^2 \theta_A$. In Fig.~\ref{fig:X} we show the value of this ratio for $\tan\beta = 1,2,5$ and $\cos^2\theta_A = 1,0.1,0.01$. From inspection of the plot we see that for $\tan\beta \gtrsim 2$ and $m_{H^\pm} \gtrsim 100 \; {\rm GeV}$ this ratio is always larger than about 95\%.

\begin{figure}
\begin{center}
\includegraphics[width=0.4 \linewidth]{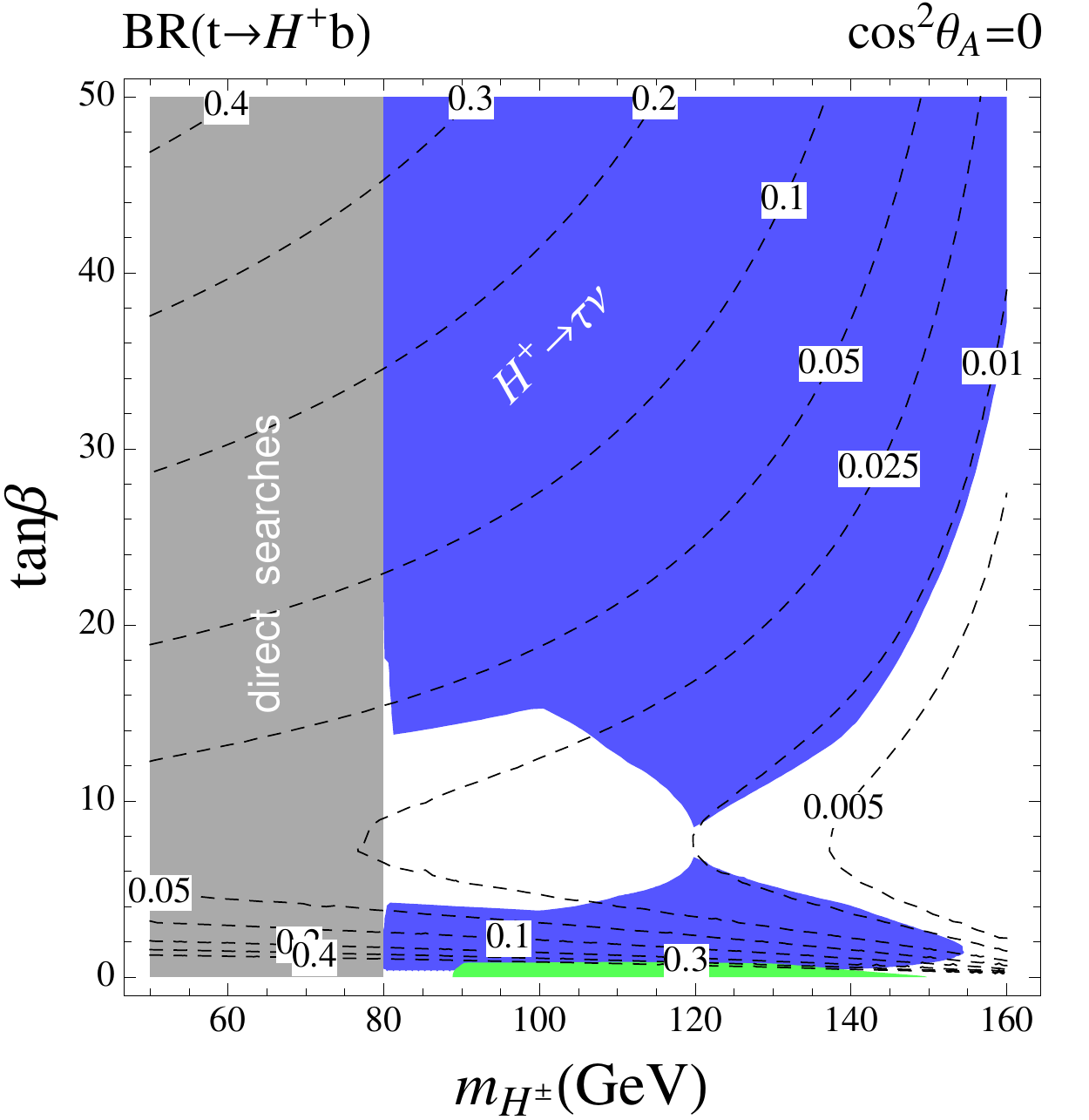}
\includegraphics[width=0.4 \linewidth]{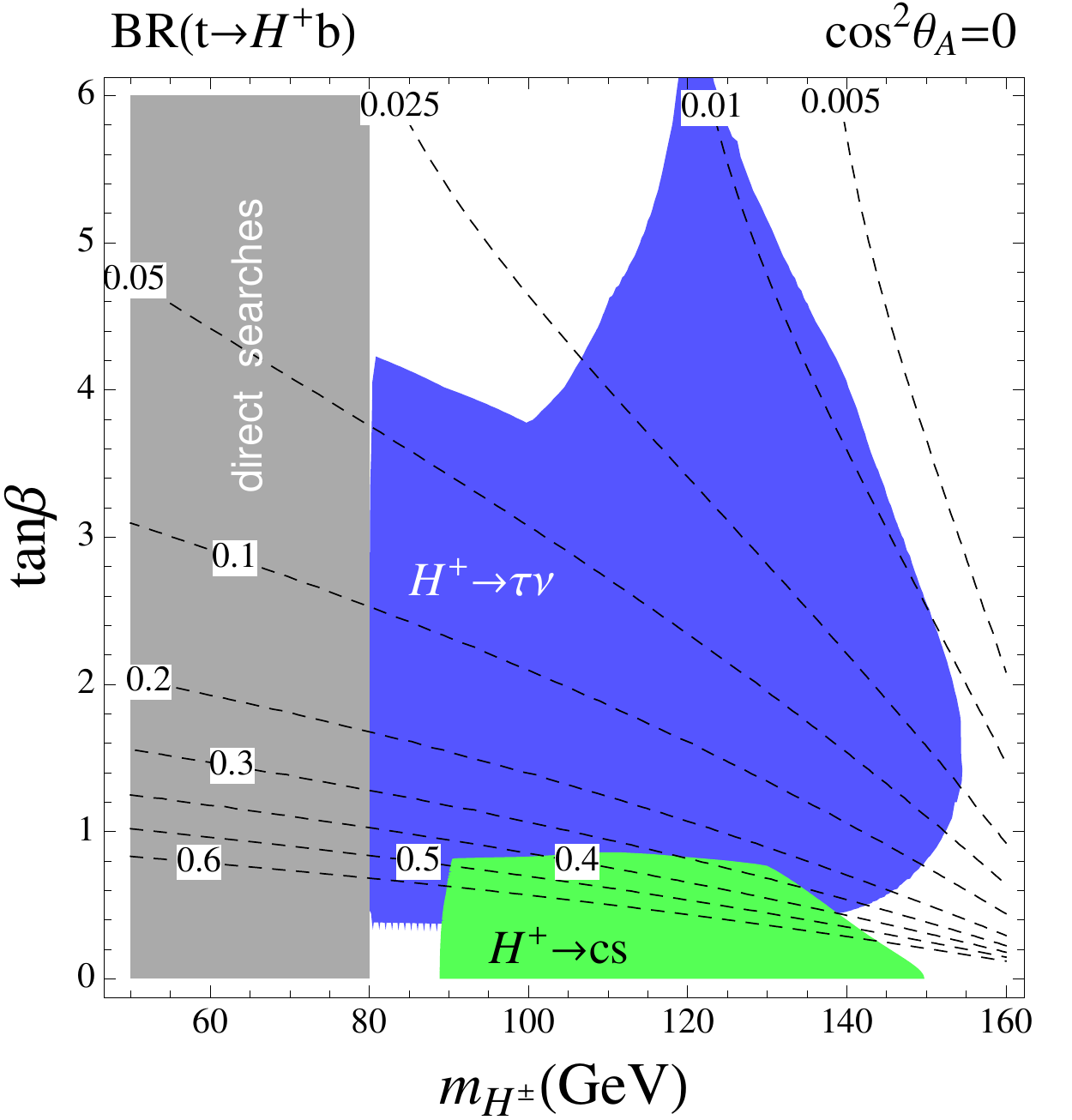}
\end{center} 
\vskip -0.7cm
\caption{\it Contours of constant ${\rm BR}(t\to H^+ b)$ (labelled). Assuming $\cos^2\theta_A = 0$, the grey region is excluded by direct $H^\pm$ searches at ALEPH~\cite{Heister:2002ev}, L3~\cite{Achard:2003gt} and DELPHI~\cite{Abdallah:2003wd}. The blue region is excluded by $H^\pm \to \tau\nu$ searches at D0~\cite{Abazov:2009wy}, ATLAS~\cite{Aad:2012tj} and CMS~\cite{:2012cw}. The green region are excluded by $H^\pm \to c s$ searches at ATLAS~\cite{ATLAS-cs-thesis}.  \label{fig:mhtb0}}
\end{figure}

In Fig.~\ref{fig:mhtb0} we show the allowed regions in the $(m_{H^\pm},\tan\beta)$ plane for $\cos^2\theta_A = 0$. We include the constraints from direct charged Higgs searches at ALEPH~\cite{Heister:2002ev}, L3~\cite{Achard:2003gt}, DELPHI~\cite{Abdallah:2003wd} and OPAL~\cite{Abbiendi:2008aa} (shaded gray), from $H^\pm \to \tau\nu$ searches at D0~\cite{Abazov:2009wy}, ATLAS~\cite{Aad:2012tj} and CMS~\cite{:2012cw} (shaded blue), and from $H^\pm \to c s$ searches at ATLAS~\cite{ATLAS-cs-thesis} (shaded green).  The dashed lines are constant ${\rm BR}(t\to H^+ b)$ contours. The combined LEP exclusion region has been presented in Ref.~\cite{Searches:2001ac} and reads $m_{H^\pm} > 78.6 \; {\rm GeV}$ at 95\% confidence level. This bound assumes $BR (H^\pm \to W A) = 0$, nevertheless we take it as an indication of the real limit. 

\begin{figure}
\begin{center}
\includegraphics[width=0.4 \linewidth]{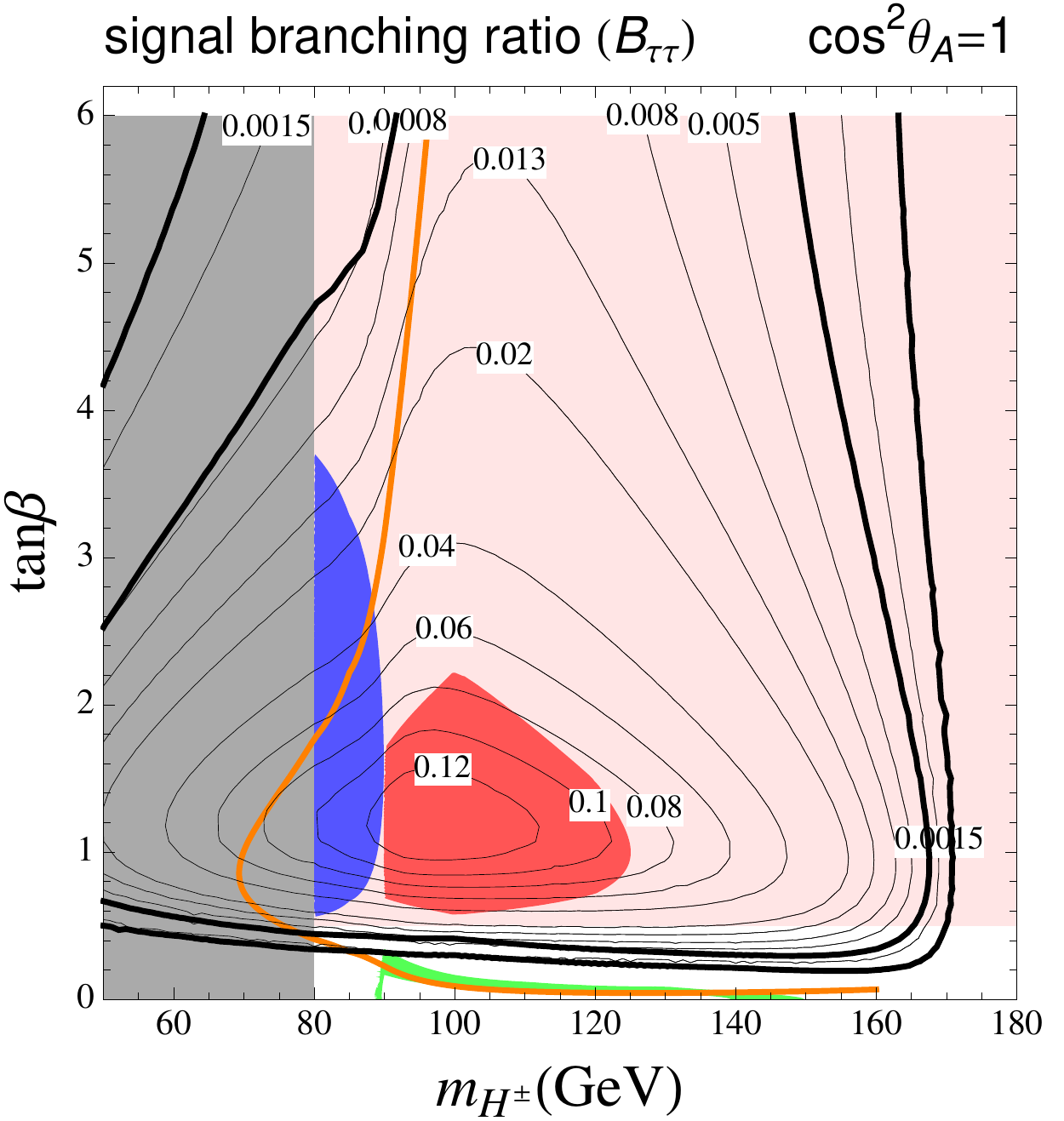}
\includegraphics[width=0.4 \linewidth]{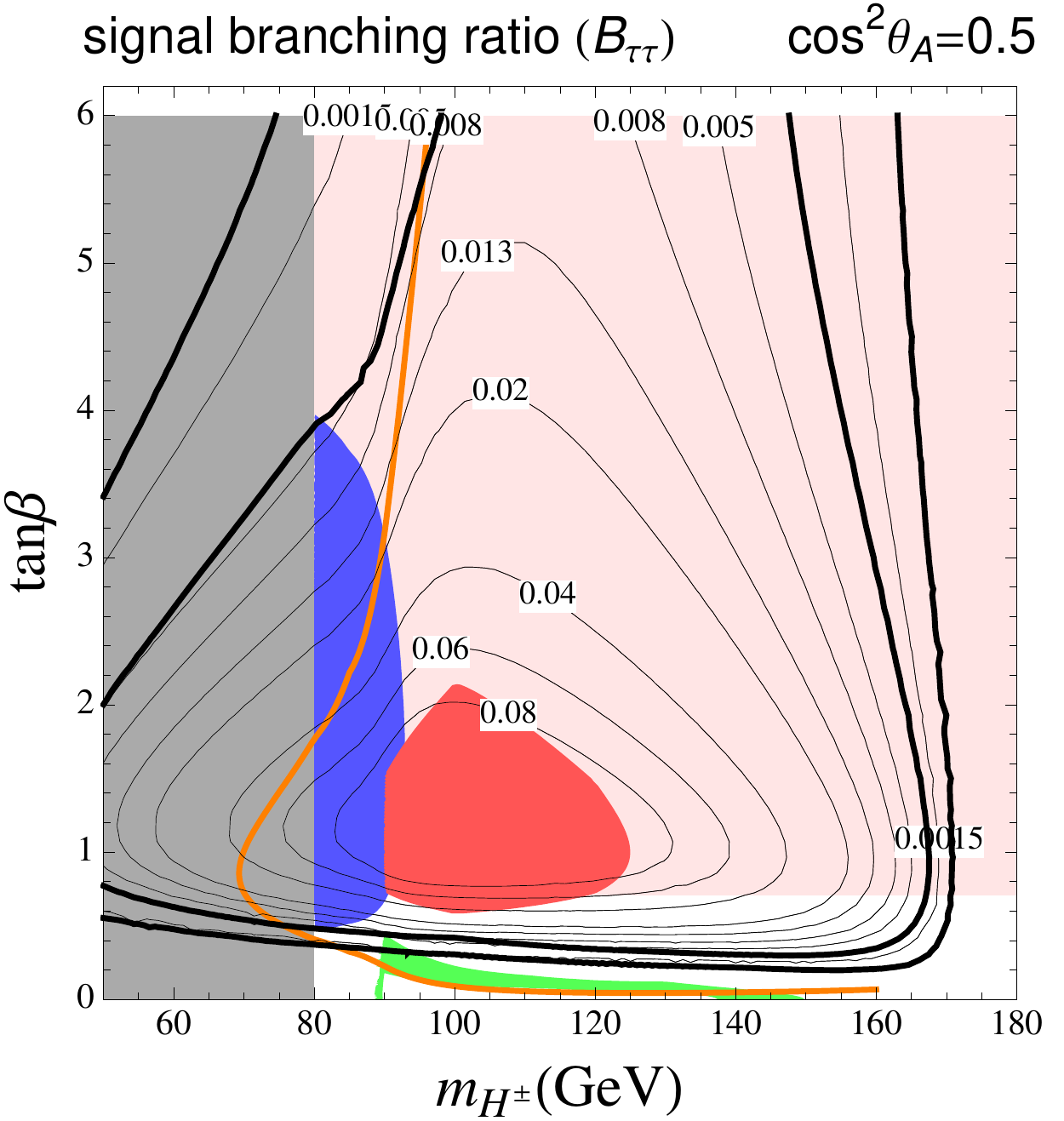}
\includegraphics[width=0.4 \linewidth]{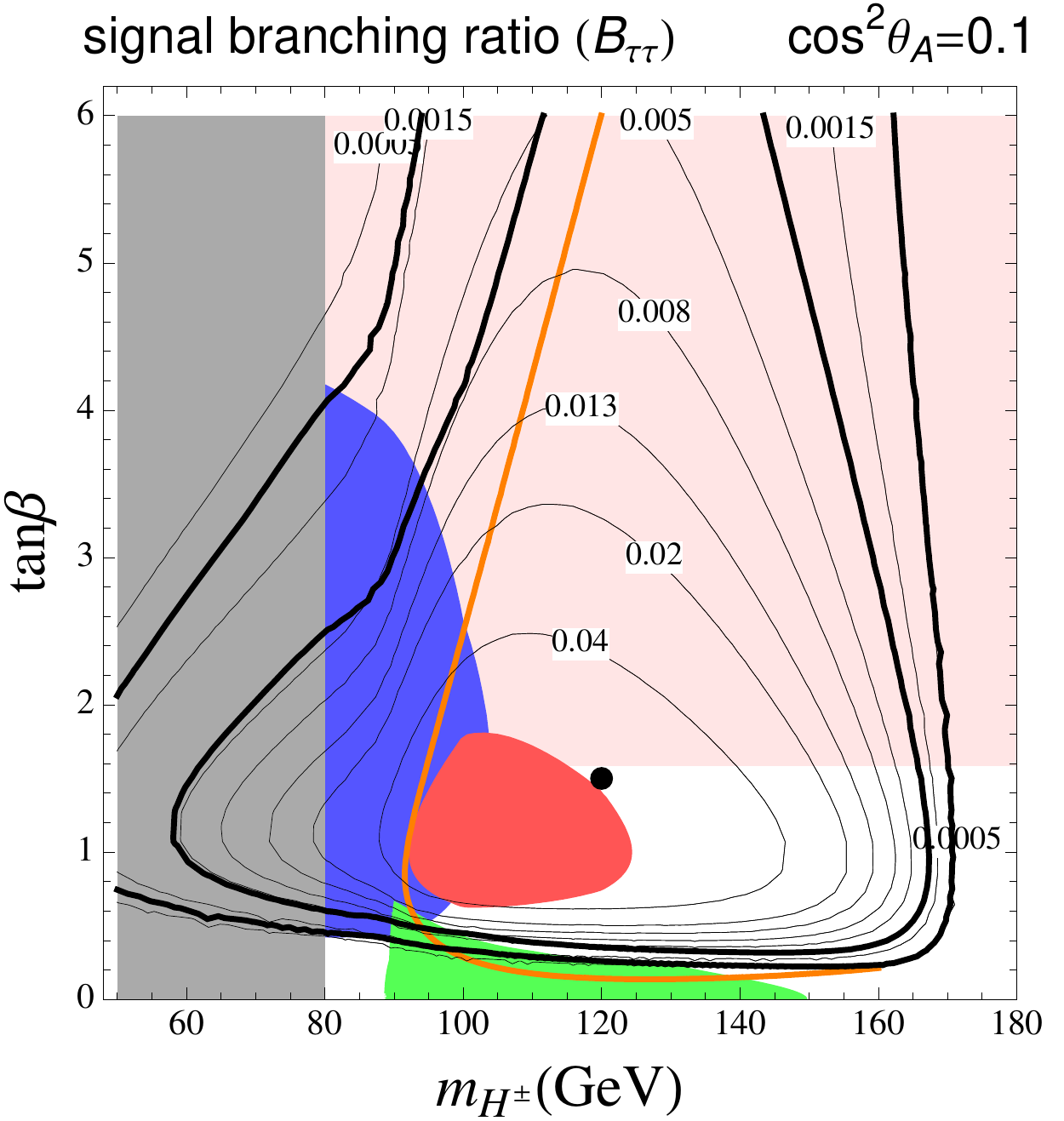}
\includegraphics[width=0.4 \linewidth]{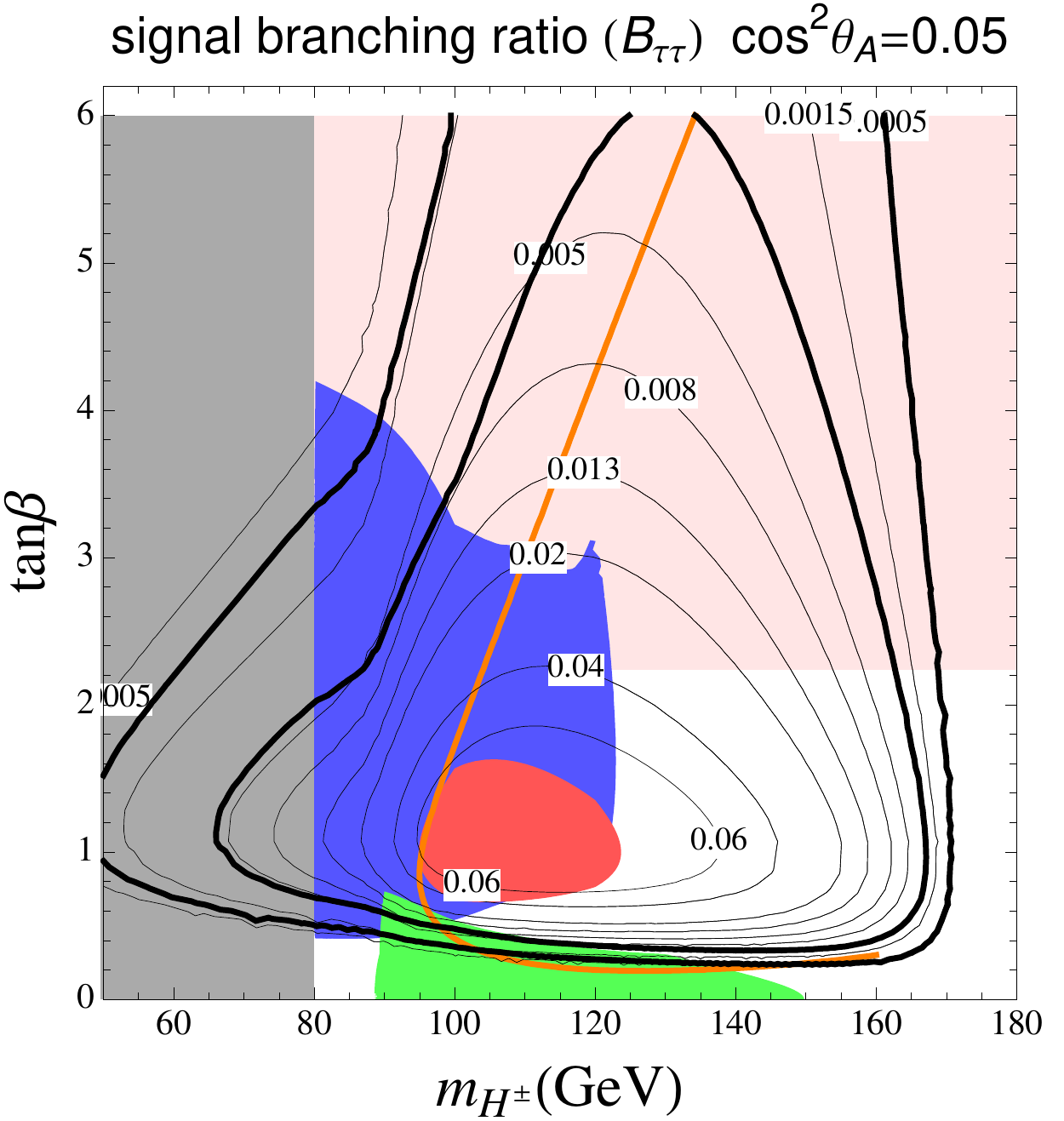}
\end{center}
\vskip -0.7cm
\caption{\it Contours of constant ${\rm BR} (t\to H^+ b) \times {\rm BR} (H^+\to W^+ A) \times {\rm BR} (A\to \tau\tau)$ values. Contours corresponding to the $A\to \mu\mu$ mode are easily obtained using the $m_\mu^2/m_\tau^2 = 3.53 \times 10^{-3}$ rescaling factor. For different values of $\cos^2\theta_A$, the gray, blue and green regions have the same meaning as in Fig.~\ref{fig:mhtb0}. The red regions are excluded by a direct $t\to b H^+ \to b W^+ A \to b W^+ \tau^+ \tau^-$ search at CDF~\cite{Aaltonen:2011aj}. The pink areas correspond to the constraints from Upsilon decays discussed in Sec.~\ref{sec:upsilon} ($\tan\beta \cos \theta_A \lesssim 0.5$). The region to the right of the single orange contour has ${\rm BR} (H^+\to W^+ A) \geq 50\%$. The black dot is the point that we consider in Sec.~\ref{sec:lhc}. The inner and outer thicker contours are the LHC reach with $20\; {\rm fb}^{-1}$ at $8 \; {\rm TeV}$ and $40\; {\rm fb}^{-1}$ at $14 \; {\rm TeV}$, respectively. \label{fig:mhtb1}}
\end{figure}

\begin{figure}
\begin{center}
\includegraphics[width=0.4 \linewidth]{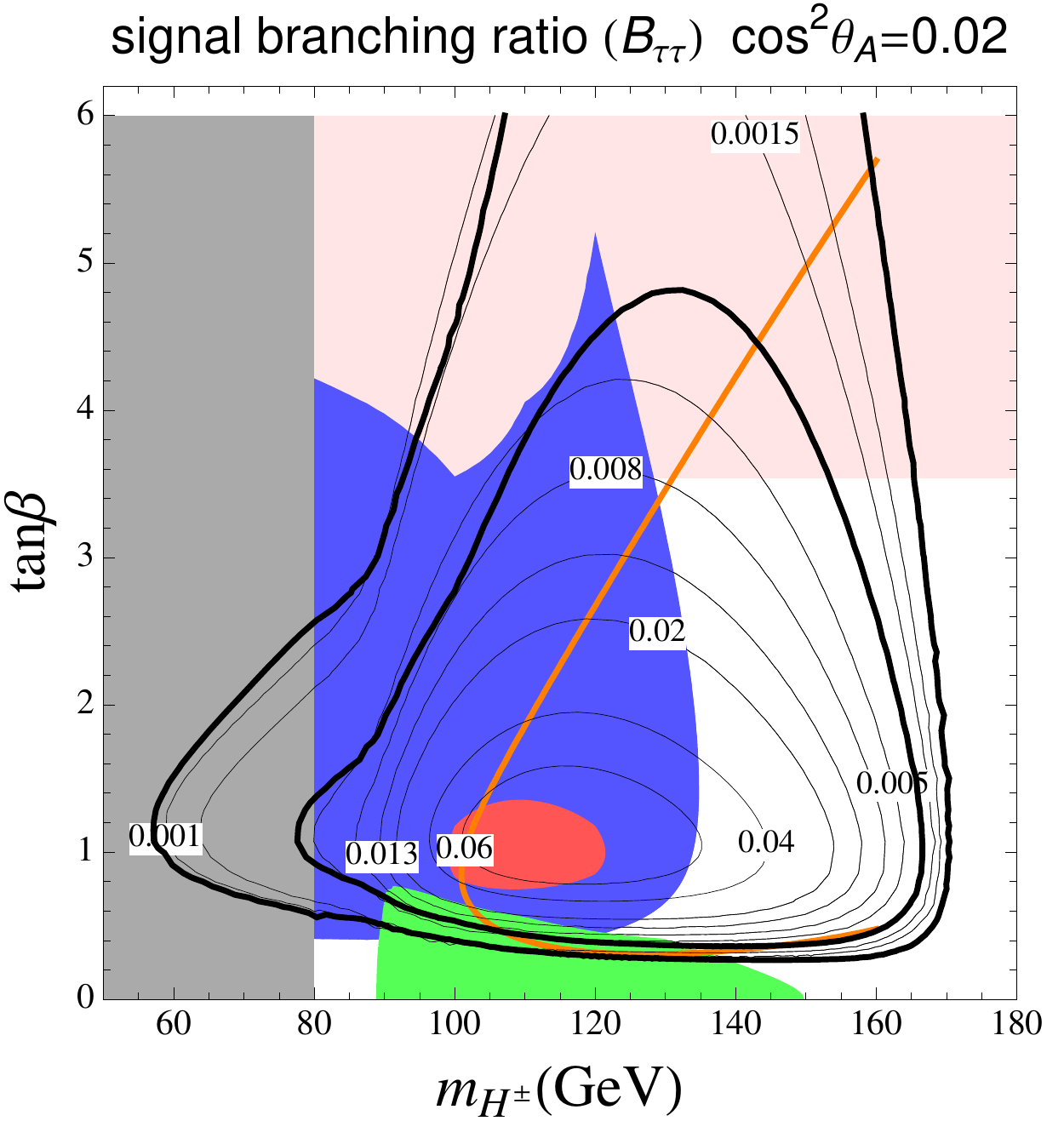}
\includegraphics[width=0.4 \linewidth]{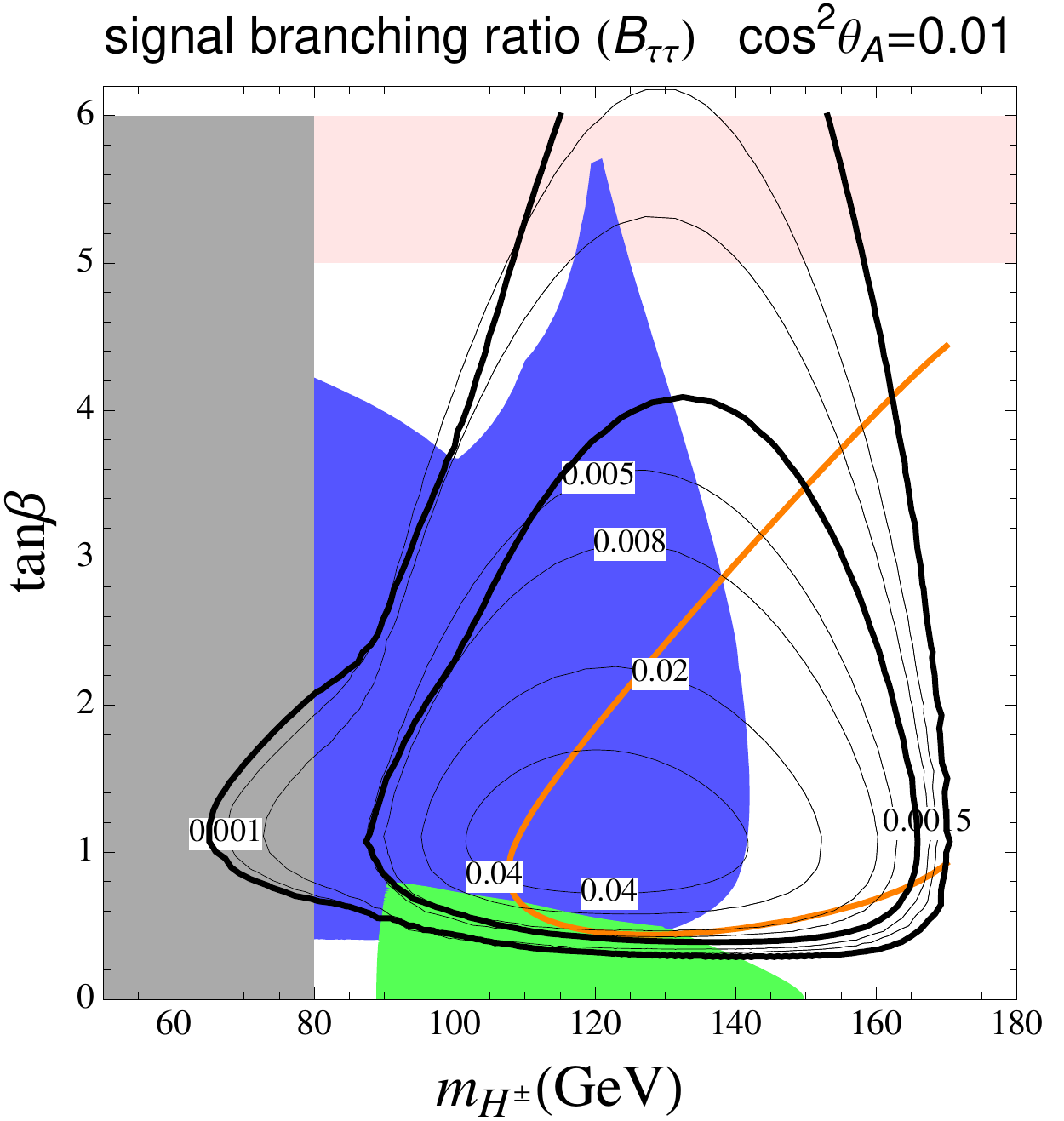}
\includegraphics[width=0.4 \linewidth]{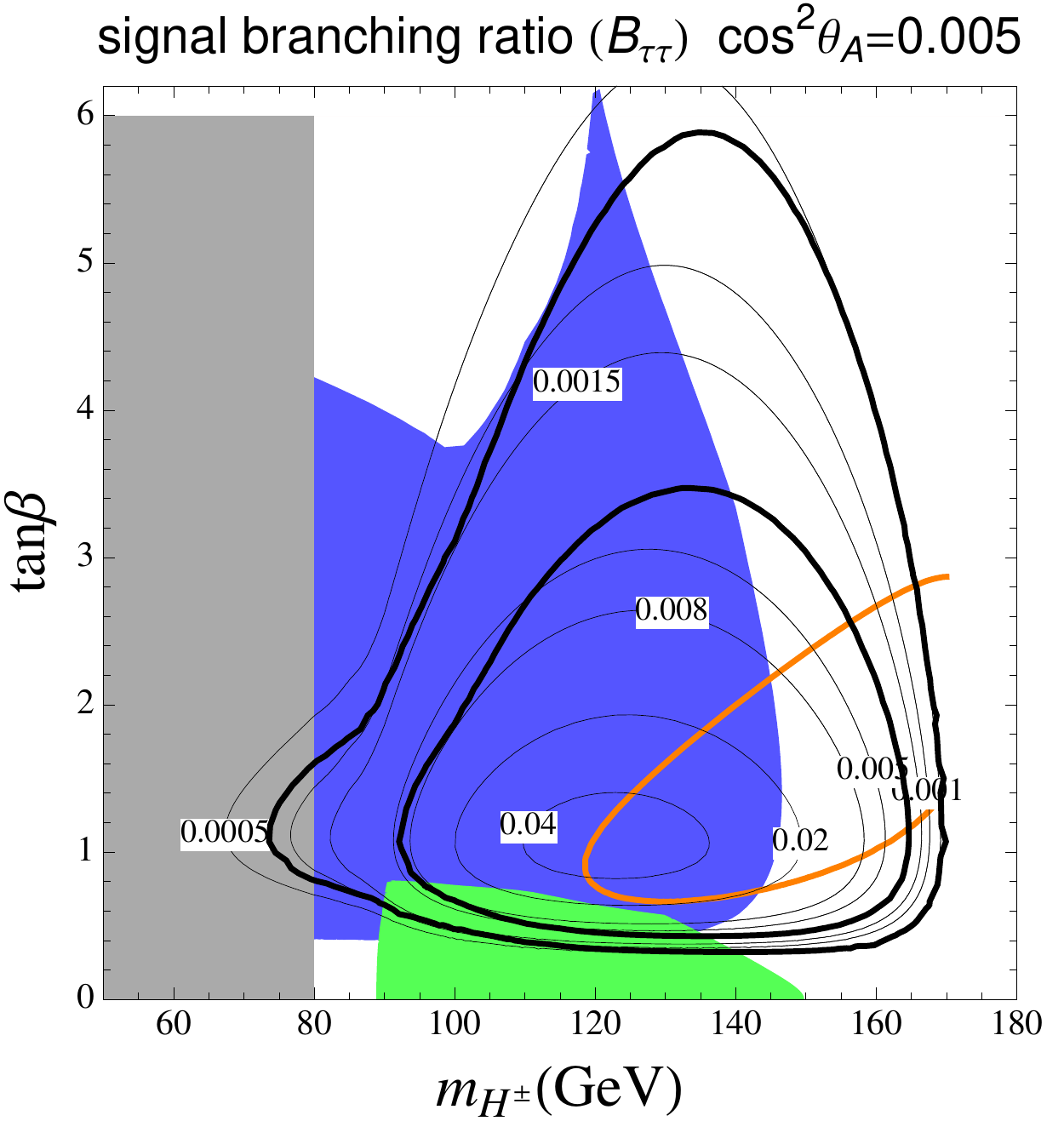}
\includegraphics[width=0.4 \linewidth]{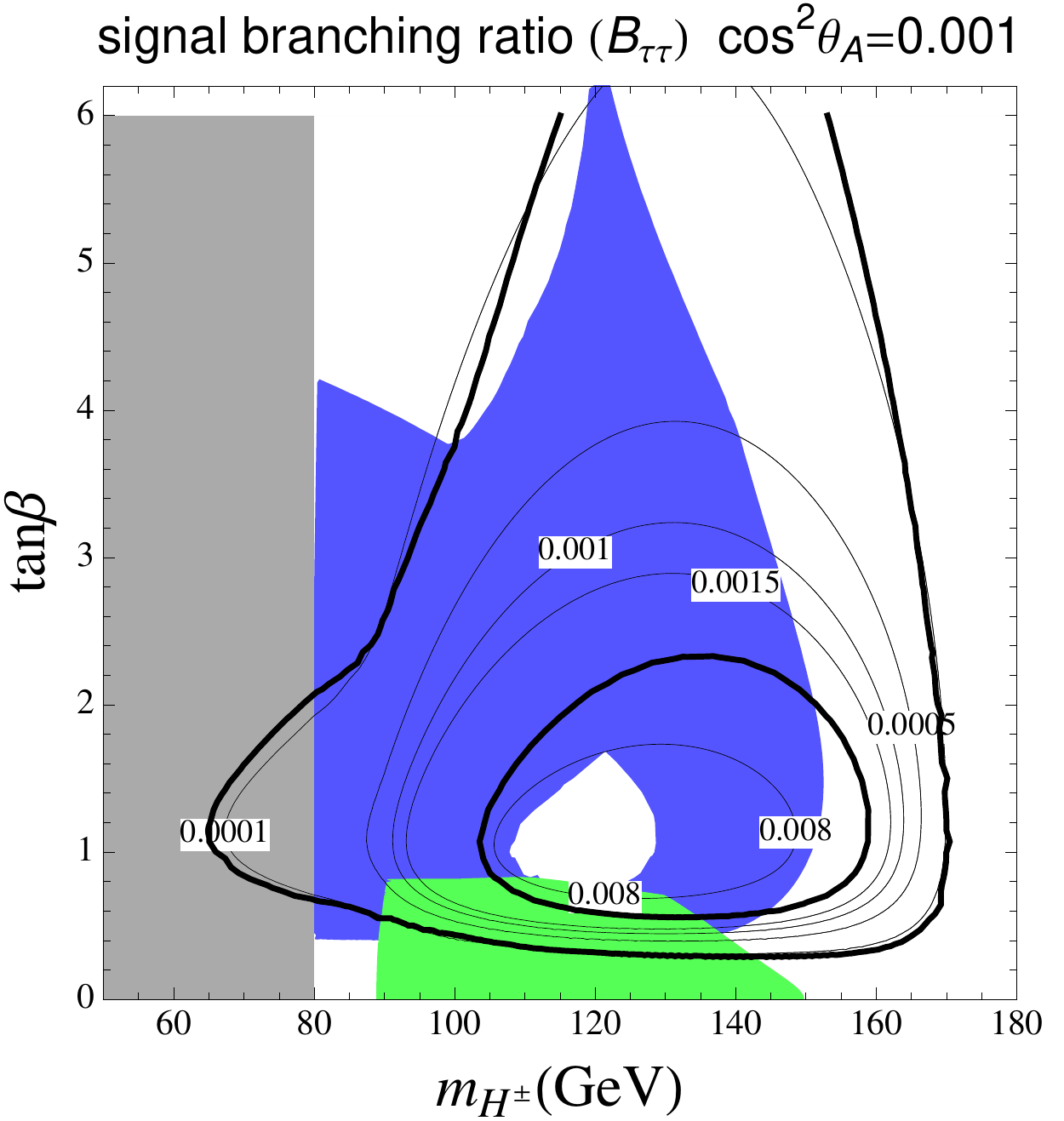}
\end{center}
\vskip -0.7cm
\caption{See the caption in Fig.~\ref{fig:mhtb1}. For $\cos^2\theta_A < 0.01$ the CDF constraint disappears. For $\cos^2\theta_A < 0.005$, the $H^+ \to W^+ A$ branching ratio is always smaller than 50\%. \label{fig:mhtb2} }
\end{figure}

If we allow a non-vanishing mixing between the singlet and doublet CP-odd Higgs bosons ($\cos^2\theta > 0$) the regions excluded by the Tevatron and LHC searches gradually shrink because the branching ratios $H^\pm \to (\tau\nu, \bar c s)$ decrease as $\cos^2\theta_A$ increases. In Figs.~\ref{fig:mhtb1} and \ref{fig:mhtb2} we show how these excluded regions change for non--zero values of the $\theta_A$ mixing angle. For $\cos^2\theta_A > 0$ some portions of the $(m_{H^\pm},\tan\beta)$ plane are excluded by the CDF~\cite{Aaltonen:2011aj} search for $t\to b H^+ \to b W^+ A \to b W^+ \tau^+ \tau^-$ that constrains the product ${\rm BR} (t\to H^+ b) \times {\rm BR} (H^+\to W^+ A) \times {\rm BR} (A\to \tau\tau)$. The size of these excluded regions (shaded red) depends on $\cos^2\theta_A$; a smaller doublet fraction implies a lower $H^\pm \to W^{(*)} A$ branching ratio and therefore a lower experimental reach. Upsilon decay studies at BaBar and direct CP-odd Higgs searches at ATLAS and CMS (see the discussion at the end of Sec.~\ref{sec:upsilon}) roughly exclude the shaded pink regions. These constraints are extremely strong for large $\cos^2\theta_A$ but weaken rapidly with decreasing $\cos^2\theta_A$. The area to the right of the orange contours in each plot corresponds to ${\cal B} (H^\pm \to W^{(*)} A) \geq 50\%$ (see also Fig.~\ref{fig:HWA}) and helps visualizing the regions in which our search strategy is expected to be the most sensitive. The black contours in Figs.~\ref{fig:mhtb1} and \ref{fig:mhtb2} correspond to constant ${\cal B}_{\tau\tau}$ (the corresponding ${\cal B}_{\mu\mu}$ values are easily obtained by multiplying by $m_\mu^2/m_\tau^2 \simeq 3.5 \times 10^{-3}$). The inner and outer thicker black contours correspond to the LHC expected sensitivity with $20\; {\rm fb}^{-1}$ at $8 \; {\rm TeV}$ and $40\; {\rm fb}^{-1}$ at $14 \; {\rm TeV}$, respectively. A detailed discussion on the expected LHC reach and on the derivation of this curves is presented in the next section (the impatient reader can skip to Sec.~\ref{sec:sensitivity} and Fig.~\ref{fig:reach}).

Recently CMS presented a search for new physics in multilepton events that is very similar in scope to the one proposed here~\cite{Chatrchyan:2012mea}. Unfortunately the results of this experimental analysis do not appear to constrain the parameter space of the class of models we consider. Events with invariant mass of opposite-sign same-flavor (i.e. $\ell^+\ell^-$) leptons below 12 GeV are rejected in order to eliminate backgrounds from charmonium and bottomonium decays. In addition, the isolation criterion adopted in the search removes other combinations of trilepton events originating from our signal. We will discuss the CMS search in more detail in Sec.~\ref{sec:sensitivity}.


%
\section{LHC study}
\label{sec:lhc}
In this section we discuss the LHC cross section for the signal in Eq.~(\ref{signal}) and the cuts that we recommend in order to reduce the irreducible tri--lepton background. We conclude with a study of the signal to background ratio.

\subsection{Collider signature}
\label{sec:signature}
\begin{figure}[t]
\begin{center}
\includegraphics[width=0.49 \linewidth]{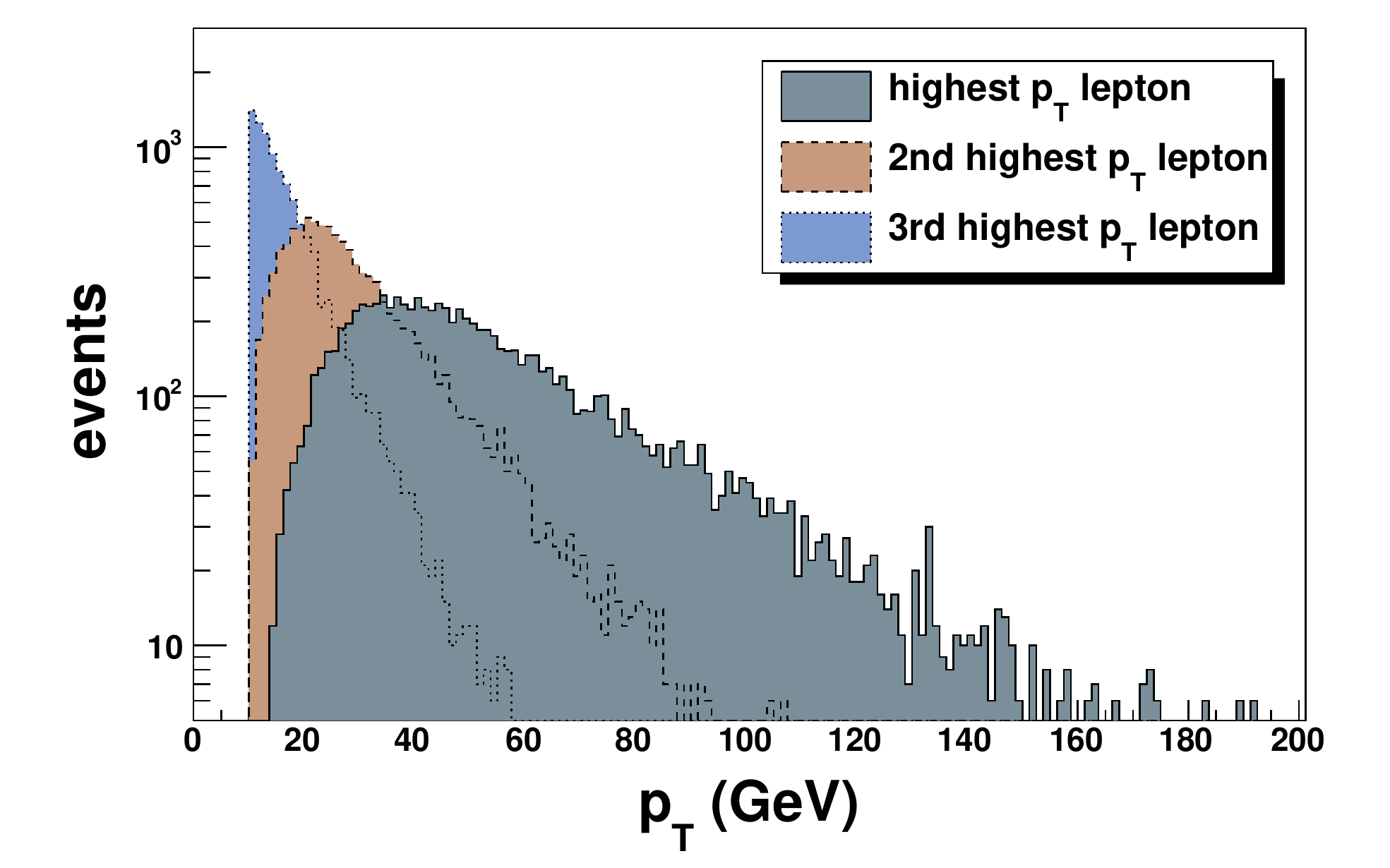}
\includegraphics[width=0.49 \linewidth]{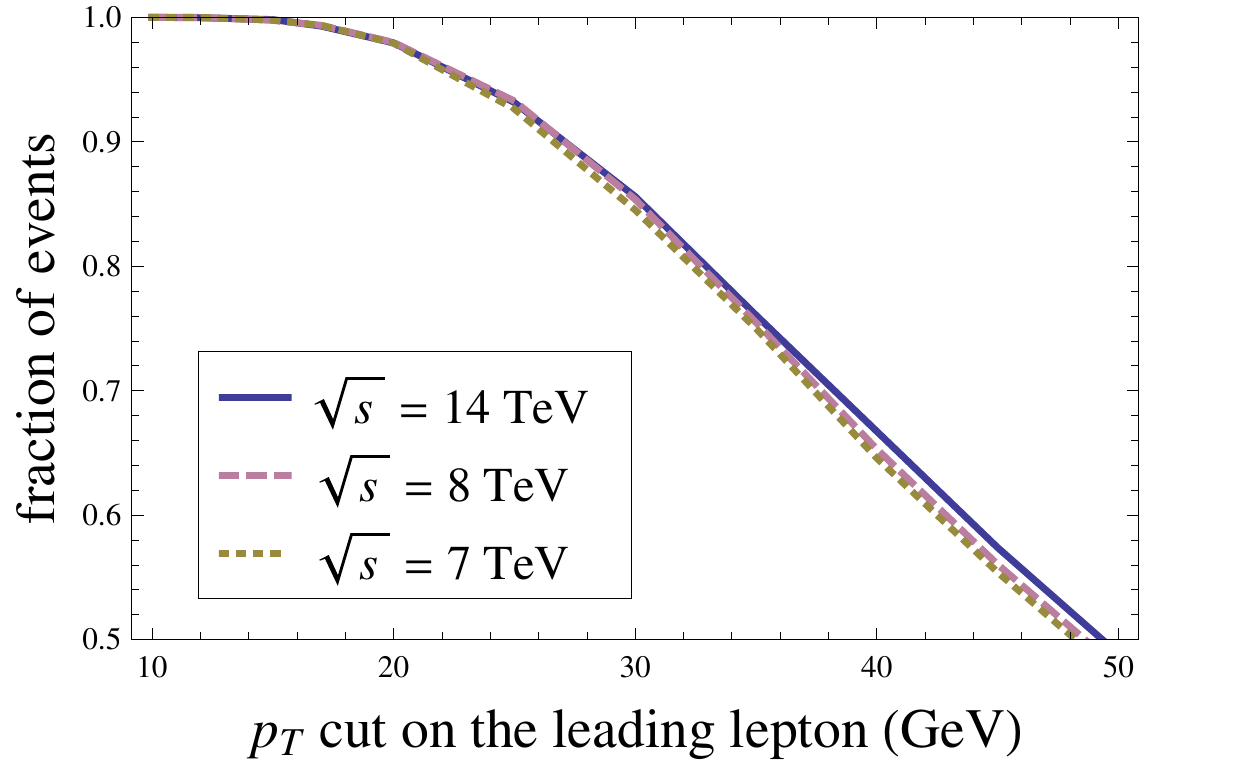}
\end{center}
\vskip -0.7cm
\caption{\it Left panel: $p_T$ distribution of the three highest $p_T$ leptons in the signal for $\sqrt{s} = 14 \; {\rm TeV}$ in a $10^4$ events sample. The cut on the minimum $p_T$ of all leptons is 10 GeV. Right panel: Fraction of events that survive an additional cut on the highest $p_T$ lepton. \label{fig:ptfrac}}
\end{figure}

Our signal can be extracted from the following distinct topologies (in parenthesis we specify the CP-odd Higgs decay that can lead to the final state):
\begin{align}
pp &\to \mu^+\mu^- e \; X  \;\;\; (A^0\to (\mu\mu,\tau\tau)) \label{mme}\\
pp &\to \mu^+\mu^- \mu \; X  \;\;\; (A^0\to (\mu\mu,\tau\tau)) \label{mmm}\\ 
pp &\to e^+e^- e \; X  \;\;\; (A^0\to \tau\tau) \label{eee}\\
pp &\to e^+ e^- \mu \; X \;\;\; (A^0\to \tau\tau) \label{eem}\\
pp &\to e^\pm e^\pm \mu^\mp \; X \;\;\; (A^0\to \tau\tau) \label{eemSS} \\
pp &\to \mu^\pm \mu^\pm e^\mp \; X \;\;\; (A^0\to \tau\tau) \label{mmeSS}
\end{align}
where $e$ and $\mu$ indicate an electron or muon with either charge. In Eqs.~(\ref{eemSS}) and
(\ref{mmeSS}) only upper or lower charges are allowed. In order to suppress the irreducible tri--lepton background (see the discussion in Sec.~\ref{sec:bkgd}) we require lepton isolation\footnote{We impose isolation by requiring absence of hadronic activity in a cone $\Delta R < 0.4$ around each lepton.} and, in addition, we impose the following cuts on the three highest $p_T$ leptons ($\ell=e,\; \mu$):
\begin{align}
& p_T (\ell) > 
\begin{cases}
20 \; {\rm GeV} & \hbox{highest $p_T$ lepton} \cr
10 \; {\rm GeV} & \hbox{remaining leptons} \cr
\end{cases}
\label{cuts1}\\
&\eta (\ell) < 2.5  \label{cuts2}\\
&m_{\ell^+ \ell^-} \in [0,12] \; {\rm GeV} \; , \label{cuts3}
\end{align}
where the choice of leptons used to calculate the invariant mass depends on the actual signal topology. Heavy hadron production with subsequent semileptonic or double--semileptonic decays is potentially a major source of irreducible trilepton background. Fortunately the $p_T$ spectrum of these leptons is quite soft and a $p_T > 20 \; {\rm GeV}$ cut on the highest $p_T$ lepton is enough to severely suppress this kind of background leptons. We will present a through discussion of this source of background in Sec.~\ref{sec:bkgd}. The impact of this cut on our signal is quite minimal (it reduces the cross section by about 5\% with respect to a flat $p_T(\ell) > 10 \; {\rm GeV}$ cut). In the left and right panels of Fig.~\ref{fig:ptfrac} we show the $p_T$ spectra of the three highest $p_T$ leptons in our signal and the fraction of events that survive after increasing the cut on the highest $p_T$ lepton, respectively.

Some considerations are common to all topologies. In trilepton events with an underlying $A\to\mu\mu$ decay we expect at least two muons in the selection which most likely are those produced in the $A$ decay. In fact, leptons produced in semileptonic $b$-hadron decays are heavily suppressed because of hadronic isolation: the probability of obtaining an isolated high $p_T$ lepton from a decaying $b$--hadron is about $1/200$ and is much smaller than the $W$ leptonic branching ratio ($\sim 10\%$). Moreover, it is possible that the second $W$ also undergoes a leptonic decay and that this lepton has larger $p_T$ than one of the muons in the $A$ decay. There are three sources of suppression of this possibility: the leptonic $W$ branching ratio ($\sim 10\%$), the $p_T$ cut and the di-muon invariant mass cut (remember that in this case the di-muon invariant mass will not typically be close to $m_A$). Therefore, this effect is very tiny and will be neglected. Another feature of the $A\to \mu\mu$ mode is that, in the three muon final state, the opposite sign muons that reconstruct the CP-odd Higgs are, in the majority of cases, the single opposite sign muon and the lowest $p_T$ same sign muon. For the benchmark point we consider this selection criterion picks up 80\% of the correct di-muon pairs.

In the $A\to\tau\tau$ case, the situation is more complicated because the $\tau$'s can decay into muons or electrons. We can again safely neglect simultaneous semileptonic decays of both $W$ bosons because very few events would have a di-lepton invariant mass that falls in the $[0,12]\; {\rm GeV}$ range. If both $\tau$'s decay to muons, and one of the $W$'s decays into an electron, the di-muon system originates entirely from the $A$ decay and its invariant mass is going to be strictly smaller than $m_A$. The interesting possibility of having each tau decaying to a different lepton, allows for the same sign and same flavor topologies in Eqs.~(\ref{eemSS}) and (\ref{mmeSS}) whose background turns out to be extremely small. As in the $A\to \mu\mu$ case, the opposite charge leptons that come from the tau decays can be identify by selecting the lowest $p_T$ same sign lepton (we don't distinguish between $e$ and $\mu$ as both can appear in tau decays). For the benchmark point we adopt this criterion picks up 89\% of the correct di-lepton pairs.

Let us now briefly discuss individual features of the signal topologies we consider:
\begin{itemize}
\item $\mu^+ \mu^- e$. Events coming from an underlying $A\to\mu\mu$ decay are trivially extracted by considering the di-muon invariant mass. The $A\to\tau\tau$ channel is more complex because of the possibility of having the two taus decaying to different leptons: in this case the di-muon invariant mass will not peak at all below $m_A$ and these effects are tiny. In conclusion, if the three highest $P_T$ leptons are $\mu\mu e$, we will simply plot the di-muon invariant mass.
\item $\mu^+\mu^-\mu$. In this case there is an intrinsic ambiguity in the choice of the ``correct'' di-muon pair. As explained above, by considering the lowest $p_T$ same sign muon we select the correct di-muon pair with a 80\% probability.
\item $e^+e^- e$ and $e^+e^-\mu$. Decays with an underlying $A\to\mu\mu$ do not sizably contribute to this selection. In fact, the $e^+e^-\mu$ final state could be obtained with two $W$ decays to electrons; the requirement of low di--lepton invariant mass severely suppresses this contribution. The $e^+e^- e$ final state is, instead, almost impossible to produce in $A\to\mu\mu$ events. In both modes each of the light leptons can originate from a tau. By selecting the lowest $p_T$ same sign lepton we reconstruct the correct di-lepton pair with a 89\% probability.
\item $\mu^\pm\mu^\pm e^\mp$ and $e^\pm e^\pm \mu^\mp$. These signals are produced by an underlying $A\to \tau\tau$ decay in which the two taus decay to different flavor leptons. The leptons originating in the $A$ decay can be either of the two opposite charge combinations. Therefore,the signal extraction is very similar to the $e^+e^- e$ and $e^+e^-\mu$ cases. The main difference is that the background for two same-sign same-flavor leptons is severely reduced.
\end{itemize}

\begin{figure}[ht]
\begin{center}
\includegraphics[width=0.49 \linewidth]{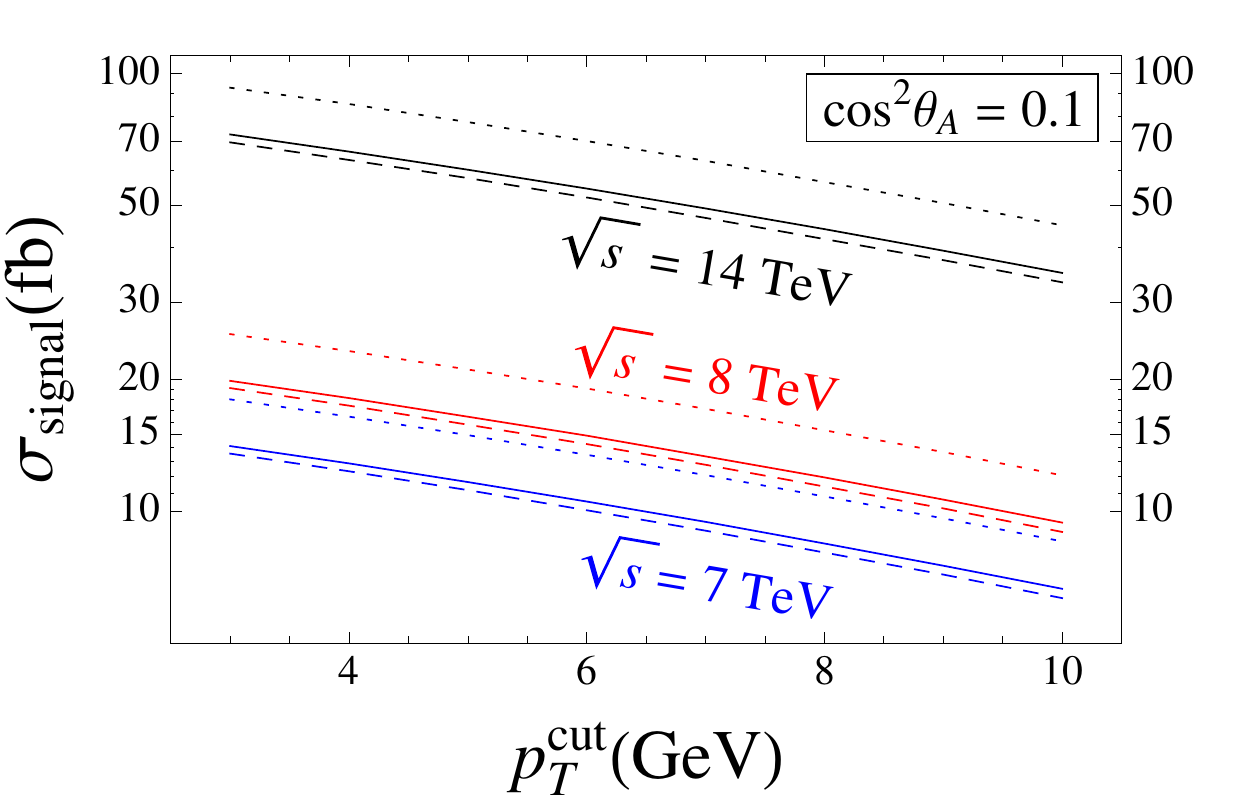}
\includegraphics[width=0.49 \linewidth]{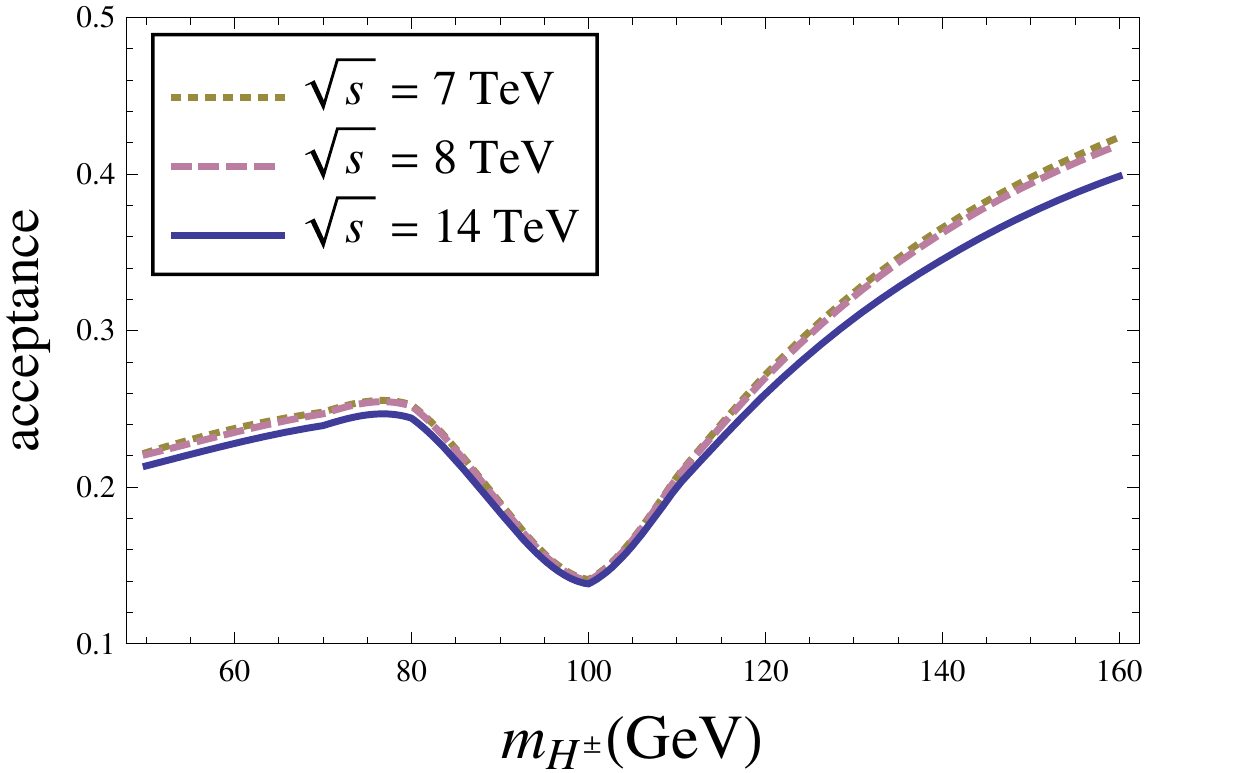}
\end{center}
\vskip -0.7cm
\caption{\it Left pane: Dependence of the signal cross section on the $p_T$ cut on the three leptons. Dotted, solid and dashed lines correspond to $m_A = 5,\; 8, \; 10 \; {\rm GeV}$, respectively. Right panel: Dependence of acceptance $r_{\mu\mu}$ on the charged Higgs mass for different center--of--mass energies. \label{fig:pt}}
\end{figure}

Before applying any cut to the final state muons, the signal cross sections can be calculated exactly in terms of the parameters of the model and the $t\bar t$ cross section. As a benchmark point we take $m_{H^\pm} = 120 \; {\rm GeV}$, $\tan\beta=1.5$, $m_A=8 \; {\rm GeV}$, and $\cos^2 \theta_A = 0.1$, see the black dot in Fig.~\ref{fig:mhtb1}, for which we have
\begin{align}
{\rm BR} (t\to b W) &= 88.7\%  \; , \label{twb}\\
{\rm BR} (t\to H^\pm W) &=11.3\% \; , \\
{\rm BR} (H^\pm \to A W) &= 99.3\% \; . 
\end{align}
We also take the NNLL $pp\to t\bar t$ cross sections from Ref.~\cite{Beneke:2012eb,Beneke:2012gn}:
\begin{align}
\sigma (pp\to t\bar t ) & = 
\begin{cases}
162.4^{+6.7}_{-6.9} {}^{+7.3}_{-6.8}\; {\rm pb} & {\rm for \;} E_{\rm CM} = 7  \; {\rm TeV} \cr
231.8^{+9.6}_{-9.9} {}^{+9.8}_{-9.1}\; {\rm pb} & {\rm for \;} E_{\rm CM} = 8 \; {\rm TeV} \cr
896^{+40}_{-37} {}^{+65}_{-64}\; {\rm pb} & {\rm for \;} E_{\rm CM} = 14 \; {\rm TeV} \cr
\end{cases} \; .
\end{align}
Using the above inputs and cross sections we obtain ${\cal B}_{\mu\mu} = 3.6 \times 10^{-4}$, ${\cal B}_{\tau\tau} = 0.092$, and the following signal cross sections:
\begin{align}
\sigma (pp\to A \ell X \to \mu \mu\ell X) &= \; 4 \times \sigma (pp\to t\bar t) \times  {\cal B}_{\mu\mu} \times {\rm BR} (W \to \ell \nu)  \nonumber\\
 &=  
\begin{cases}
24.5 \; {\rm fb} & {\rm for \;} E_{\rm CM} = 7 \; {\rm TeV} \cr
35.0 \; {\rm fb} & {\rm for \;} E_{\rm CM} = 8 \; {\rm TeV} \cr
135.1 \; {\rm fb} & {\rm for \;} E_{\rm CM} = 14 \; {\rm TeV} \cr
\end{cases} \; ,
\end{align}
\begin{align}
\sigma (pp\to A \ell X \to \tau\tau\ell X \to \ell\ell\ell X) 
&  = \; 4 \times \sigma (pp\to t\bar t) \times  {\cal B}_{\tau\tau} \times {\rm BR} (\tau\to\ell\nu\bar\nu)^2  \times {\rm BR} (W \to \ell \nu)  \nonumber\\
 &    =  
\begin{cases}
187.1 \; {\rm fb} & {\rm for \;} E_{\rm CM} = 7 \; {\rm TeV} \cr
267.0 \; {\rm fb} & {\rm for \;} E_{\rm CM} = 8 \; {\rm TeV} \cr
1032.1 \; {\rm fb} & {\rm for \;} E_{\rm CM} = 14 \; {\rm TeV} \cr
\end{cases} \; ,
\end{align}
where $\ell$ is either $e$ or $\mu$. The factor of 4 comes from having two top quarks that can decay to $H^\pm b$ and two $W$ bosons that can produce the charged lepton. In order to reduce the background to manageable levels, we require the cuts in Eqs.~(\ref{cuts1}-\ref{cuts3}). Using SHERPA v.1.4.0~\cite{Gleisberg:2008ta, Schumann:2007mg, Krauss:2001iv, Gleisberg:2008fv}, we extract the ratio of the signal cross section calculated with and without these cuts. We obtain the following acceptances:
\bea
r_{\mu\mu} & = & \frac{\sigma (pp\to A \ell X \to \mu \mu\ell X)_{\rm SHERPA}^{\rm cut}}{\sigma (pp\to A \ell X \to \mu \mu\ell X)_{\rm SHERPA}} = 
\begin{cases}
0.27 & {\rm for \;} E_{\rm CM} = 7 \; {\rm TeV}  \cr
0.27 & {\rm for \;} E_{\rm CM} = 8 \; {\rm TeV} \cr
0.26 & {\rm for \;} E_{\rm CM} = 14 \; {\rm TeV} \cr
\end{cases} \; ,
\label{rescalemu}\\
r_{\tau\tau} & = & \frac{\sigma ( pp\to A \ell X \to \tau\tau\ell X \to \ell\ell\ell X )_{\rm SHERPA}^{\rm cut}}{\sigma ( pp\to A \ell X \to \tau\tau\ell X \to \ell\ell\ell X )_{\rm SHERPA}} = 
\begin{cases}
0.034 & {\rm for \;} E_{\rm CM} = 7 \; {\rm TeV} \cr
0.036 & {\rm for \;} E_{\rm CM} = 8 \; {\rm TeV} \cr
0.039 & {\rm for \;} E_{\rm CM} = 14 \; {\rm TeV} \cr
\end{cases} 
 \; . \label{rescaletau}
\eea
The final signal cross sections that we obtain are:
\bea
\sigma (pp\to A \ell X \to \mu \mu\ell X)^{\rm cut} & = & 
\begin{cases}
6.7 {\rm fb}& {\rm for \;} E_{\rm CM} = 7 \; {\rm TeV} \cr
9.4 \; {\rm fb} & {\rm for \;} E_{\rm CM} = 8 \; {\rm TeV} \cr
35.0  \; {\rm fb}& {\rm for \;} E_{\rm CM} = 14 \; {\rm TeV} \cr
\end{cases} 
\label{signalmu}\\
\sigma (pp\to A \ell X \to \tau\tau\ell X \to \ell\ell\ell X )^{\rm cut} &=&
\begin{cases}
6.4  \; {\rm fb}& {\rm for \;} E_{\rm CM} = 7 \; {\rm TeV} \cr
9.5 \; {\rm fb} & {\rm for \;} E_{\rm CM} = 8 \; {\rm TeV} \cr
40.3  \; {\rm fb}& {\rm for \;} E_{\rm CM} = 14 \; {\rm TeV}\cr
\end{cases} 
 \; . \label{signaltau}
\eea
The signal cross section (where we include only the $A\to\mu\mu$ channel) as a function of the lepton $p_T$ cut and for $m_A = 5,8,10 \; {\rm GeV}$ is shown in the left panel of Fig.~\ref{fig:pt} and Table~\ref{tab:signalcut}. Note that here we show the results we obtain imposing the same $p_T$ cut on all final state leptons. The cuts that we actually propose ($p_T > 20 \; {\rm GeV}$ for the leading lepton and $p_T > 10 \; {\rm GeV}$ for the other two) reduce the $p_T^{\rm cut} = 10 \; {\rm GeV}$ case by 5\% (this reduction is implemented in Eqs.~(\ref{rescalemu}--\ref{signaltau})). 

Finally, we investigate the dependence of the signal acceptance (defined as the fraction of events that survive the cuts we impose) on the model parameters. The only parameter whose impact on the acceptance is quite relevant is the charged Higgs mass. In the right panel of Fig.~\ref{fig:pt} we show the dependence of the acceptance $r_{\mu\mu}$ on $m_{H^\pm}$. The behavior of the acceptance in the range $80 \; {\rm GeV} < m_{H^\pm} < 100 \; {\rm GeV}$ is a result of the transition between the on-shell $H^\pm \to W^\pm A$ and the off-shell three-body  $H^\pm \to W^{\pm *} A \to f_i \bar f_j A$ decays. Just above the threshold the momentum of the CP-odd Higgs in the charged Higgs rest-frame is small. Below the threshold we have an effective three body decay and the mass of the charged Higgs is distributed amont the final state particles.

\begin{table}[ht]
\begin{center}
$$\sigma (pp\to A \ell X \to \mu \mu\ell X)^{\rm cut} \; {\rm [fb]}$$
\begin{tabular}{|cc||c|c|c|c|c|c|c|c|}\hline
$m_A$ & $\sqrt{s}$ & $p_T> 3$ & $p_T> 4 $ & $p_T> 5 $ & $p_T> 6 $ & $p_T> 7 $ & $p_T> 8 $ & $p_T> 9 $ & $p_T> 10 $ \\ 
\scriptsize [GeV] &  \scriptsize[TeV] & \scriptsize[GeV] &\scriptsize [GeV] &\scriptsize [GeV] & \scriptsize[GeV] & \scriptsize[GeV] &\scriptsize [GeV] & \scriptsize[GeV] & \scriptsize[GeV]  \\ 
\hline
\hline
 5 & 7 & 18.0 & 16.5 & 14.9 & 13.5 & 12.1 & 10.8 & 9.6 & 8.5 \\
 5 & 8 & 25.4 & 23.2 & 21.1 & 19.1 & 17.1 & 15.3 & 13.7 & 12.1 \\
 5 & 14 & 92.9 & 85.2 & 77.5 & 70.2 & 63.2 & 56.5 & 50.6 & 45.0 \\ \hline
 8 & 7 & 14.1 & 12.9 & 11.7 & 10.5 & 9.5 & 8.4 & 7.5 & 6.7 \\
 8 & 8 & 19.9 & 18.1 & 16.5 & 14.9 & 13.4 & 12.0 & 10.6 & 9.4 \\
 8 & 14 & 72.6 & 66.4 & 60.3 & 54.6 & 49.2 & 44.1 & 39.4 & 35.1 \\ \hline
 10 & 7 & 13.6 & 12.3 & 11.2 & 10.1 & 9.0 & 8.1 & 7.2 & 6.3 \\
 10 & 8 & 19.1 & 17.4 & 15.8 & 14.2 & 12.8 & 11.4 & 10.2 & 9.0 \\
 10 & 14 & 69.7 & 63.5 & 57.7 & 52.1 & 46.8 & 41.9 & 37.5 & 33.4 \\ \hline
\multicolumn{2}{|c||}{$r_{\mu\mu}$}  & 0.54 & 0.49 & 0.45 & 0.40 & 0.36 & 0.33 & 0.29 & 0.26 \\ \hline
\end{tabular}
\caption{\it Cross sections (in fb) for $pp\to A \ell X \to \mu \mu\ell X$. In order to compare different $p_T$ cuts, we require the three charged leptons to have the same minimum $p_T$. Our optimal selection requires a hard lepton with $p_T > 20 \; {\rm GeV}$ and two softer ones with $p_T > 10 \; {\rm GeV}$. The last row of the table gives the acceptance $r_{\mu\mu}$; this quantity is essentially independent of $m_A$ and $\sqrt{s}$.  \label{tab:signalcut}}
\end{center}
\end{table}

\subsection{Background}
\label{sec:bkgd}
We classify the background to our signal (three charged leptons, with low invariant di-lepton mass and relatively low $p_T$) into two main groups:
\begin{itemize}
\item {\bf Electroweak.} In this group the underlying hard scattering has at least one electroweak (EW) gauge boson ($W$, $Z$ or $\gamma^*$) in the final state. We can further divide these processes according to the number of leptons that originate in single and double semileptonic heavy hadron decays. The only process in which all three leptons originates from EW bosons is $pp\to W Z/\gamma^*$, and it has a negligible contribution to our background (see Table~\ref{tab:background}). Processes in which only one lepton is produced in heavy meson decays are $pp\to (bZ,\; c Z,\; b\bar bZ,\; c\bar cZ,\; t\bar t,\; t W)$ and represent the dominant source of EW background. Finally the processes $pp\to (t\bar b,\; b\bar b W,\; c\bar c W)$ require two single or one double semileptonic heavy hadron decay. 
\item {\bf QCD.} In this group all three leptons come from single or double semileptonic decays of heavy final state hadrons. The underlying processes in this case are $p\to (b\bar b,\; c\bar c,\;b\bar bb\bar b,\; c\bar cc\bar c,\; b\bar bc\bar c )$.
\end{itemize}
Electroweak background topologies have been extensively studied in Ref.~\cite{Sullivan:2008ki}. The results presented in the second column of Table II of Ref.~\cite{Sullivan:2008ki} have been calculated for $\sqrt{s}$ = 14 TeV. We use Madgraph 5.2~\cite{Alwall:2011uj} to calculate appropriate rescaling factors for each background and obtain the corresponding results at $\sqrt{s}$ = 7 and 8 TeV. Moreover, we limit the integration to the $m_{\ell\ell} \in [0,12] \; {\rm GeV}$ range\footnote{The complete histograms of the various contributions to trilepton production at low di--lepton invariant mass have been kindly provided to us by Zack Sullivan.}. 

The estimates in Ref.~\cite{Sullivan:2008ki} combine together samples with $\ell = e, \mu$; therefore, in order to extract the backgrounds to the signatures in Eqs.~(\ref{mme}--\ref{mmeSS}) we have to apply appropriate combinatorial correction factors. We summarize our findings in Table~\ref{tab:background}. Here we give explicit cross sections for the $\mu^+ \mu^- e$ and $e^+ e^- \mu$ and specify the combinatorial factors that have to be applied in order to get the remaining topologies. Heavy hadron production in which the final state b- and c-hadrons decay semileptonically have not been studied in Ref.~\cite{Sullivan:2008ki} but are relevant in the phase space region we are interested in. The procedure we followed to estimate these purely hadronic backgrounds is presented below.
\begin{table}[t]
\begin{center}
\begin{tabular}{|c|l|c|c|c|c|c|c|c|c|c|}\hline
\multirow{2}{*}{} & mode & \multicolumn{3}{|c|}{$\mu^+ \mu^- e$, $e^+ e^- \mu$} & 
       \multicolumn{3}{|c|}{$\mu^+ \mu^- \mu$, $e^+ e^- e$} & 
       \multicolumn{3}{|c|}{$\mu^\pm \mu^\pm e^\mp$, $e^\pm e^\pm \mu^\mp$} \cr \cline{2-11}
\multirow{10}{*}{\rotatebox{90}{\mbox{EW (fb)}}} & $\sqrt{s}$ & 7 TeV & 8 TeV & 14 TeV & 7 TeV & 8 TeV & 14 TeV & 7 TeV & 8 TeV & 14 TeV \cr \hline
& $W Z/\gamma$ & 0.7 & 0.8 & 2.6 & \multicolumn{3}{|c|}{1} & \multicolumn{3}{|c|}{0} \cr 
& $bZ$ & 5.4 & 7.0 & 19.0 & \multicolumn{3}{|c|}{1} & \multicolumn{3}{|c|}{0} \cr
& $cZ$ & 1.4 & 1.7 & 4.7 & \multicolumn{3}{|c|}{1} & \multicolumn{3}{|c|}{0} \cr
& $bbZ$ & 3.7 & 4.7 & 12.2 & \multicolumn{3}{|c|}{1} & \multicolumn{3}{|c|}{0} \cr
& $ccZ$ & 2.1 & 2.6 & 6.4 & \multicolumn{3}{|c|}{1} & \multicolumn{3}{|c|}{0} \cr
& $t\bar t$ & 0.1 & 0.1 & 0.4 & \multicolumn{3}{|c|}{1/2} & \multicolumn{3}{|c|}{1/2} \cr
& $t W$ & 0.01 & 0.02 & 0.08 & \multicolumn{3}{|c|}{1/2} & \multicolumn{3}{|c|}{1/2} \cr
& $ t\bar b$ & $1.3 \times 10^{-4}$ & $ 1.5 \times 10^{-4}$ & $3/1 \times 10^{-4}$ & \multicolumn{3}{|c|}{1/2} & \multicolumn{3}{|c|}{1/2} \cr
& $b\bar b W $ & $1.3 \times 10^{-3}$ & $1.5 \times 10^{-3}$ & $2.6 \times 10^{-3}$ & \multicolumn{3}{|c|}{1/2} & \multicolumn{3}{|c|}{1/2} \cr
& $c\bar c W $ & $2.6 \times 10^{-5}$ & $2.9 \times 10^{-5}$ & $5.4 \times 10^{-5}$ & \multicolumn{3}{|c|}{1/2} & \multicolumn{3}{|c|}{1/2} \cr \hline
\multirow{5}{*}{\rotatebox{90}{\mbox{QCD (fb)}}} & $ b \bar b$ & 6.1 & 7.1 & 13.3 & \multicolumn{3}{|c|}{1/2} & \multicolumn{3}{|c|}{1/2} \cr
& $ c \bar c$ & 7.5 & 8.8 & 16.3 & \multicolumn{3}{|c|}{1/2} & \multicolumn{3}{|c|}{1/2} \cr
& $ b \bar b b \bar b$ & 0.2 & 0.2 & 0.6 & \multicolumn{3}{|c|}{1/2} & \multicolumn{3}{|c|}{1/2} \cr
& $ c \bar c c \bar c$ & 0.5 & 0.6 & 1.5 & \multicolumn{3}{|c|}{1/2} & \multicolumn{3}{|c|}{1/2} \cr 
& $ b \bar b c \bar c$ & 0.6 & 0.7 & 1.9 & \multicolumn{3}{|c|}{1/2} & \multicolumn{3}{|c|}{1/2} \cr \hline
\multirow{3}{*}{\rotatebox{90}{\mbox{total (fb)}}} & EW   & 13.2 & 16.9 & 45.3 & 13.2 & 16.9 & 45.1 & 0.05 & 0.07 & 0.3 \cr 
& QCD  & 14.9 & 17.5 & 33.7 & 7.4 & 8.7 & 16.8 & 7.4 & 8.7 & 16.8 \cr 
& EW+QCD    & 28.1 & 34.4 & 79.0 & 20.6 & 25.6 & 61.9 & 7.5 & 8.8 & 17.1 \cr \hline
\end{tabular}
\caption{\it Background for trilepton production with lepton isolation, the $p_T$ of the three leptons are bigger than $(20,10,10) \; {\rm GeV}$ ($p_T$ ordered) and $m_{\ell\ell} < 12 \; {\rm GeV}$. The mode label refers to the three different topologies we discuss in Sec.~\ref{sec:signature}; we give explicit cross sections for the $\ell_i^\pm \ell_i^\mp \ell_j$ topology and just present the rescaling factors one has to apply to the other modes. \label{tab:background}}
\end{center}
\end{table}

Trilepton events originated by heavy flavor production ($pp\to b\bar b, \;c\bar c, \; b\bar b b\bar b, \; c\bar c c\bar c, \; b\bar b c\bar c$) followed by single and/or double semileptonic decays of heavy mesons are negligble at large dilepton invariant mass but are the dominant source of background at low $p_T$ and low invariant mass. The first step towards an estimate of these backgrounds is to calculate cross sections for these five channels with $p_T (b,c) > 10 \; {\rm GeV}$ (we use SHERPA v.1.4.0~\cite{Gleisberg:2008ta, Schumann:2007mg, Krauss:2001iv, Gleisberg:2008fv}). Following Ref.~\cite{Sullivan:2008ki}, the probability of emitting an isolated lepton with $p_T > 10 \; {\rm GeV}$ from an heavy quark with $p_T > 10 \; {\rm GeV}$ is roughly obtained by multiplying the cross section for heavy flavor production by a suppression factor $\sim 1/200$ (for each required isolated lepton). Finally we extract the fraction of events in the dilepton invariant mass region from a Drell-Yan sample and assume a degree of universality in the shape of the dilepton invariant mass distribution (our study finds that the fraction of events with invariant mass below 12 GeV is about 30\%). The results we obtain indicate that these processes are the dominant source of background for $p_T > 10 \; {\rm GeV}$ leptons at low invariant mass. 

In order to keep these backgrounds under control we note that lepton isolation implies that only b- and c-hadrons with $p_T\lesssim 35 \; {\rm GeV}$ have an appreciable rate for production of muons with $p_T > 10 \; {\rm GeV}$ (see Ref.~\cite{Sullivan:2008ki}). We take advantage of this feature by requiring an additional cut of $20\; {\rm GeV}$ on the highest $p_T$ lepton (that, as discussed in the previous section and in Fig.~\ref{fig:ptfrac}, reduces the signal by about 5\%). From the results in Ref.~\cite{ Sullivan:2008ki} we found that this requirement causes an additional suppression factor (that we estimated to be 0.12). 

Finally we note that underlying hadronic processes with only two heavy hadrons ($b\bar b,\; c\bar c$) have a huge cross section of order $O(10 \; \mu b)$ but require double semileptonic decays of one heavy hadron. An appropriate convolution of the probability distributions presented in Ref.~\cite{Sullivan:2008ki} leads us to estimate the additional suppression associated with a double semileptonic decay at the 5\% level.

Note that the procedure outlined above leads only to a rough estimate of these backgrounds and should be trusted only if the resulting rates are well below our signal. In any case, pushing the cut on the highest $p_T$ lepton to $30 \; {\rm GeV}$ would essentially eliminate these hadronic backgrounds at the cost of a 20\% signal suppression (see Fig.~\ref{fig:ptfrac}). 

Our estimates for $b\bar b$ and $c\bar c$ production with $p_T(b,c)>10 \; {\rm GeV}$ are:
\begin{align}
\sigma_{b\bar b} =
\begin{cases}
7.3 \; {\rm \mu b} & 7 \; {\rm TeV} \cr
8.6  \; {\rm \mu b} & 8 \; {\rm TeV} \cr
16.2  \; {\rm \mu b} & 14 \; {\rm TeV} \cr
\end{cases} 
\;\;
{\rm ,}
\;\;
\sigma_{c\bar c} =
\begin{cases}
9.0 \; {\rm \mu b} & 7 \; {\rm TeV} \cr
10.6  \; {\rm \mu b} & 8 \; {\rm TeV} \cr
19.6  \; {\rm \mu b} & 14 \; {\rm TeV} \cr
\end{cases}  \; .
\end{align}
Each process carries a suppression factor $\frac{4 \times 0.3 \times 0.12 \times 0.05}{200^3} = 2.3 \times 10^{-10}$ where the factor of 4 takes into account all possible final state leptons combinations, each $1/200$ factor accounts for the emission of an isolated lepton, the 0.3, 0.12 and 0.05 factors account for the $m_{\ell\ell} < 12\; {\rm GeV}$ cut, the effects of the $p_T>20\; {\rm GeV}$ cut and the extra suppression due to the required double semileptonic decay, respectively. The results we obtain are of order $O(10\; {\rm fb})$ and are presented in table~\ref{tab:background}.  

Estimates for leptons from $b\bar b b\bar b$, $c\bar c c\bar c$ and $b\bar b c\bar c$ with $p_T(b,c)>10 \; {\rm GeV}$ are:
\begin{align}
\sigma_{b\bar bb\bar b} =
\begin{cases}
4.7 \; {\rm n b} & 7 \; {\rm TeV} \cr
6.2  \; {\rm n b} & 8 \; {\rm TeV} \cr
16.4  \; {\rm n b} & 14 \; {\rm TeV} \cr
\end{cases} 
{\rm ,}
\;
\sigma_{c\bar c c\bar c}  =
\begin{cases}
13.7 \; {\rm n b} & 7 \; {\rm TeV} \cr
17.8  \; {\rm n b} & 8 \; {\rm TeV} \cr
44.3  \; {\rm n b} & 14 \; {\rm TeV} \cr
\end{cases} 
{\rm ,}
\;
\sigma_{b\bar b c\bar c}  =
\begin{cases}
16.5 \; {\rm n b} & 7 \; {\rm TeV} \cr
20.8  \; {\rm n b} & 8 \; {\rm TeV} \cr
53.6  \; {\rm n b} & 14 \; {\rm TeV} \cr
\end{cases}
 . 
\end{align}
Each process carries a suppression factor $\frac{8 \times 0.3 \times 0.12}{200^3} = 3.6 \times 10^{-8}$ where the factor of 8 takes into account all possible final state leptons combinations, and the other factors have the same meaning as above. The results we obtain are of order few fb and are presented in table~\ref{tab:background}.  

\subsection{Extraction of the signal and LHC sensitivity}
\label{sec:sensitivity}
\begin{figure}
\begin{center}
\includegraphics[width=0.45 \linewidth]{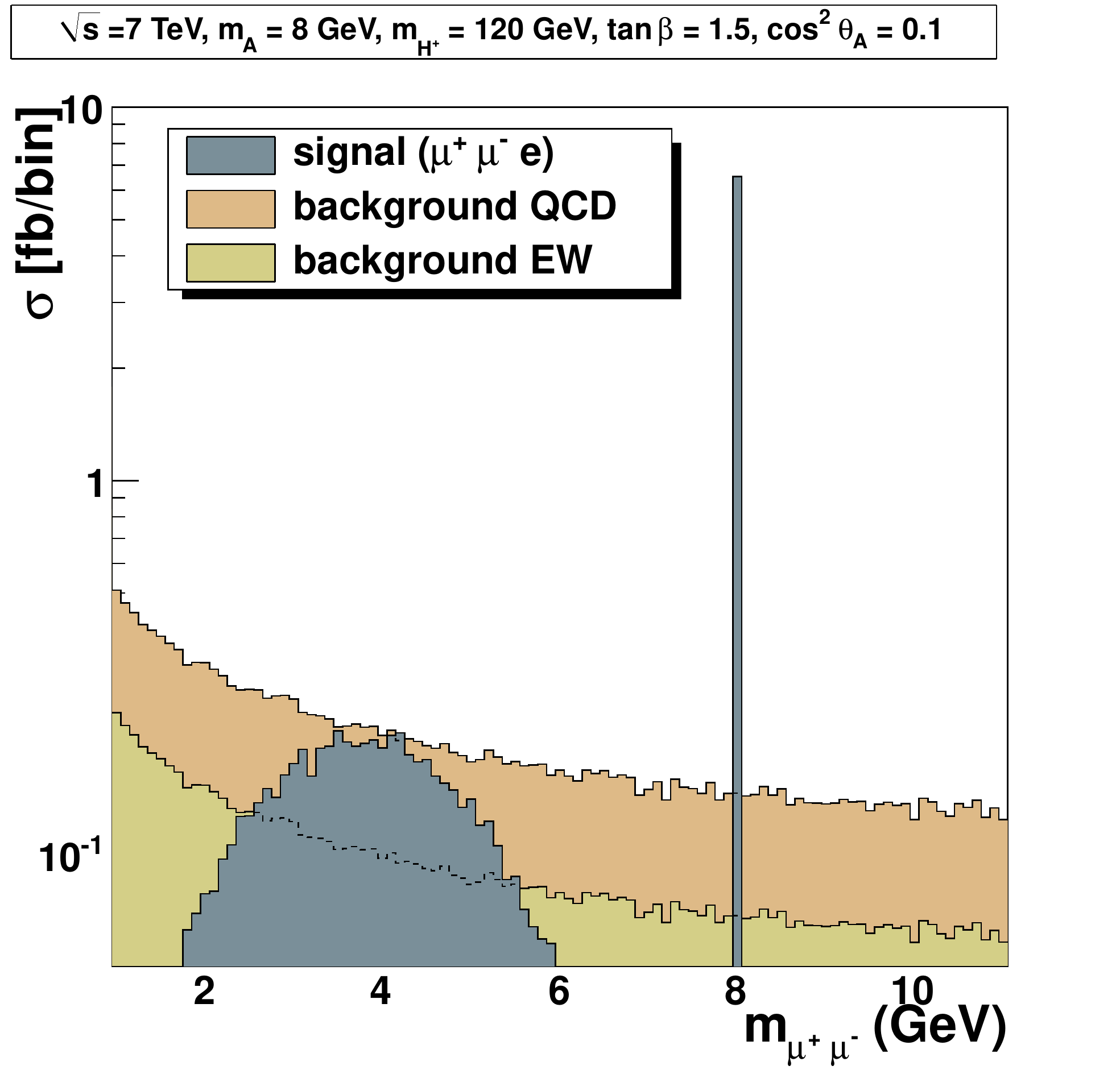}
\includegraphics[width=0.45 \linewidth]{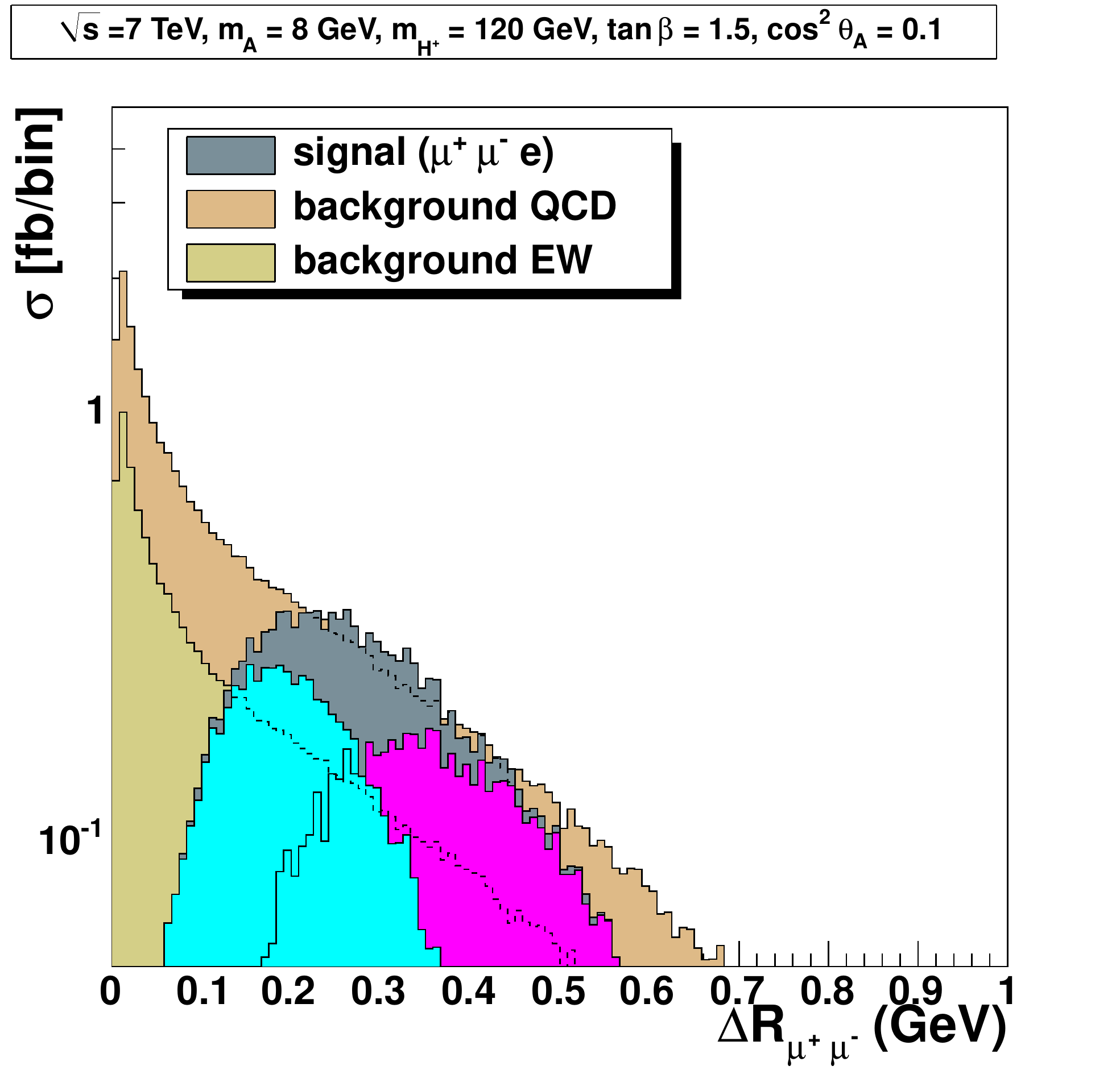}
\includegraphics[width=0.45 \linewidth]{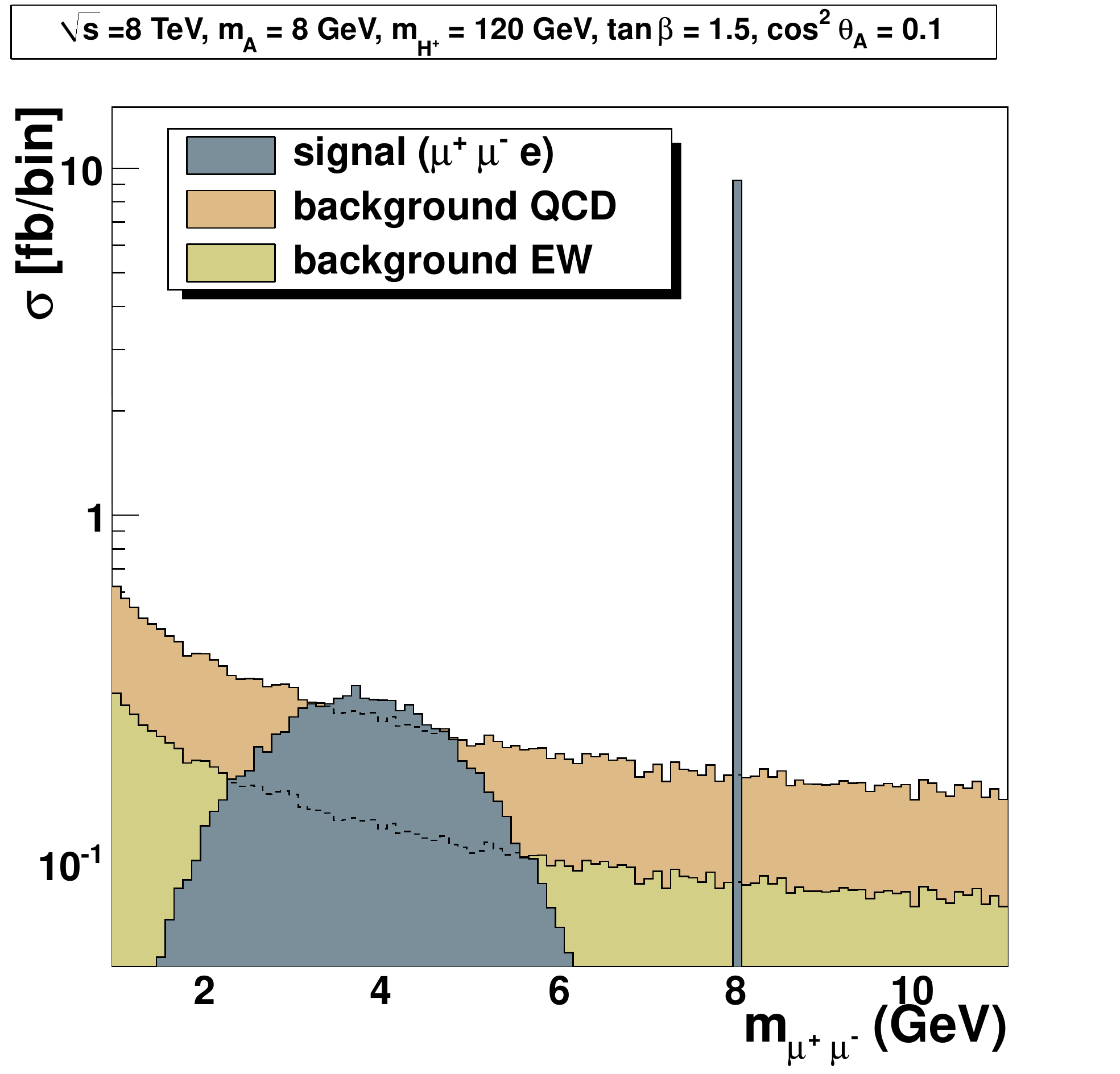}
\includegraphics[width=0.45 \linewidth]{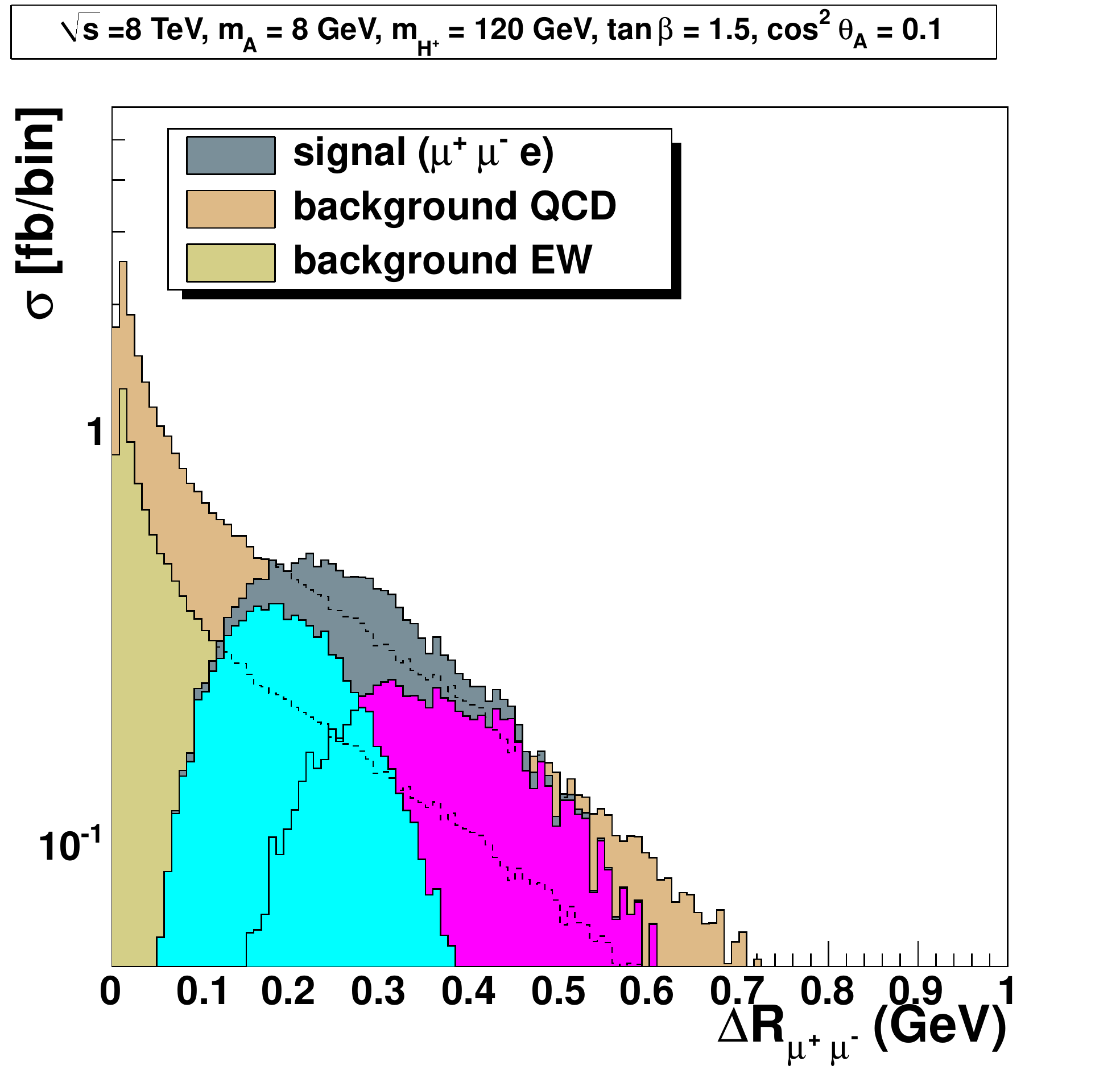}
\includegraphics[width=0.45 \linewidth]{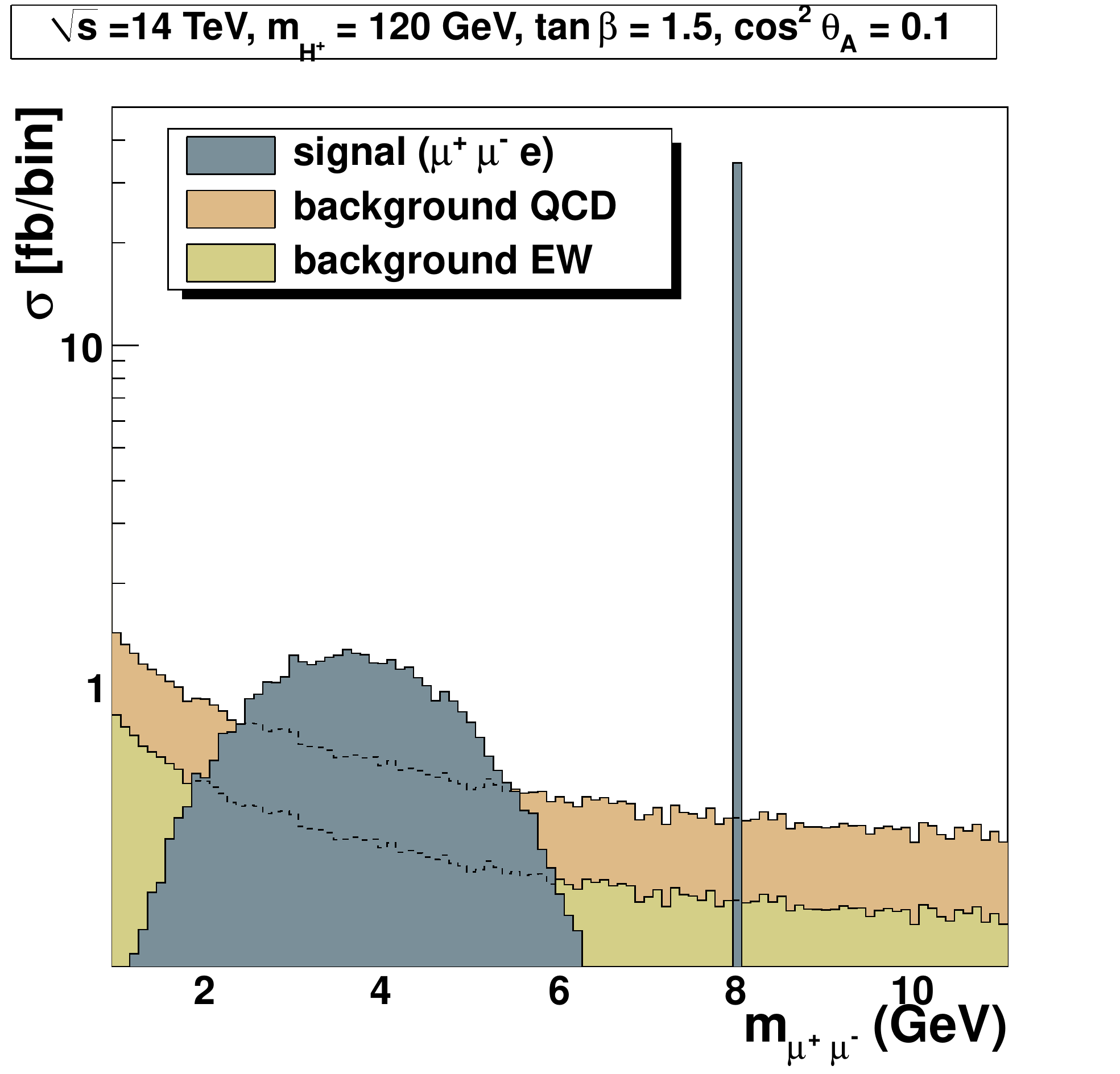}
\includegraphics[width=0.45 \linewidth]{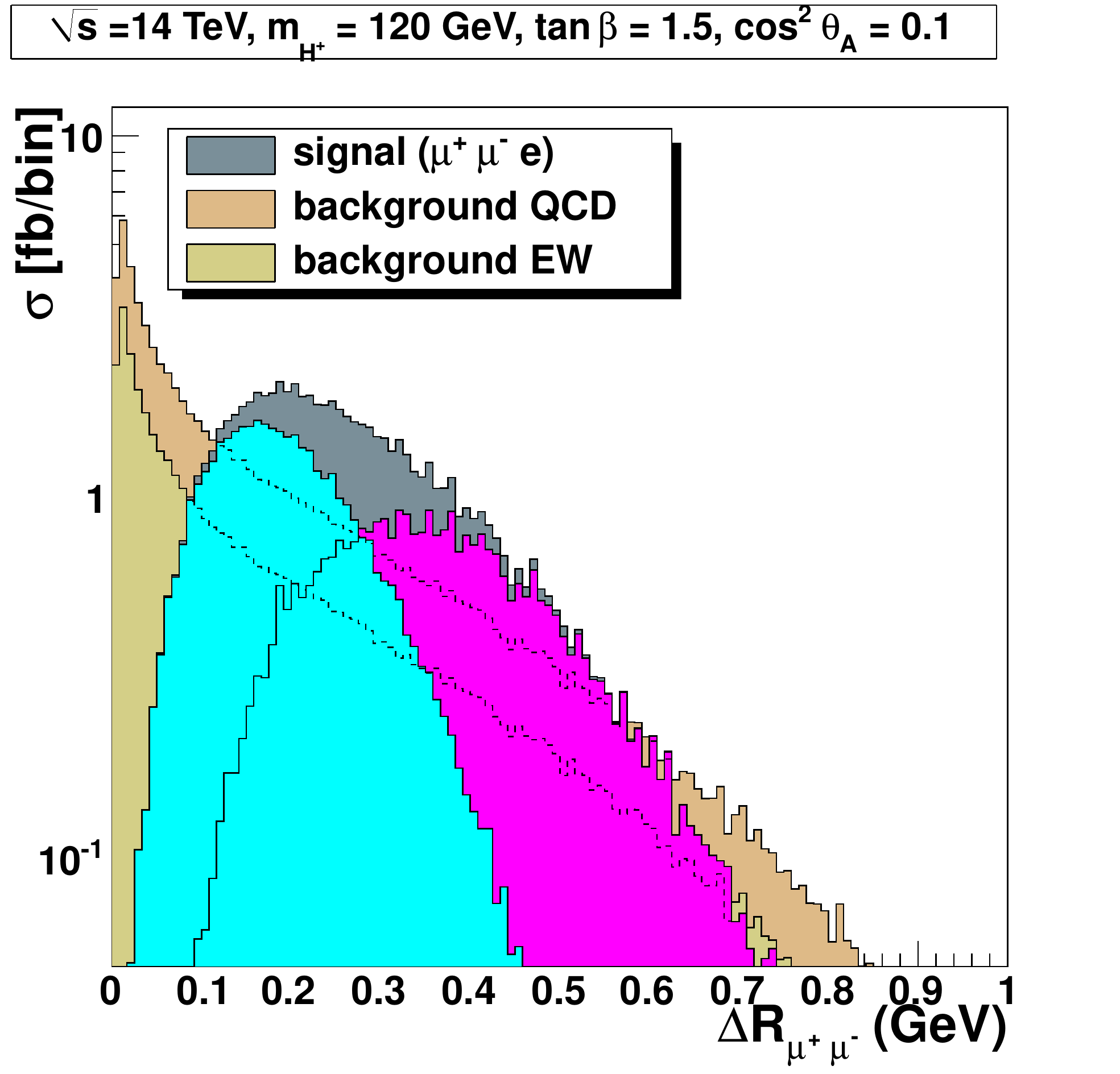}
\end{center}
\vskip -0.7cm
\caption{di-muon invariant mass and $\Delta R_{\mu\mu}$ distributions for the $\mu^+\mu^- e$ topology. In the $\Delta R_{\mu\mu}$ distribution we further separate contributions originating from the $A\to\tau\tau$ (blue) and $A\to \mu\mu$ (magenta). \label{fig:mme}}
\end{figure}

\begin{figure}
\begin{center}
\includegraphics[width=0.45 \linewidth]{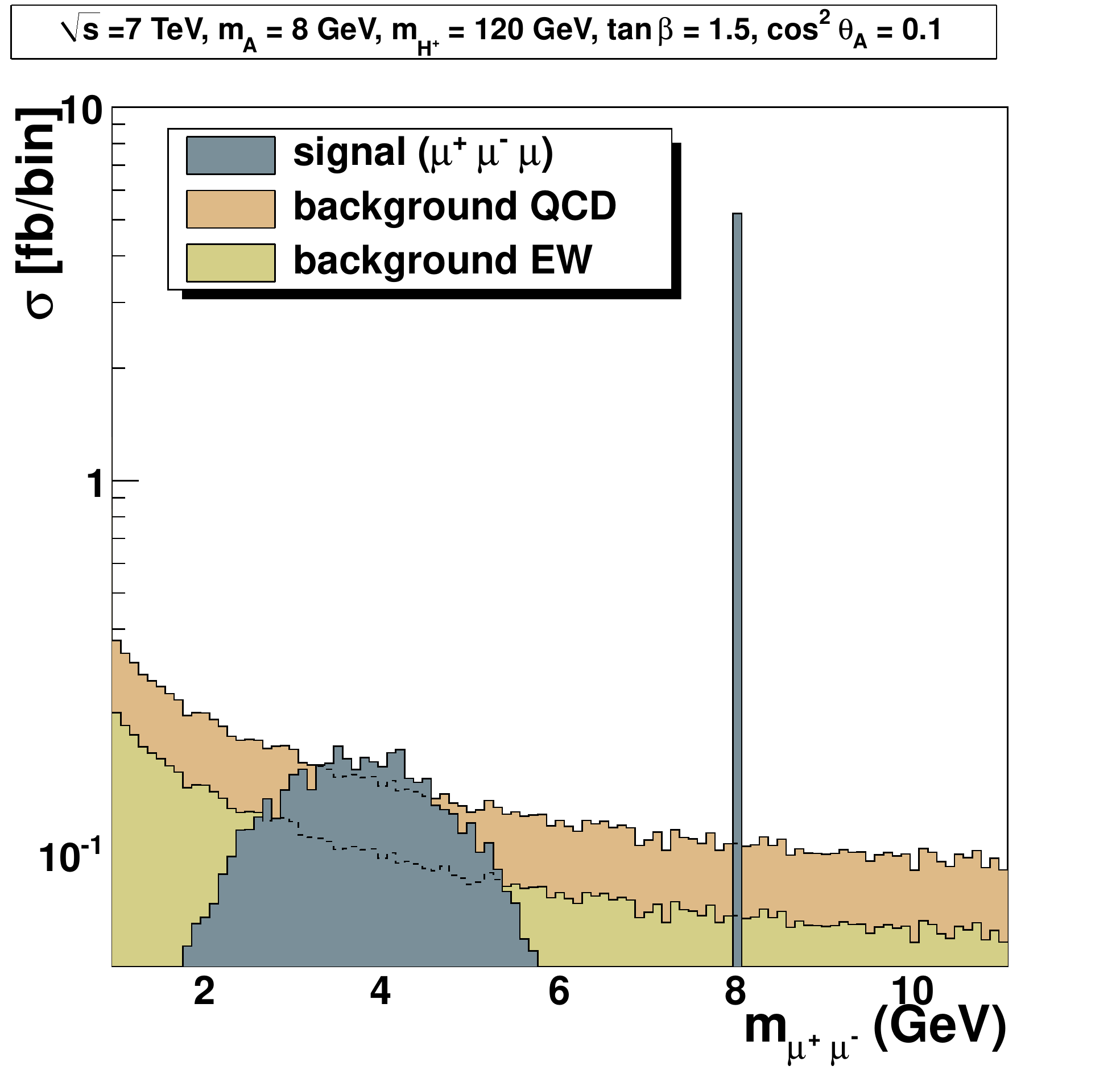}
\includegraphics[width=0.45 \linewidth]{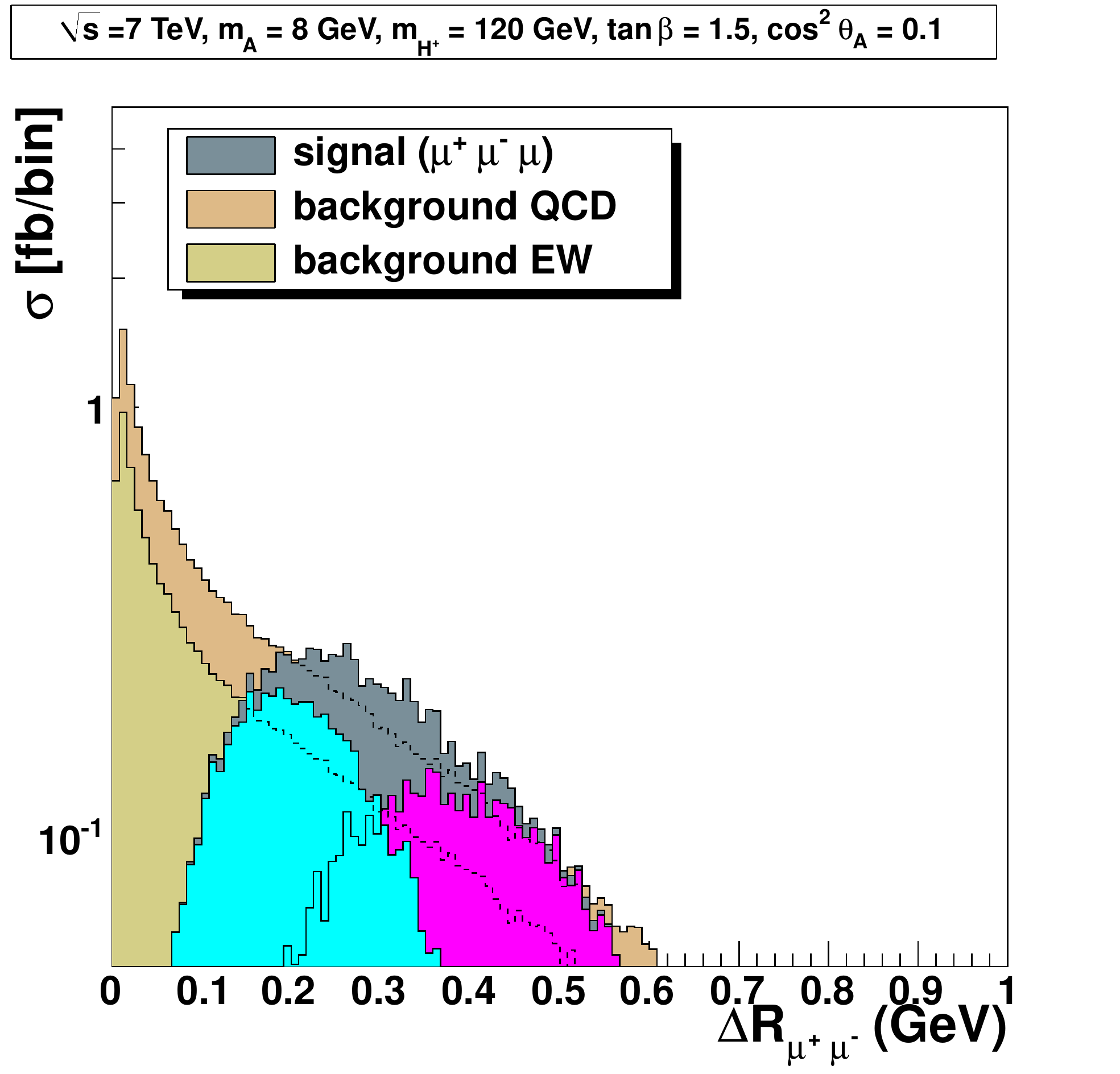}
\includegraphics[width=0.45 \linewidth]{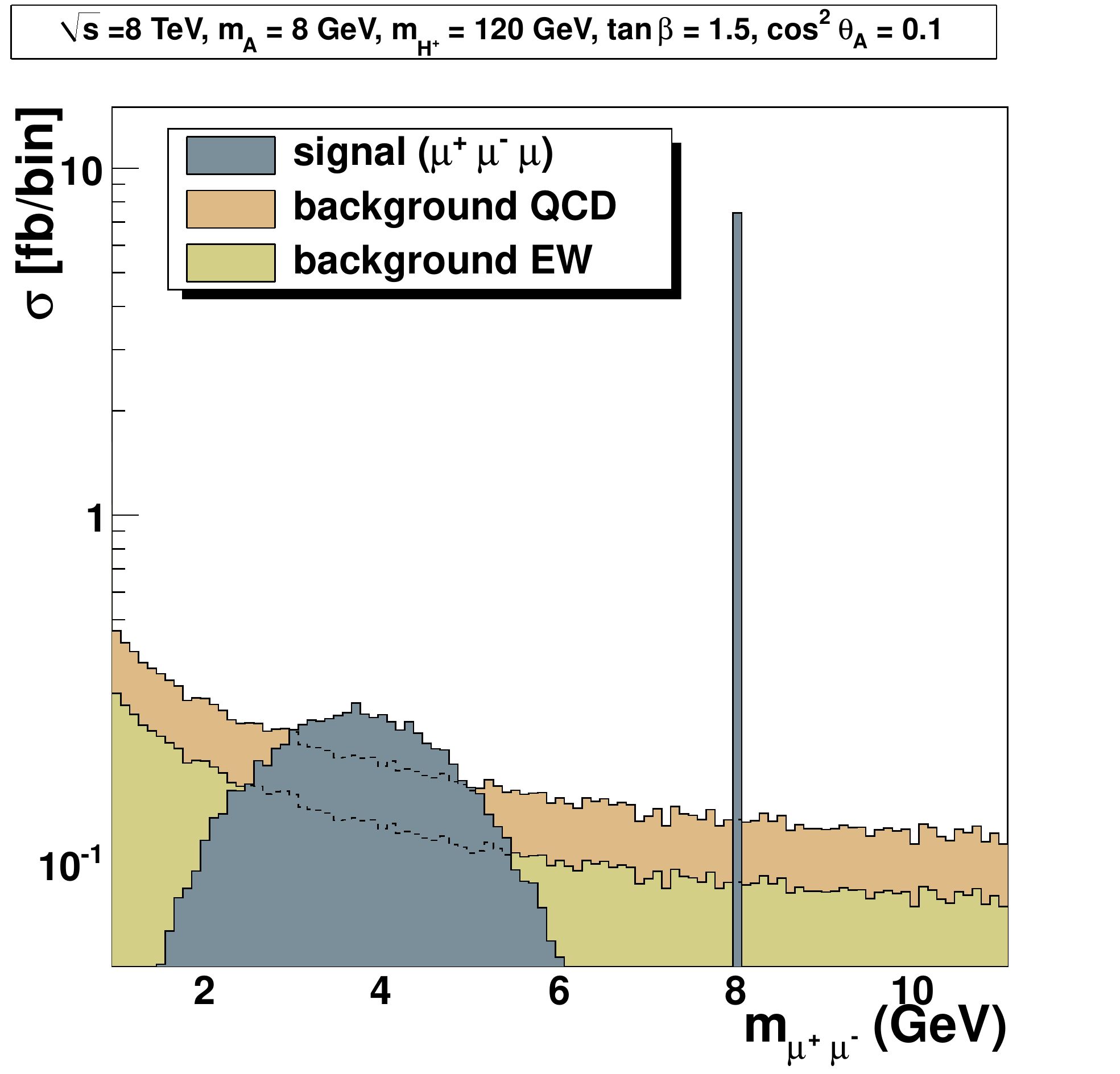}
\includegraphics[width=0.45 \linewidth]{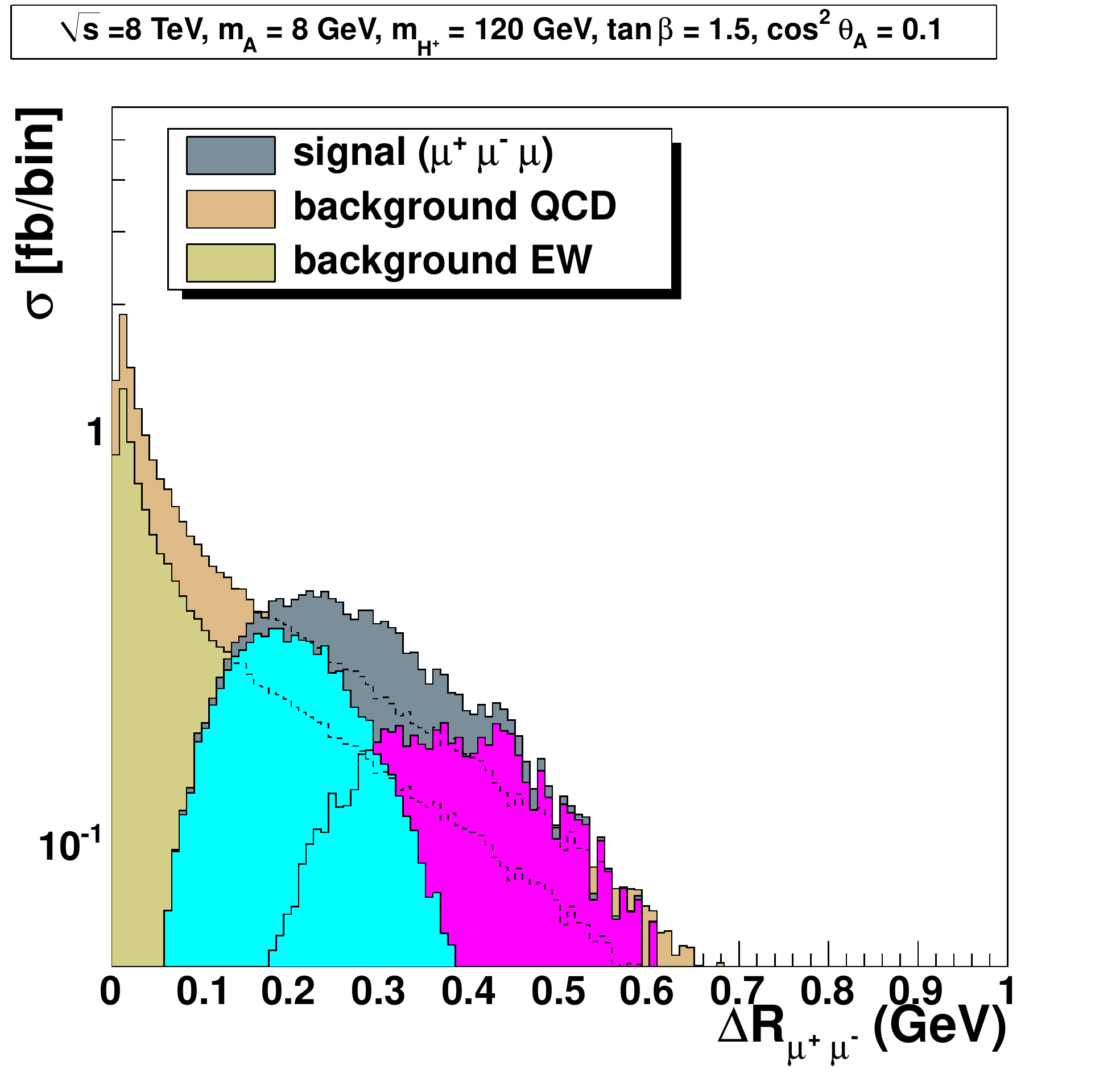}
\includegraphics[width=0.45 \linewidth]{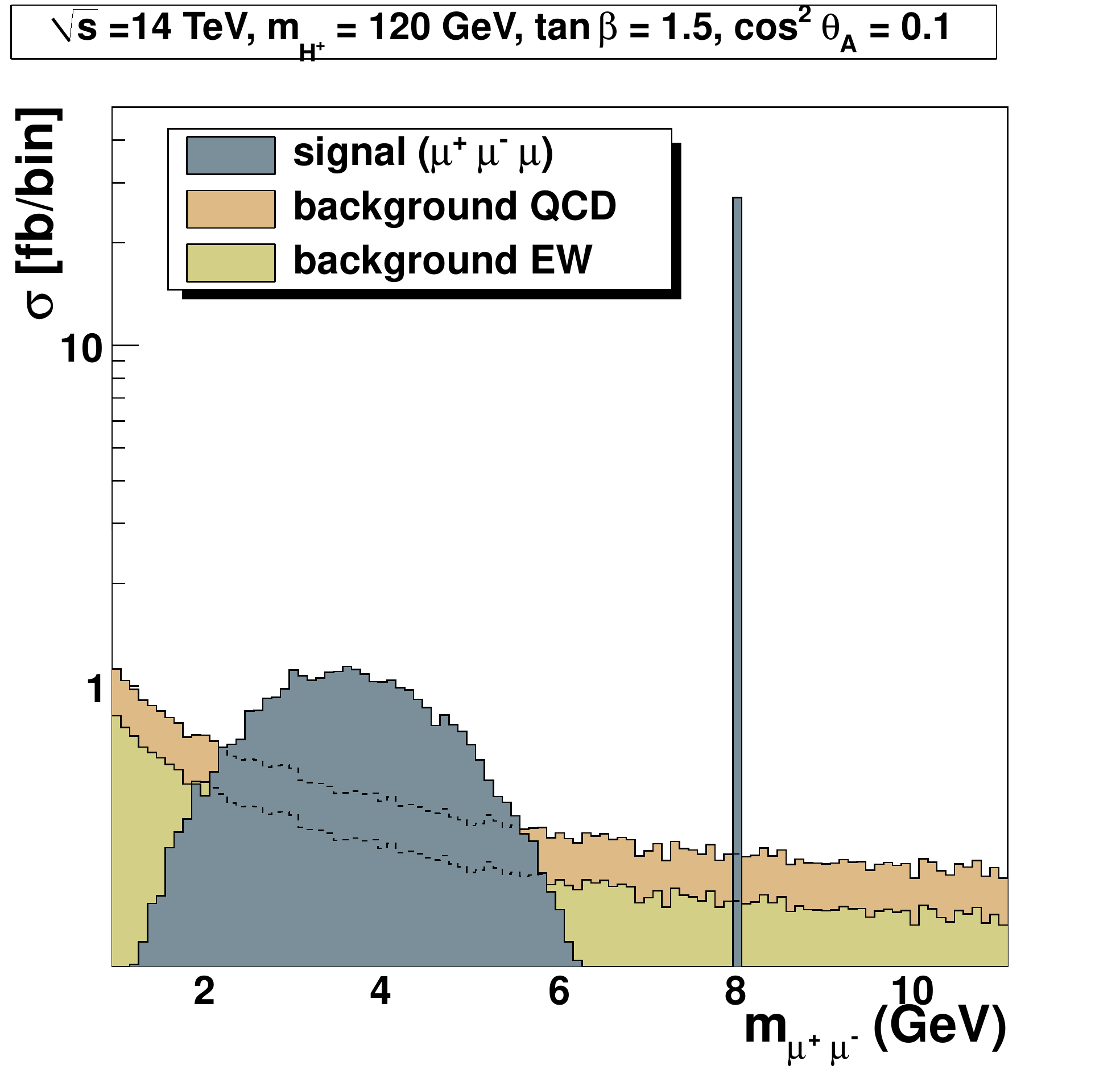}
\includegraphics[width=0.45 \linewidth]{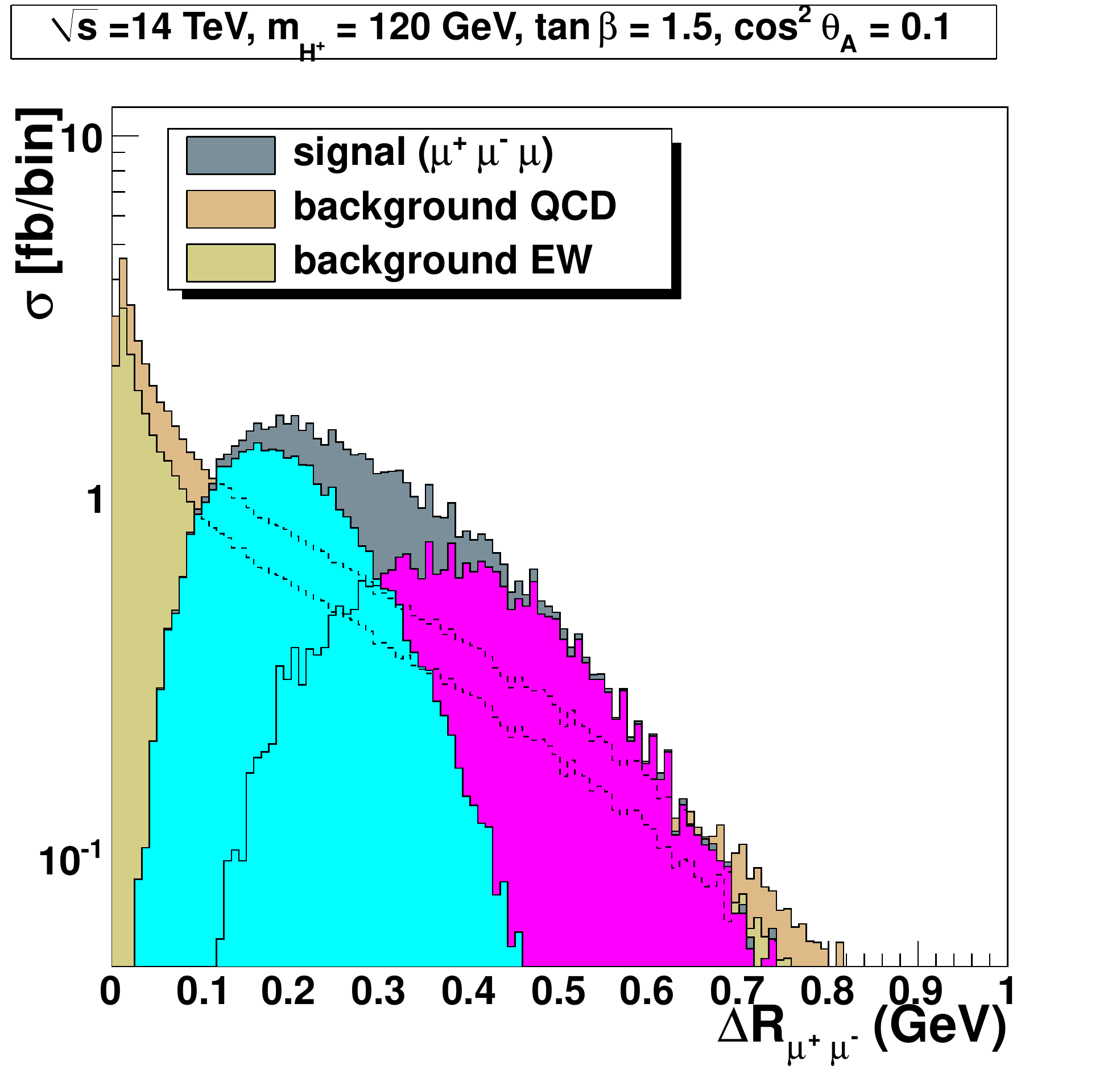}
\end{center}
\vskip -0.7cm
\caption{Di-muon invariant mass and $\Delta R_{\mu\mu}$ distributions for the $\mu^+\mu^- \mu$ topology. In the $\Delta R_{\mu\mu}$ distribution we further separate contributions originating from the $A\to\tau\tau$ (blue) and $A\to \mu\mu$ (magenta). \label{fig:mmm}}
\end{figure}

\begin{figure}
\begin{center}
\includegraphics[width=0.45 \linewidth]{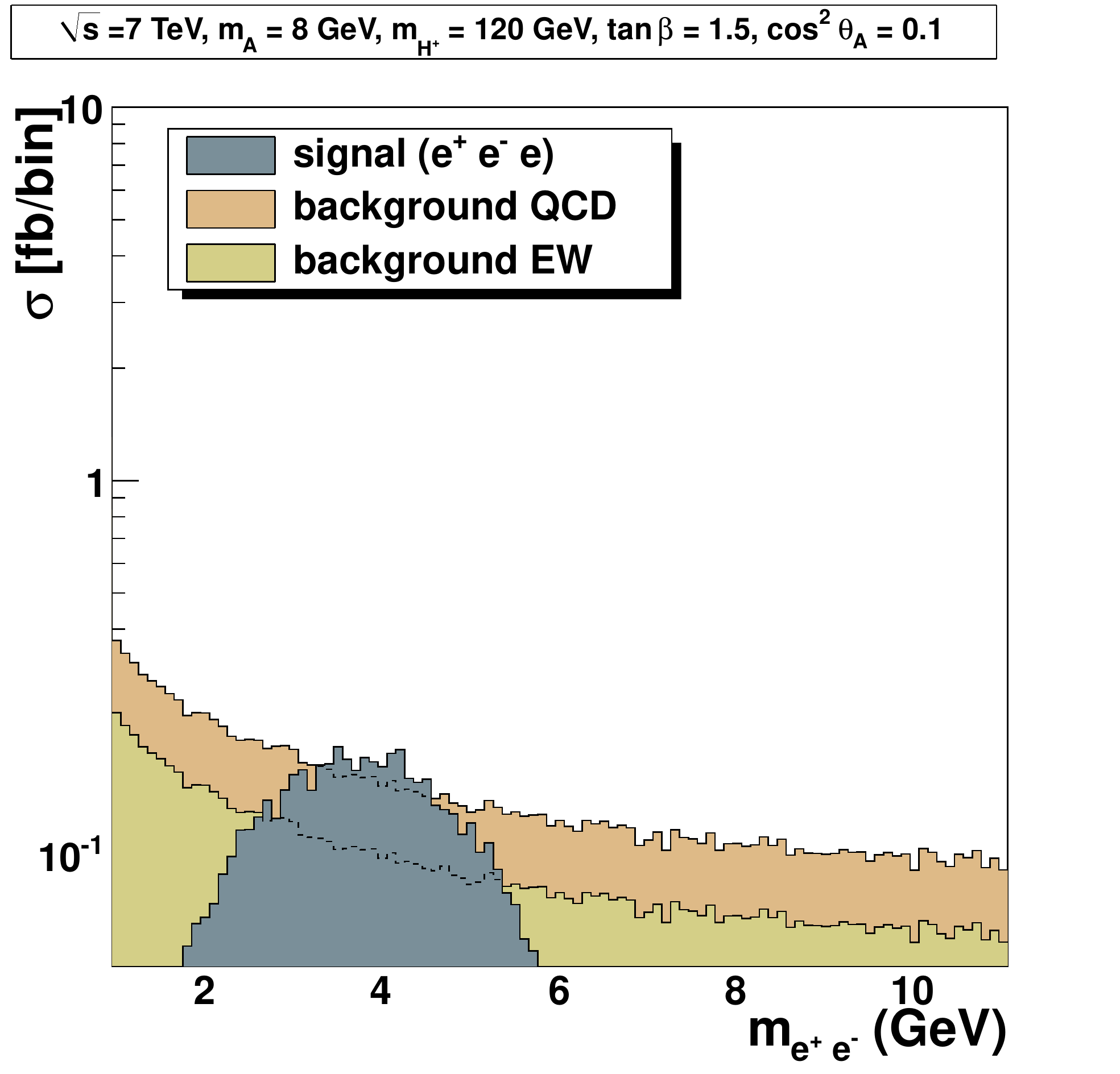}
\includegraphics[width=0.45 \linewidth]{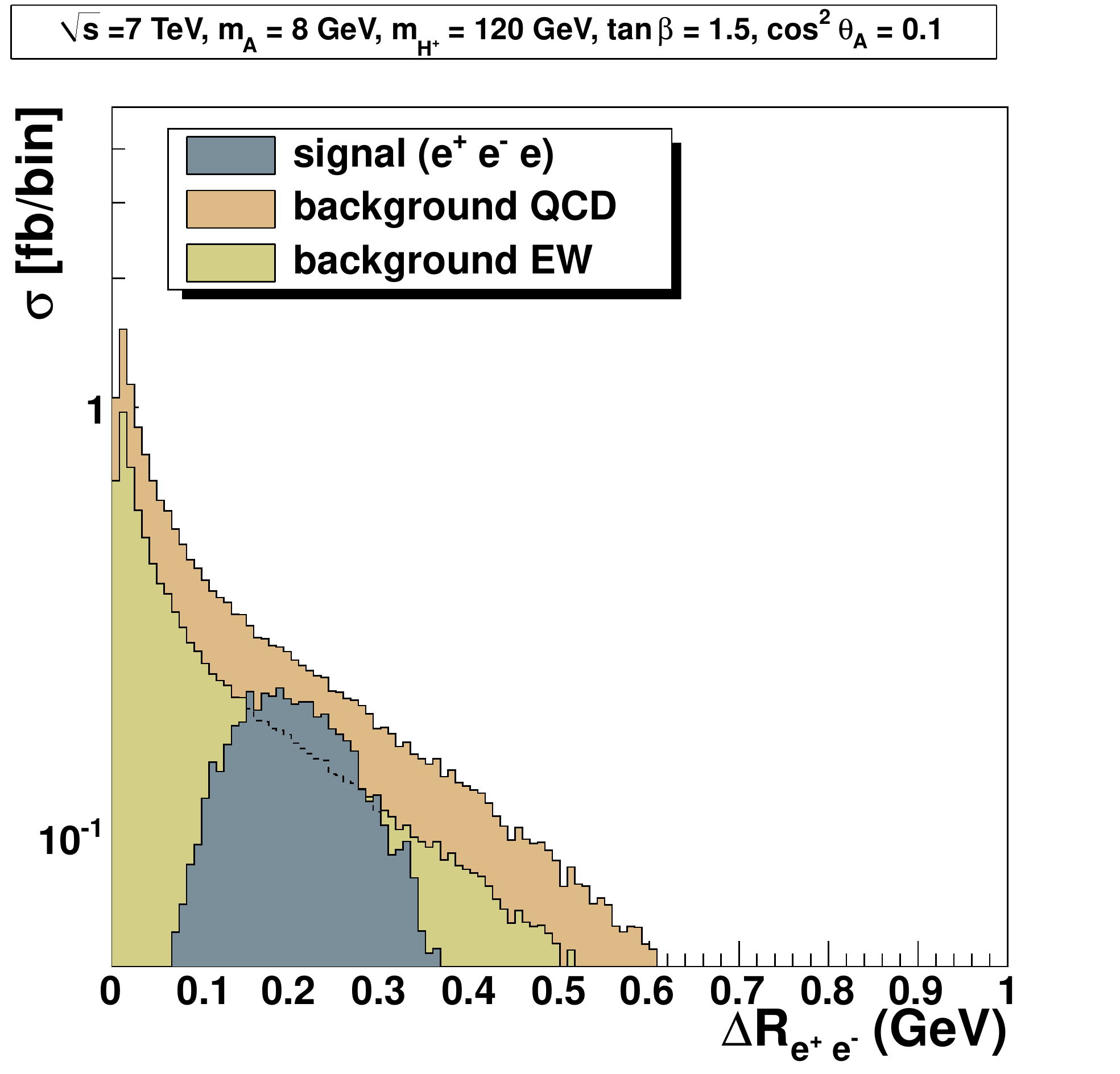}
\includegraphics[width=0.45 \linewidth]{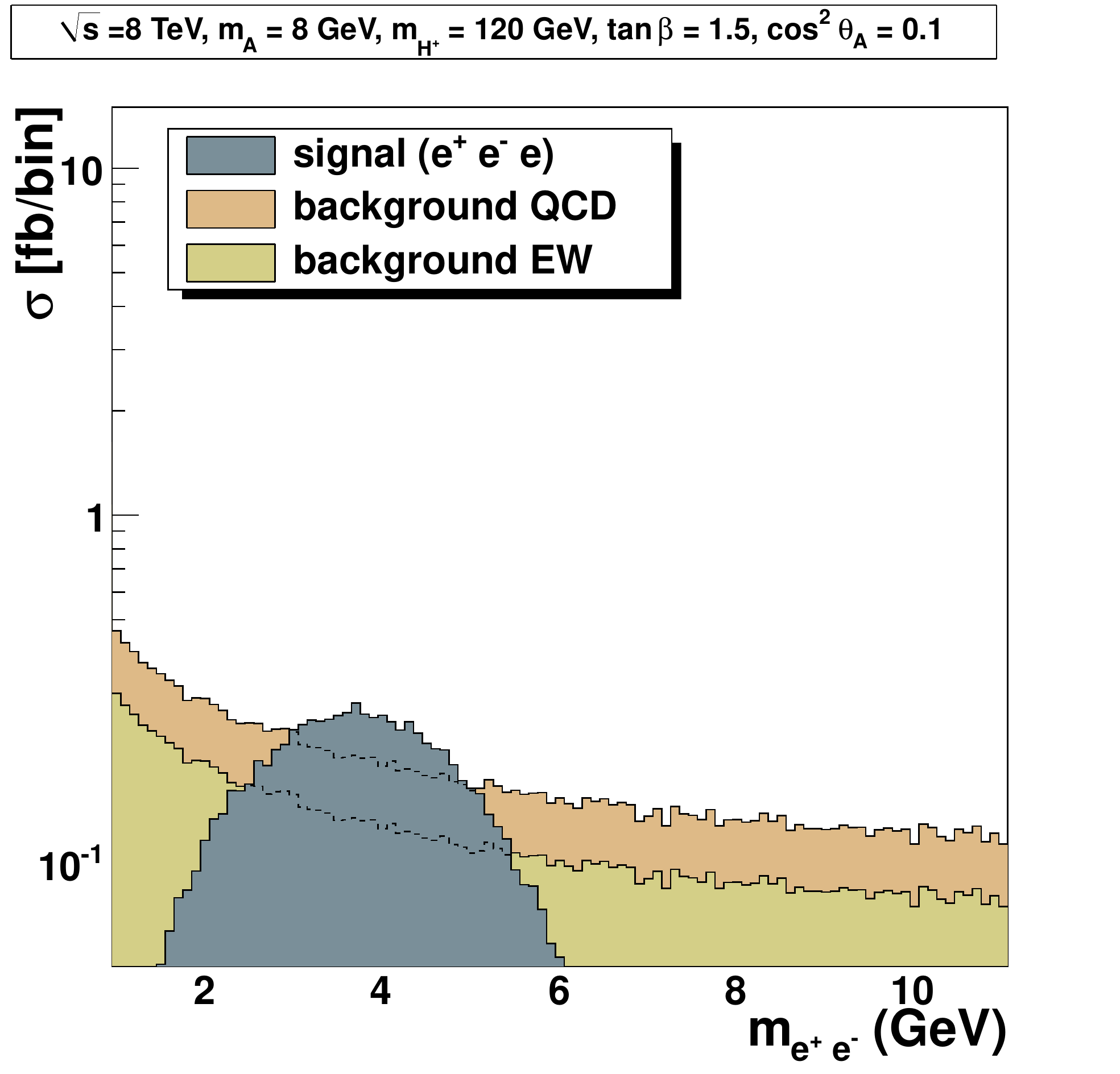}
\includegraphics[width=0.45 \linewidth]{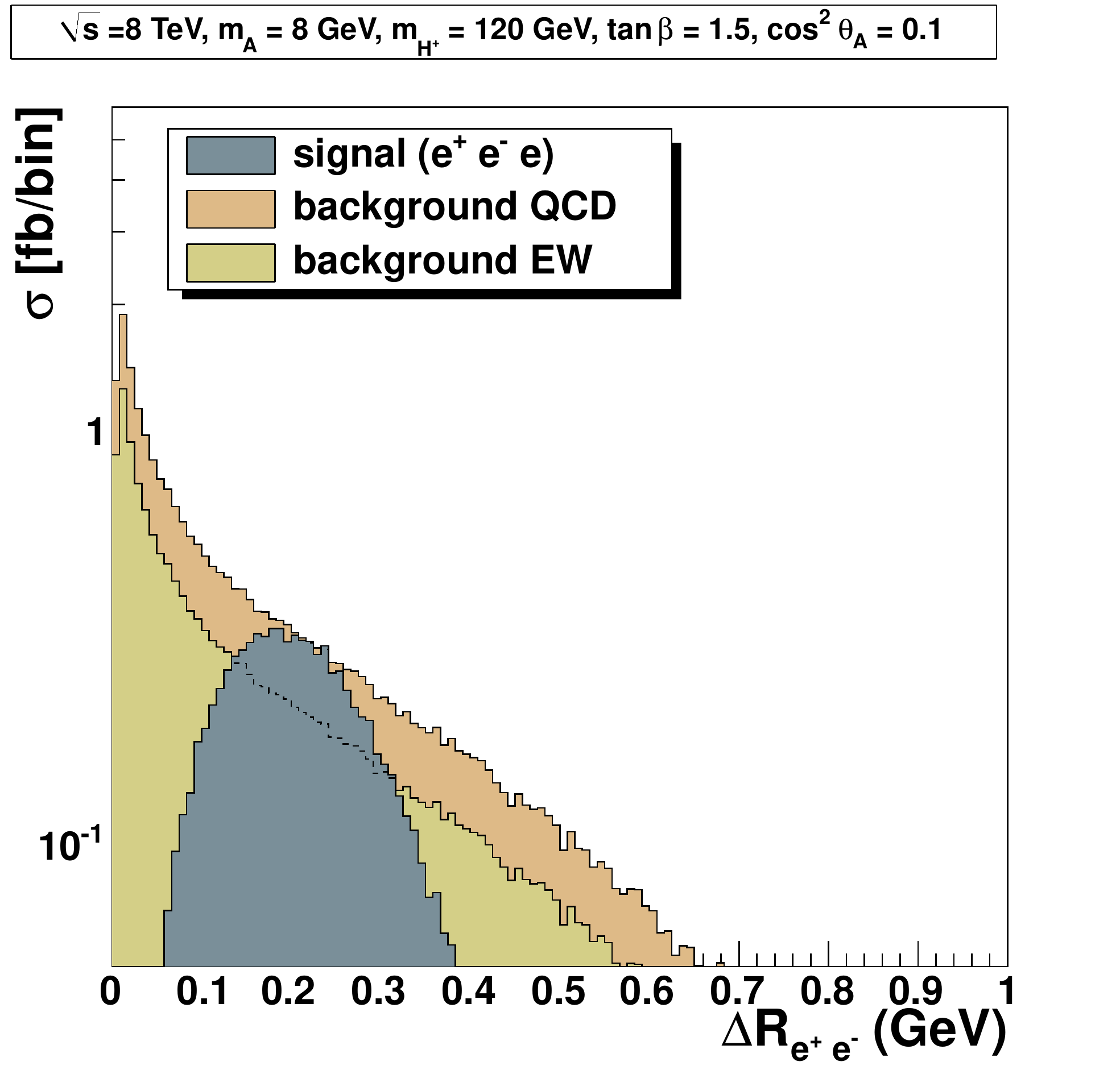}
\includegraphics[width=0.45 \linewidth]{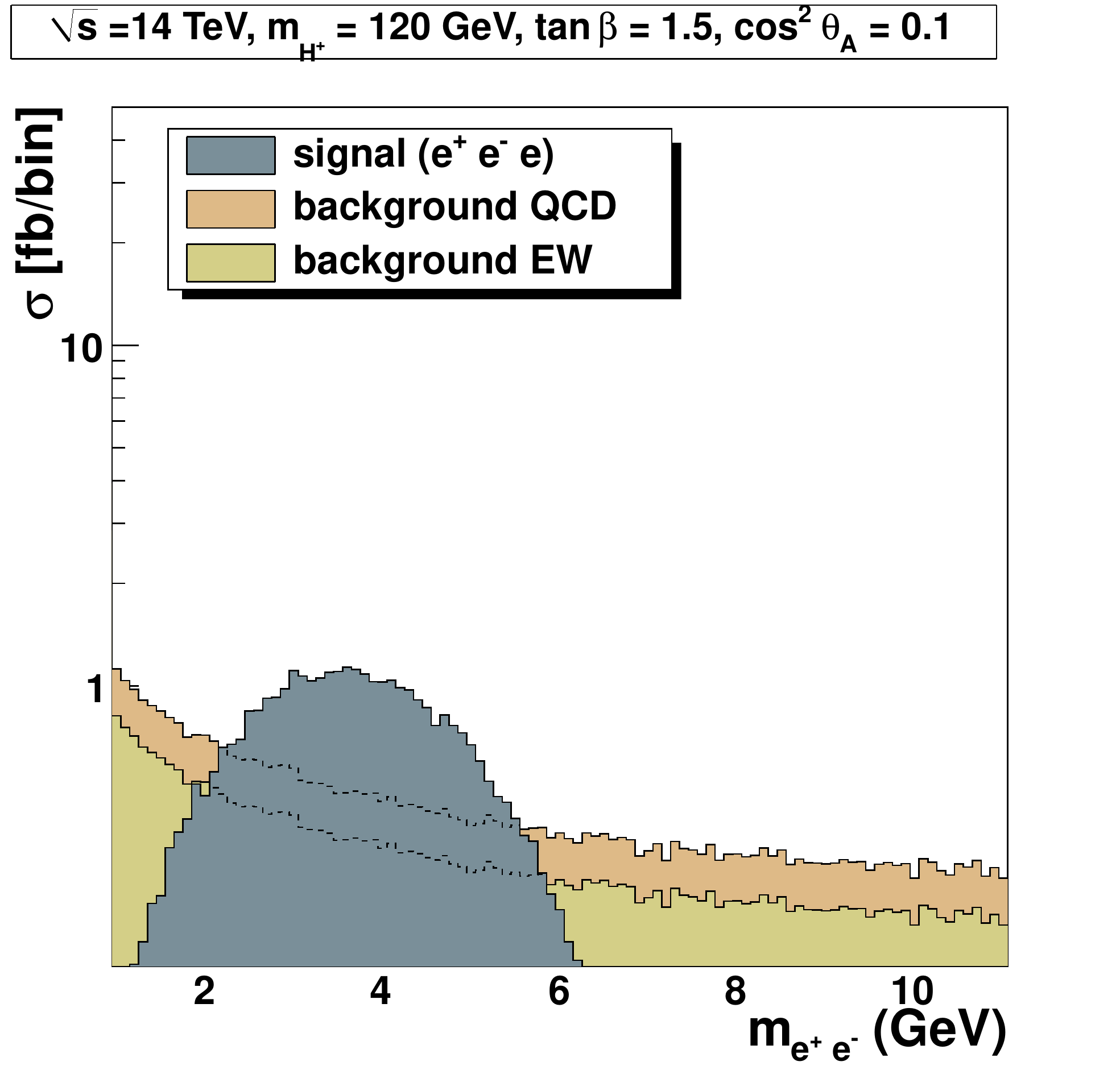}
\includegraphics[width=0.45 \linewidth]{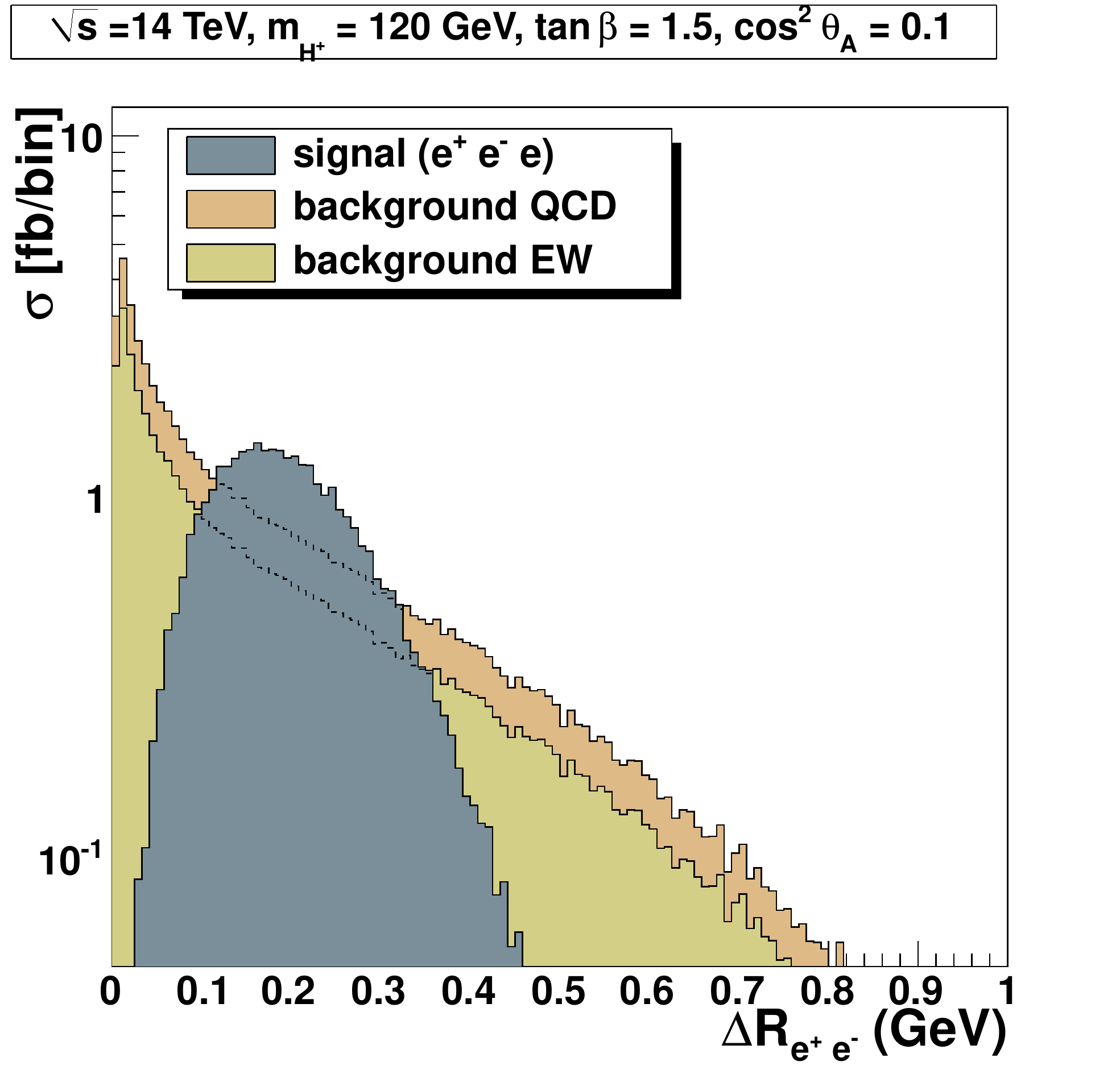}
\end{center}
\vskip -0.7cm
\caption{Di-electron invariant mass and $\Delta R_{ee}$ distributions for the $e^+e^- e$ topology. The $e^+ e^- \mu$ topology yields identical distributions.\label{fig:eee}}
\end{figure}

\begin{figure}
\begin{center}
\includegraphics[width=0.45 \linewidth]{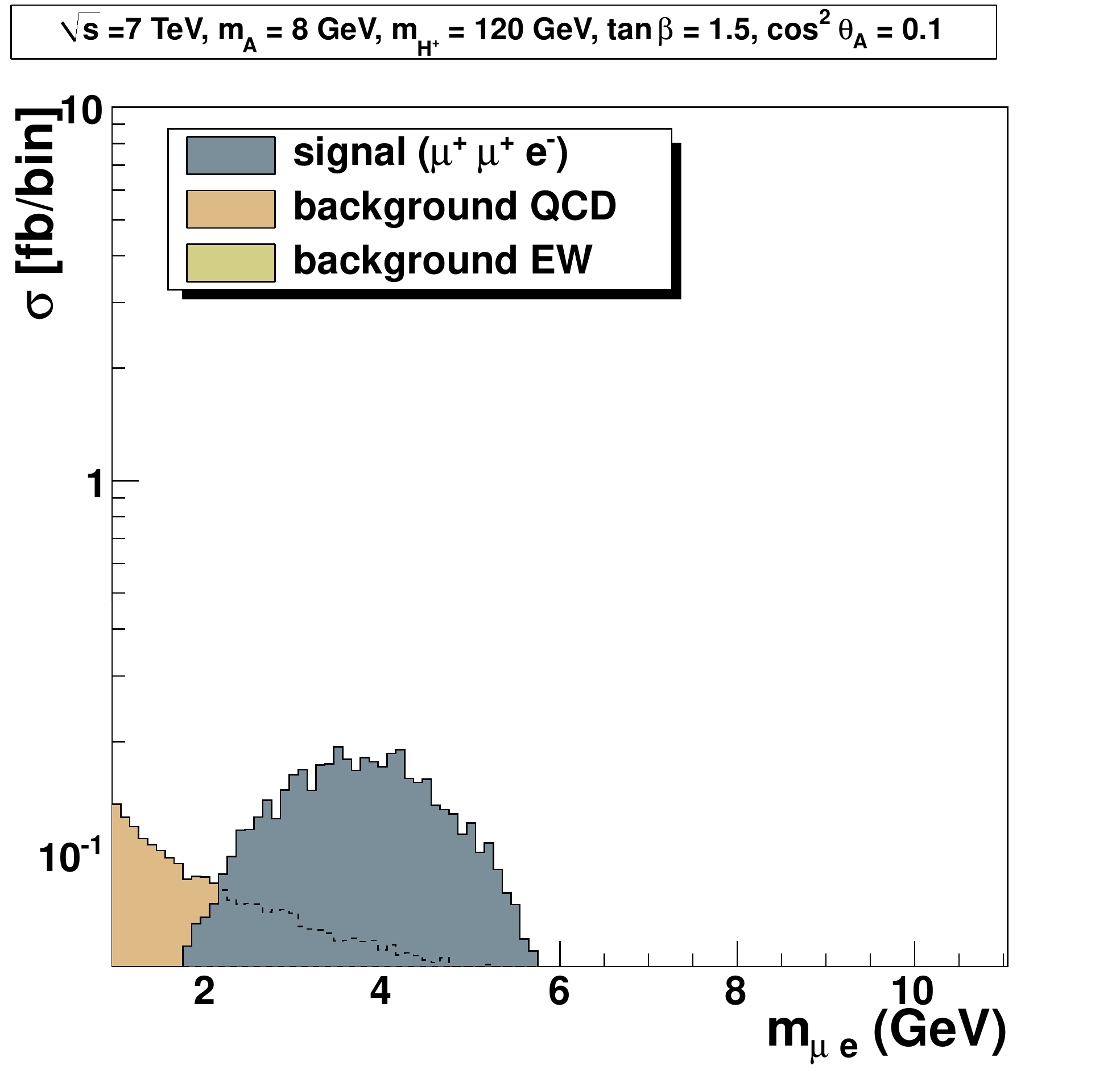}
\includegraphics[width=0.45 \linewidth]{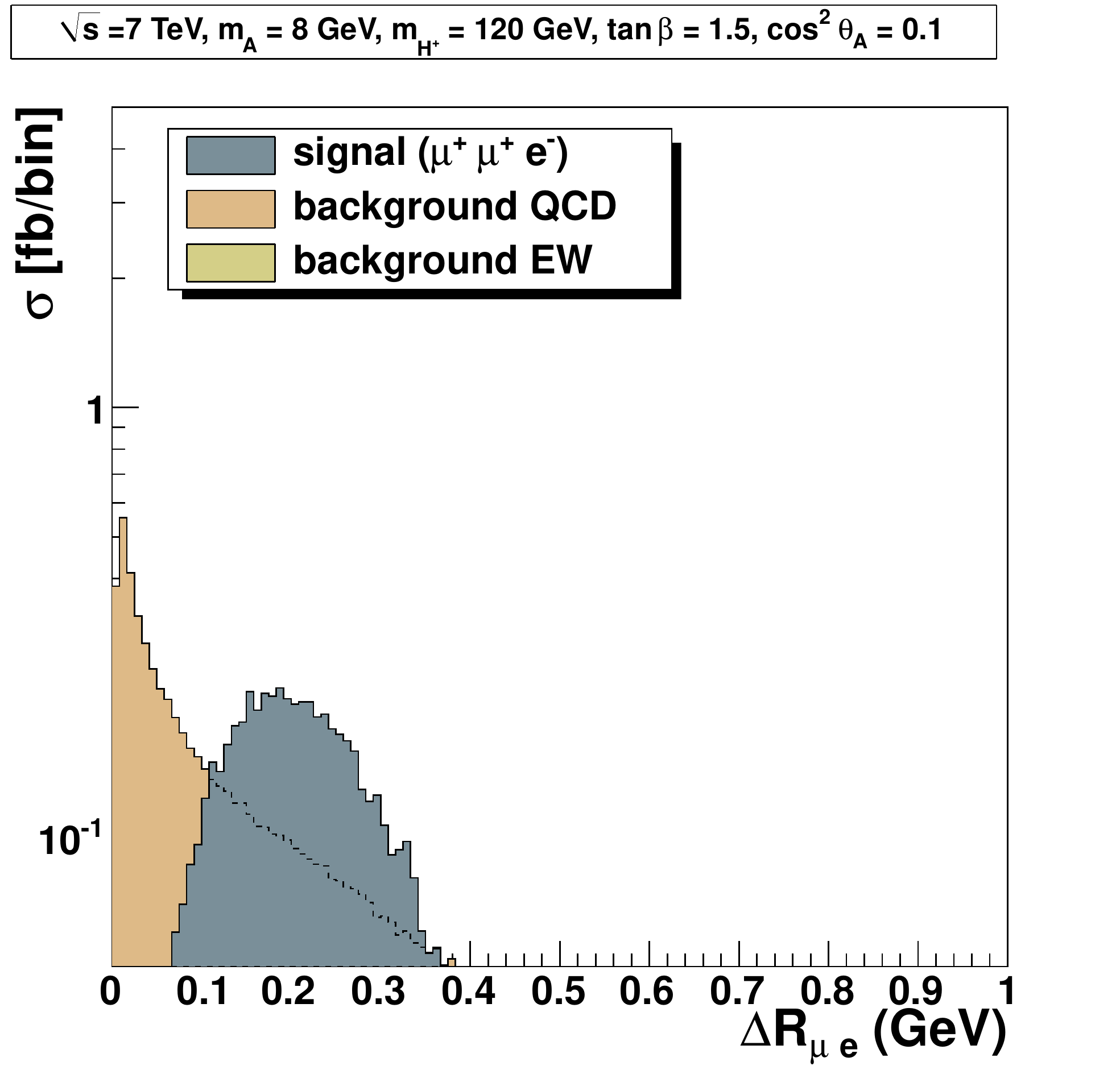}
\includegraphics[width=0.45 \linewidth]{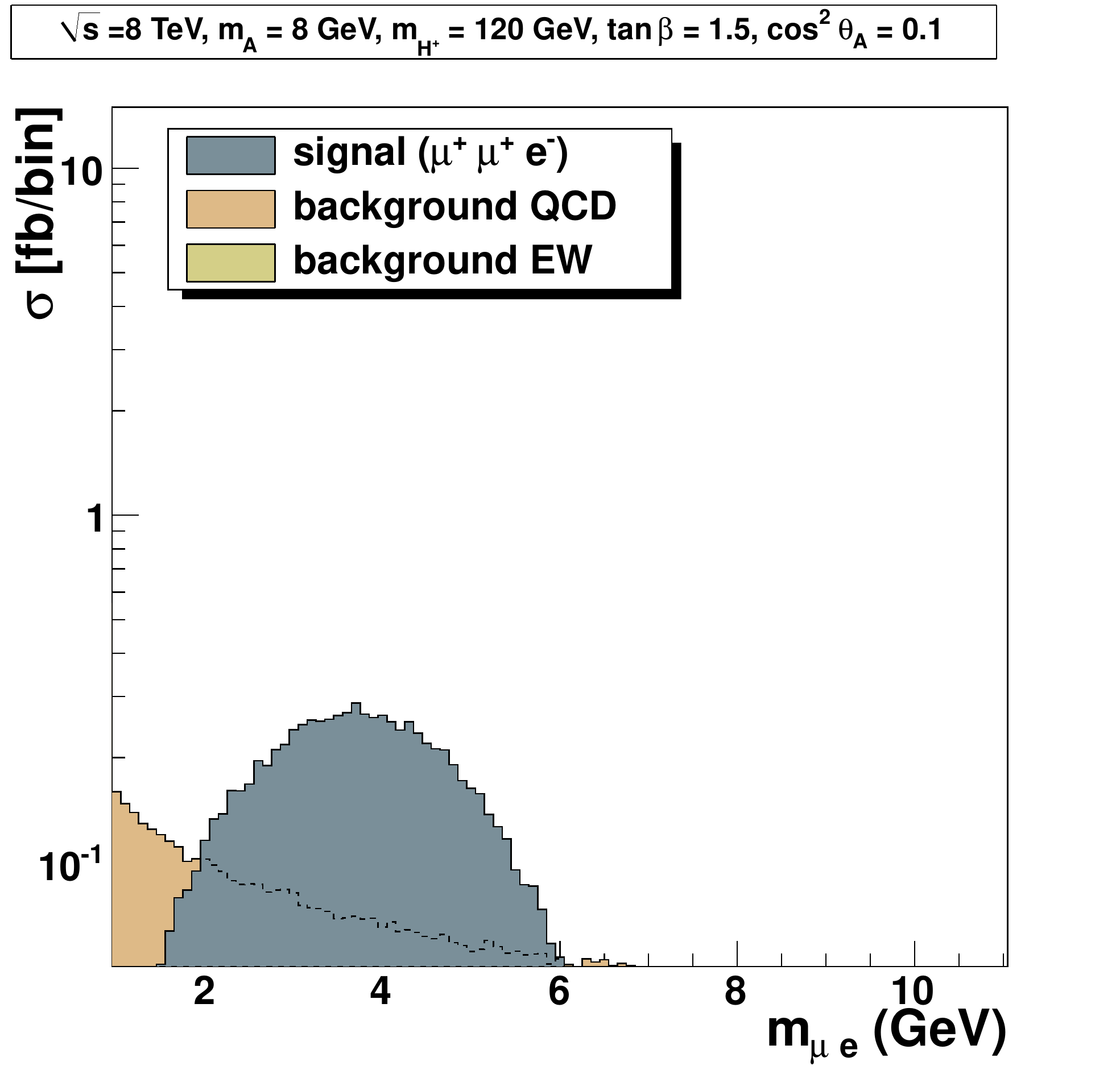}
\includegraphics[width=0.45 \linewidth]{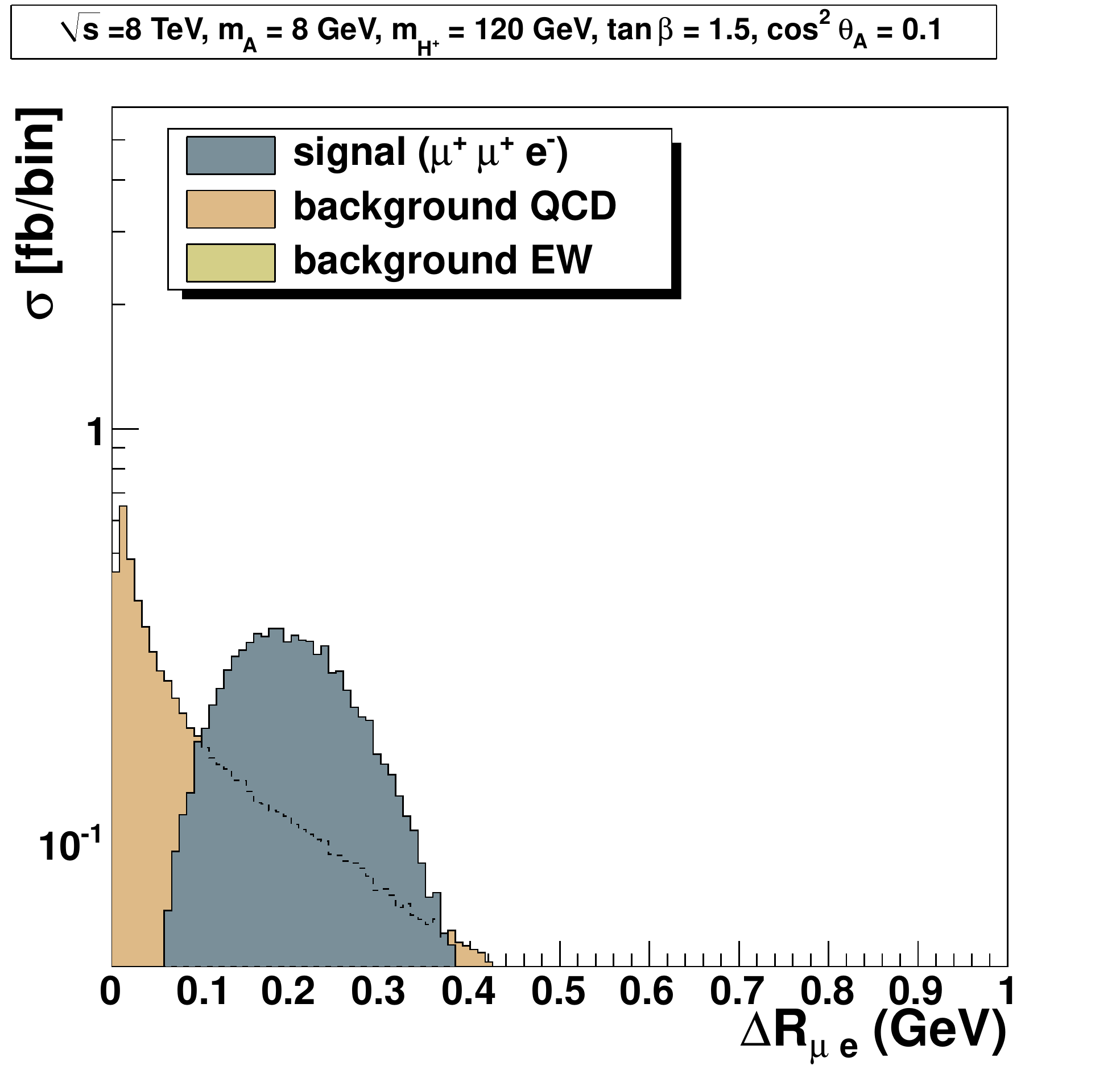}
\includegraphics[width=0.45 \linewidth]{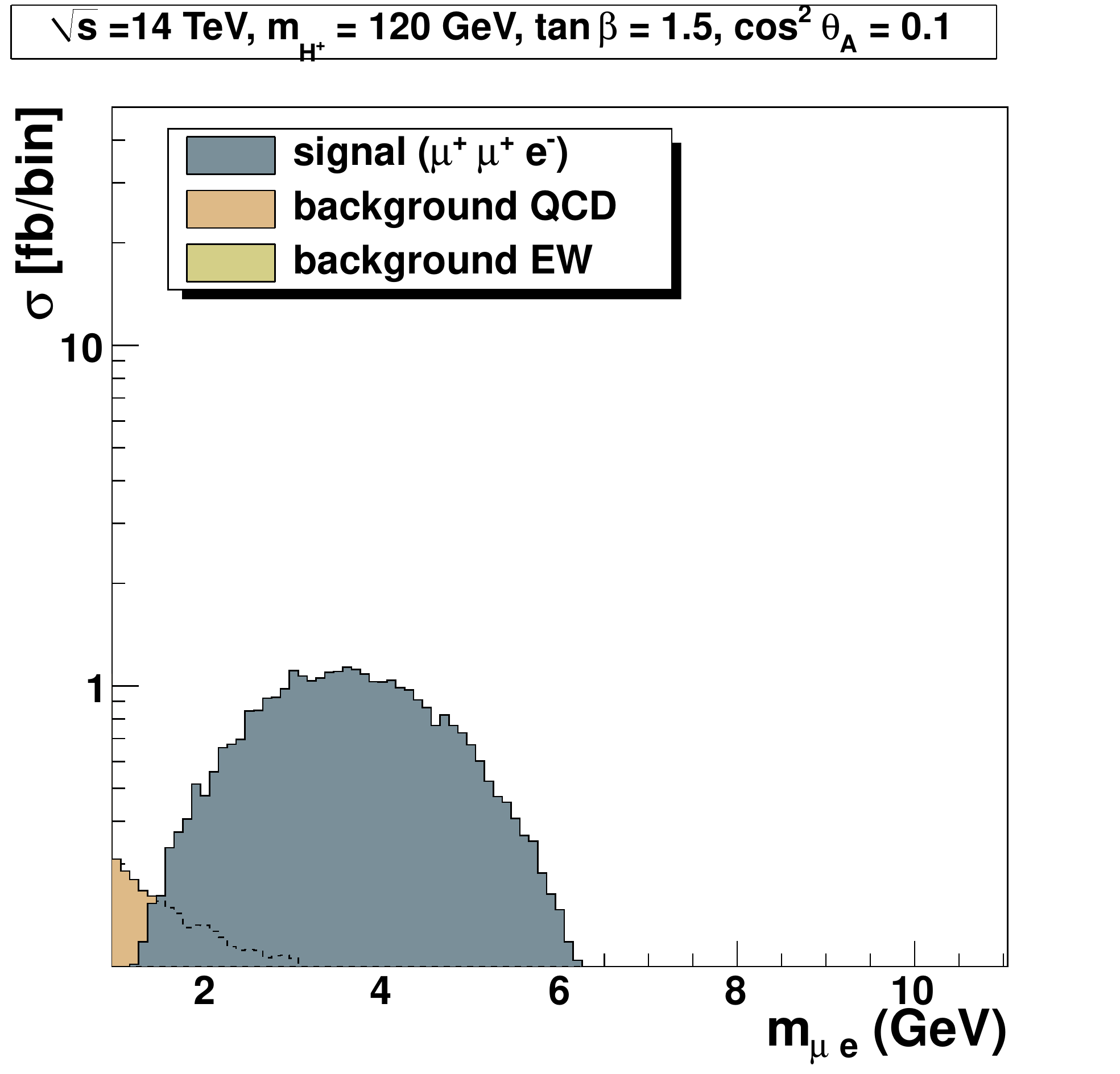}
\includegraphics[width=0.45 \linewidth]{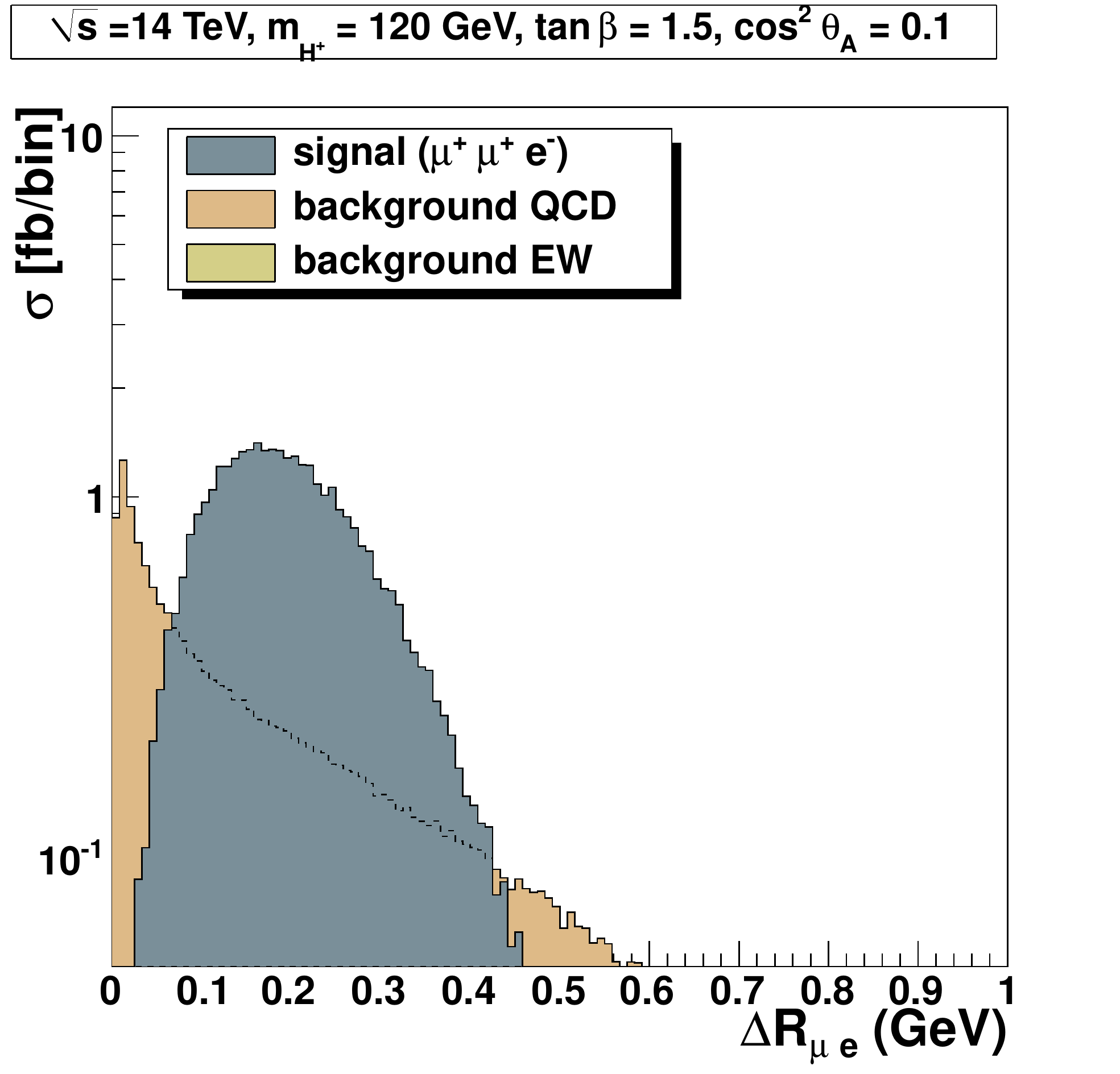}
\end{center}
\vskip -0.7cm
\caption{$m_{\mu^\pm e^\mp}$ and $\Delta R_{\mu^\pm e^\mp}$ distributions for the $\mu^\pm \mu^\pm e^\mp$ topology. \label{fig:SS}}
\end{figure}

In this section we combine the signal and background estimates discussed in Secs.~\ref{sec:signature} and \ref{sec:bkgd} and extract the expected LHC sensitivity. In Figs.~\ref{fig:mme}-\ref{fig:SS} we show signal and background for various topologies and center--of--mass energies. The benchmark point that we consider is given above Eq.~(\ref{twb}), the integrated signals are presented in Eqs.~(\ref{signalmu}) and (\ref{signaltau}) and the background is summarized in Table~\ref{tab:background}.

In Fig.~\ref{fig:mme} we show the di-muon invariant mass, $m_{\mu\mu}$, and separation, $\Delta R_{\mu\mu} = \sqrt{\Delta \phi^2 + \Delta \eta^2}$, for the $\mu^+\mu^- e$ topology and for $\sqrt{s}$ = 7, 8 and 14 TeV. The sharp peak and the broad bump correspond to underlying $A\to \mu\mu$ and $A\to\tau\tau$ decays. As seen in Sec.~\ref{sec:signature} the large enhancement of $A\to\tau\tau$ modes is compensated by the suppression implied by the two semileptonic $\tau\to\mu\nu_\mu\nu_\tau$ decays. On top of this, the broadening of the $\tau\tau$ peak due to the large amount of missing energy makes the identification of $A\to\tau\tau$ modes rather difficult. We comment on this later in this section. 

Following the expected mass resolution of the ATLAS detector, we choose to present the histograms in $100$ MeV bins; as seen in $\Upsilon\to\mu\mu$ studies at ATLAS (see, for instance, Ref.~\cite{atlasmumu}), this is indeed the resolution for low $p_T$ and low invariant mass di-muons emitted in the central region. In this study we assume an exact 100 MeV sensitivity but will conservatively double the background in our estimate of the experimental sensitivity.

In each plot we separate the total background into its EW and QCD components. As discussed in Sec.~\ref{sec:bkgd} the latter can be reduced further by requiring a harder cut on the highest $p_T$ lepton: for $p_T \gtrsim 30 $ GeV we loose about 10\% of the signal (see Fig.~\ref{fig:ptfrac}) but the QCD background becomes essentially negligible.

Another important feature of our signal (that is common to di-muon and di--tau events) is the low $\Delta R_{\mu\mu}$ we find between the two muons that reconstruct the CP-odd Higgs $A$. This feature is common for both signal and background and has a simple kinematical origin, namely high $p_T$ and low invariant mass of the lepton pair. In Fig.~\ref{fig:mme} we separately show the $\Delta R_{\mu\mu}$ distributions for events originated from the $A\to\tau\tau$ bump (blue) and from the $A\to \mu\mu$ peak (magenta). We note that in the $A\to\tau\tau$ case the upper limit on the angular separation is about 0.35 at 7 TeV (and it increases to 0.45 at 14 TeV), and in the $A\to\mu\mu$ case we find slightly larger separations (0.55 at 7 TeV up to 0.7 at 14 TeV). This demonstrates that the recent CMS search for new physics in multilepton events~\cite{Chatrchyan:2012mea} is not sensitive to these scenarios because our signal events would be removed by the isolation criterion adopted in the search. In this CMS analysis leptons are rejected if the ratio of the total transverse energy (sum of track $p_T$ and calorimeter-tower $E_T$ in a cone of $\Delta R< 0.3$ for muons and $\Delta R < 0.4$ for electrons) to the lepton $p_T$ is smaller than 15\%. In our case, as can be seen from the blue region in the top right panel in Fig.~\ref{fig:mme}, these $\Delta R$ cuts essentially eliminate all signal events. Remember that the red region corresponding to $A\to\mu\mu$ is also rejected in the CMS analysis because of the lower cut on the opposite-sign same-flavor dilepton invariant mass. The important consequence of this observations is that only hadronic lepton isolation should be applied. 

Finally, let us comment on one particular search that was presented in Ref.~\cite{Chatrchyan:2012mea} and that could possibly constrain our parameter space. CMS looked for four-lepton events (two hadronic taus and two light leptons) and even have a slight excess (see the 7th entry in Table 1 of Ref.~\cite{Chatrchyan:2012mea}). In our class of models this exact signature is easily obtained by having the CP-odd Higgs decay to taus and both $W$ bosons decay leptonically. The cross section associated to this process can be very large; in fact, we replace the $\sim 3\%$ suppression due to leptonic decays of the taus with a milder suppression from an extra leptonic $W$ decay. The two (hadronic) taus are produced in a small $\Delta R \lesssim 0.35$ cone and are mostly cut by the CMS isolation criterion. The question whether enough events survive the isolation cut at a level compatible with the results presented in Ref.~\cite{Chatrchyan:2012mea} is very interesting and a detailed analysis will be presented elsewhere~\cite{Dermisek:2020xx}. This highlights the importance of understanding pairs of hadronic taus at small angular separation. The issue of isolation in this context has also been recently discussed in Ref.~\cite{Chang:2012qu}.

For convenience, in Figs.~\ref{fig:mmm}--\ref{fig:SS}, we show the di-lepton invariant mass and $\Delta R$ distributions of the remaining signal topologies for $\sqrt{s}=$ 7, 8 and 14 TeV. As discussed in Sec.~\ref{sec:signature}, only the $\mu^+\mu^-\mu$ mode displays the $A\to\mu\mu$ peak. Same sign topologies (the $\mu^\pm \mu^\pm e^\mp$ and $e^\pm e^\pm \mu^\mp$ distributions are identical) receive background only from purely hadronic underlying processes. We further remind the reader that all QCD backgrounds can be further reduced by increasing the cut on the highest $p_T$ lepton: for the $\mu^\pm \mu^\pm e^\mp$ and $e^\pm e^\pm \mu^\mp$ modes this opens the possibility of a nearly background free search.

We define the 95\% experimental sensitivity as the cross section required to generate (at a given luminosity) twice the square root of the number of background events. We focus on two experimental scenarios: $\sqrt{s} = 8$ TeV with a luminosity ${\cal L} = 20 \; {\rm fb}^{-1}$ (end-of-2012 data set) and $\sqrt{s} = 14$ TeV with a luminosity ${\cal L} = 40 \; {\rm fb}^{-1}$ (expected end-of-2015 data set). For these two scenarios we take the background per 100 MeV bin to be about 1 fb (that is about twice our estimate as can be easily seen in Figs.~\ref{fig:mme}--\ref{fig:SS}).

\begin{figure}
\begin{center}
\includegraphics[width=0.4 \linewidth]{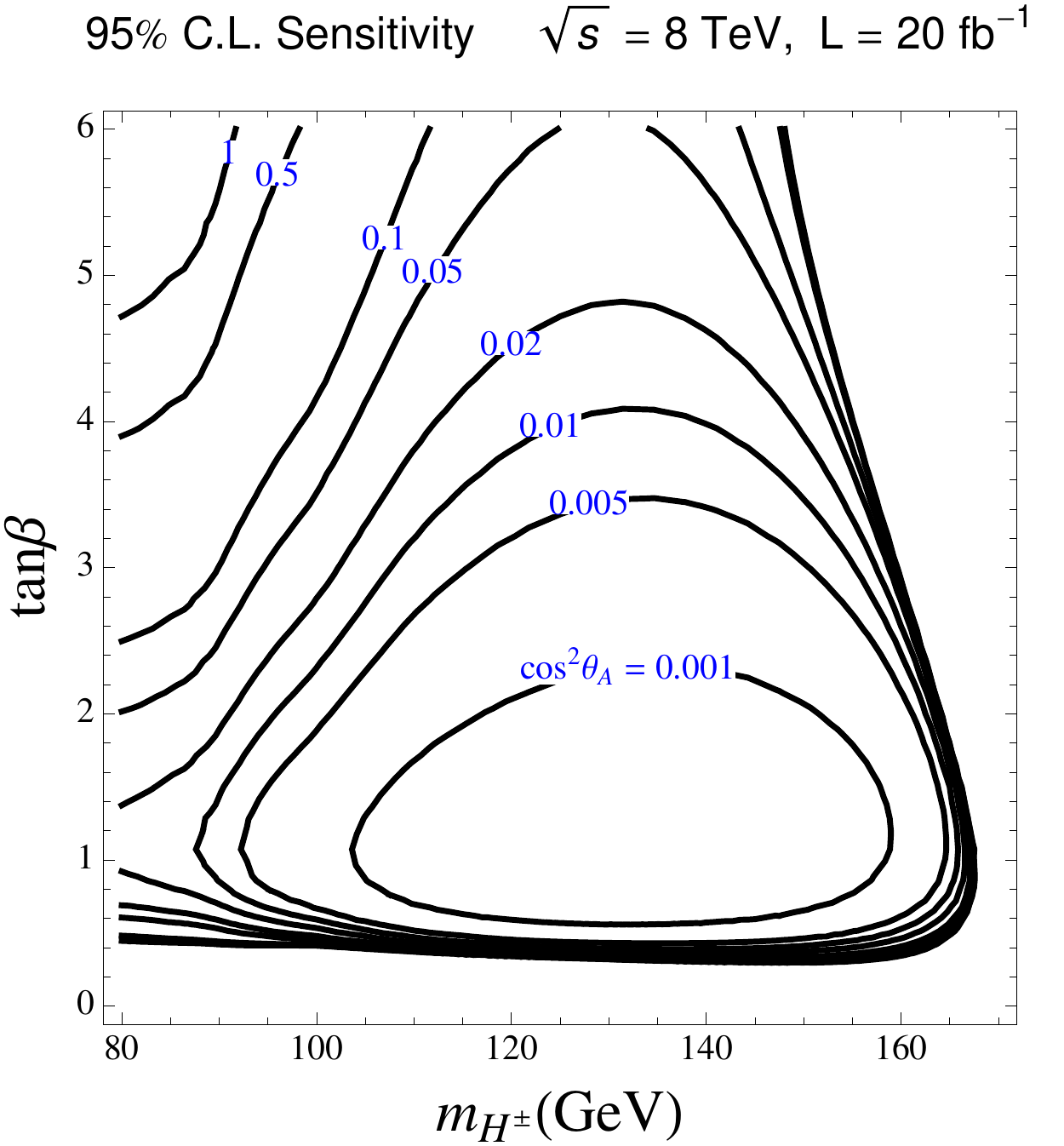}
\includegraphics[width=0.4 \linewidth]{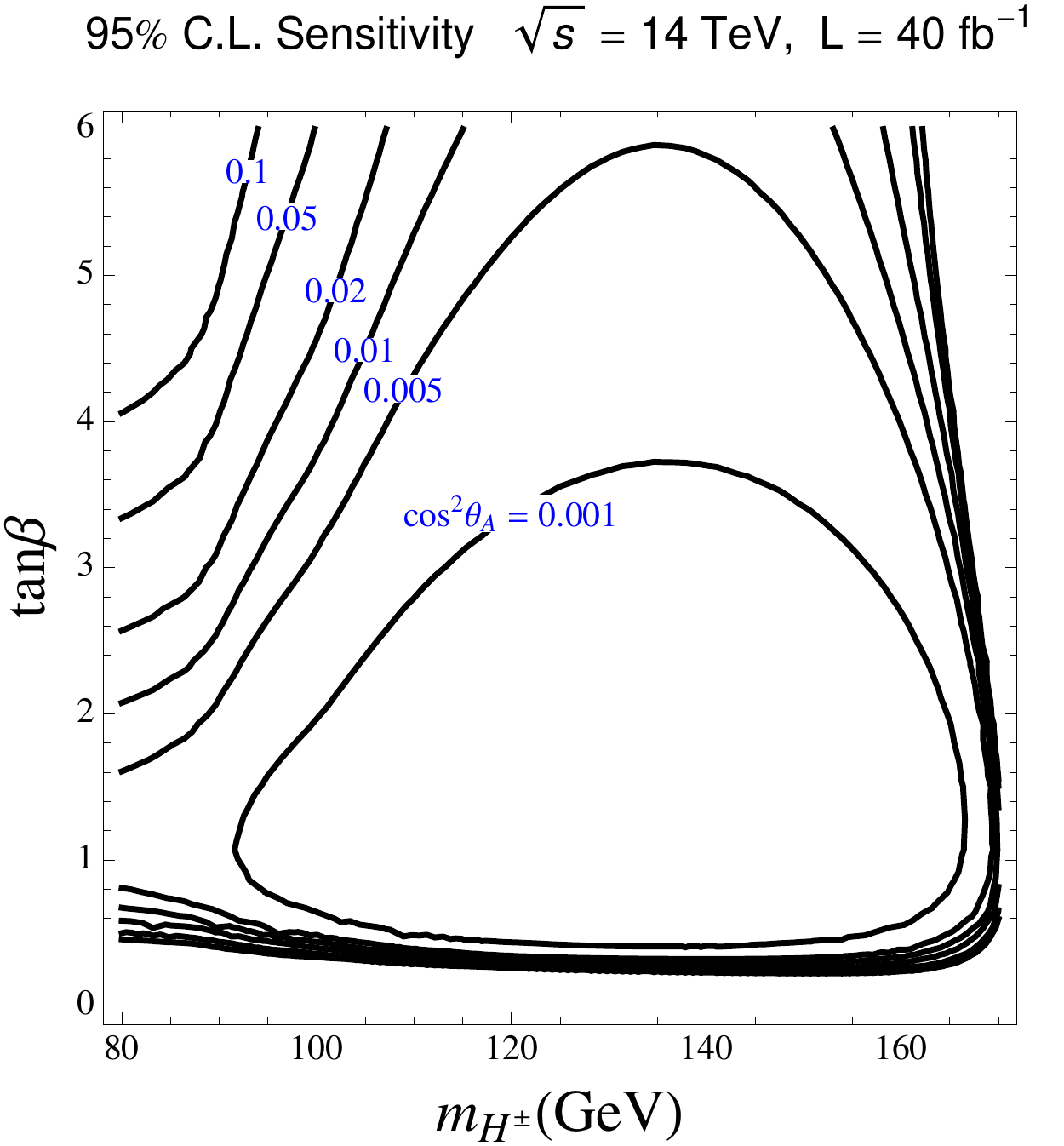}
\end{center}
\vskip -0.7cm
\caption{
Expected LHC sensitivity for the 2012 ($\sqrt{s} =$ 8 TeV and ${\cal L}= 20 \; {\rm fb}^{-1}$) and 2015 ($\sqrt{s} =$ 14 TeV and ${\cal L}= 40 \; {\rm fb}^{-1}$) scenarios. The contours are labeled by the value of $\cos^2\theta_A$.\label{fig:reach}}
\end{figure}

In Fig.~\ref{fig:reach} we present the LHC reach for these two scenarios in the $[m_{H^\pm},\tan\beta]$ plane. The contours correspond to the experimentally accessible region for different values of $\cos^2 \theta_A$ (used as label). Note that in drawing the contours we include the effects of the $m_{H^\pm}$ dependence of the acceptance (see Fig.~\ref{fig:pt}). These contours are also plotted in Figs.~\ref{fig:mhtb1} and \ref{fig:mhtb2}. It is clear that this search strategy allows to almost completely cover all the parameter space that is currently left open by present experimental results.

\section{Conclusions}
\label{sec:conclusions}
In singlet extensions of the two Higgs doublet model, the decay mode of the charged Higgs boson into $W$ and a CP-odd Higgs, $H^\pm \to W^\pm A$, dominates in a large range of charged Higgs boson mass and $\tan \beta$ even if the light CP-odd Higgs boson is $\sim 99\%$ singlet-like. For larger doublet fraction of the CP-odd Higgs the branching ratio of this decay mode is close to $100\%$. Thus the usual searches for the charged Higgs boson in top quark decays,  $H^\pm \to \tau \nu$ and  $H^\pm \to c s$, should be expanded to account for this possibility. 

We study this scenario assuming couplings of Higgs bosons to fermions, gauge bosons, and other Higgses according to the two Higgs doublet model Type II. The two CP-odd Higgses originating from doublets and the singlet can mix, and thus the light CP-odd Higgs is allowed to have an arbitrary doublet fraction. A light CP-odd Higgs boson can be motivated by approximate global symmetries and thus it occurs in a variety of models. A large singlet fraction is sufficient not to significantly modify the decay modes of the SM-like Higgs boson. It also easily allows the CP-odd Higgs to avoid detection in Upsilon decays or in direct searches, which roughly translate into limit the $\tan \beta \cos \theta_A < 0.5$.  Thus this scenario is relevant the most for small  to medium $\tan \beta$.

In the three dimensional parameter space, $m_{H^\pm}$, $\tan \beta$, and $\cos^2 \theta_A$, we summarize experimental constraints from direct searches for $H^\pm \to \tau \nu$, $H^\pm \to c s$, and $H^\pm \to W^\pm A$ followed by $A \to \tau^+ \tau^- $, map out the region where   $H^\pm \to W^\pm A$ dominates, and calculate the LHC reach of the search based on  $A \to \tau^+ \tau^-$ (with both $\tau$s decaying into leptons) and  $A \to \mu^+ \mu^-$. This search is especially advantageous if the CP-odd Higgs is below the $b \bar b $ threshold, that we assume, but it is also applicable to heavier CP-odd Higgses.

 To suppress background, especially from semileptonic bottom and charm decays, we require an additional lepton ($e$ or $\mu$) from one of the two $W$s. In addition to the $A \to \mu^+ \mu^- $ case, in which the mass of the CP-odd Higgs can be reconstructed, there is a complementary signal when $e$ and $\mu$ leptons originate from $A \to \tau^+ \tau^-$. In the special case in which we have two same sign same flavor leptons, {\it e.g.}  $\mu^+ \mu^+ e^-$, the background is very low. We find that this search  in  20 fb$^{-1}$ of data from the 8 TeV run of the LHC can constrain a large region of the parameter space to which other searches are not sensitive. We also discussed the reach of the LHC with 40 fb$^{-1}$ of data at 14 TeV center-of-mass energy. These are summarized in Figs.~\ref{fig:mhtb1} and~\ref{fig:mhtb2} by black thick lines, or in a different way in Fig.~\ref{fig:reach}. 

We showed that existing trilepton searches are not sensitive to our signal because the adopted isolation criterion effectively removes high $p_T$ muons (electrons) with $\Delta R \lesssim 0.3$ (0.4) and, as can be seen in Figs.~\ref{fig:mme}-\ref{fig:SS}, this essentially eliminates all our signal events. This shows the limitations of general trilepton searches and highlights the importance of dedicated searches that would pick up events with di-leptons ($e$, $\mu$, hadronic $\tau$) at small angular separation. 

The analysis strategy presented in this work can be further improved with the addition of b-tagging. Since the signal contains two b-jets, requiring b-tagging would not significantly reduce it while the background, especially that from semileptonic bottom and charm decays would be highly suppressed. Moreover, there are interesting signal topologies that we have not considered in detail. For example, the scenario we consider can also lead to a sizable excess of four lepton events with two hadronic taus and two light leptons in which the two taus are very close to each other. This again stresses the importance of understanding pairs of leptons at small $\Delta R$.

\section*{Acknowledgments}
\noindent
R.D. is supported in part by the  Department of Energy under grant number DE-FG02-91ER40661. E.L. thanks Zack Sullivan for discussions on the trilepton background, Frank Siegert for help with Sherpa and Niels Krumnack for many conversations on statistics.

\appendix
\section{Partial Widths}
\label{sec:formulae}
In this appendix we collect the expressions for various partial widths of the top quark, CP-odd and charged Higgs bosons. The two dominant partial widths of the top quark are:
\begin{align}
\Gamma(t\to b W^+) &= \frac{G_F}{\sqrt{2}} \frac{m_{t,pole}^3}{8 \pi} \left( 1- x_{Wt} \right)^2 \left( 1 + 2 \; x_{Wt} \right) \; ,\\
\Gamma(t\to b H^+) &= \frac{G_F}{\sqrt{2}} \frac{m_{t,pole}}{8\pi} \left( 1 - x_{Ht} \right) 
\left( \left[m_b^{\overline{MS}} (\mu_t)\right]^2 \tan^2\beta + \frac{\left[m_t^{\overline{MS}} (\mu_t)\right]^2}{\tan^2\beta}
\right) \; ,
\end{align}
where $x_{Wt} = m_W^2/m_{t,pole}^2$, $x_{Ht} = m_{H^\pm}^2/m_{t,pole}^2$ and $\mu_t = O(m_t)$. 

The three largest partial widths of the CP-odd Higgs are:
\begin{align}
\Gamma(A\to \tau^+\tau^-) &= \frac{G_F}{\sqrt{2}} \frac{m_A}{4 \pi} \sqrt{1- 4 x_{\tau A}}\; m_\tau^2 \tan^2\beta \; \cos^2\theta_A \; ,\\
\Gamma(A\to \mu^+\mu^-) &= \frac{G_F}{\sqrt{2}} \frac{m_A }{4 \pi} \sqrt{1- 4 x_{\mu A}}\; m_\mu^2\tan^2\beta\; \cos^2\theta_A \; ,\\
\Gamma(A\to c \bar c) &=  \frac{G_F}{\sqrt{2}} \frac{m_A}{4 \pi} \sqrt{1- 4 x_{c A}}\; \frac{\left[m_c^{\overline{MS}} (\mu_A)\right]^2}{\tan^2\beta}\; \cos^2\theta_A  \; ,
\end{align}
where $x_{\tau A} = m_\tau^2/m_A^2$, $x_{\mu A} = m_\mu^2/m_A^2$, $x_{c A} = m_{c,pole}^2/m_A^2$ and $\mu_A = O(m_A)$.

For a charged Higgs lighter than the top quark, the allowed on-shell two body decays are~\cite{HplusTN,Djouadi:1994gf,Hplusqq}:
\begin{align}
\Gamma(H^+\to \tau^+ \nu_\tau) &= \frac{G_F}{\sqrt{2}} \frac{m_{H^\pm}}{4 \pi} (1 - x_{\tau H})^3\; m_\tau^2 \tan^2\beta \; ,\\
\Gamma(H^+ \to c\bar s) &= N_c \; \frac{G_F}{\sqrt{2}} \frac{m_{H^\pm}}{4 \pi} (1 - x_{c H})^3\;
\left(  \frac{\left[m_c^{\overline{MS}} (\mu_H)\right]^2}{\tan^2\beta} + \left[m_s^{\overline{MS}} (\mu_H)\right]^2 \tan^2\beta 
\right) \; \Delta_{qq}  \; ,\\
\Gamma (H^+ \to W^+ A) &= \cos^2\theta_A  \frac{G_F}{\sqrt{2}} \frac{m_{H^\pm}^3}{8 \pi} \Big[
x_{AH}^2 + (1- x_{WH})^2 - 2 x_{AH} (1 + x_{WH})
\Big]^{3/2} \; ,\label{WA} 
\end{align}
where $x_{\tau H} = m_\tau^2/m_{H^\pm}^2$, $x_{c H} = m_{c,pole}^2/m_{H^\pm}^2$, $\mu_H = O(m_{H^\pm})$, $x_{W H} = m_W^2/m_{H^\pm}^2$ and $x_{AH} = m_A^2/m^2_{H^\pm}$. The factor $\Delta_{qq}$ encapsulates QCD corrections and its approximate form is~\cite{deltaqq0,deltaqq1,deltaqq2}
\begin{equation}
\Delta_{qq} \simeq 5.67 \frac{\alpha_s (m_{H^\pm})}{\pi} + (35.94 - 1.36 N_f)  \frac{\alpha_s^2 (m_{H^\pm})}{\pi^2} \; .
\end{equation}
It is also important to include decays mediated by a virtual top or W. In the scenario that we discuss in this paper ($m_{H^\pm} < m_t$), the top in the $H^+ \to t^* b \to W^+ b \bar b$ mode is always off mass-shell. In this limit the phase space integration can be easily done analytically and, in the limit in which we neglect the $m_b^2/m_t^2$ ratio, one obtains~\cite{wbb}:
\begin{align}
\Gamma(H^+ \to W^+ b\bar b) &= 
3 G_F^2 \frac{\left[m_t^{\overline{MS}} (\mu_H)\right]^4}{64 \pi^3 \tan^2\beta} m_{H^\pm} \Big[ \nonumber \\ 
& \hskip -2cm
    \frac{x_{WH}^2}{x_{tH}^3} ( 4 x_{WH} \; x_{tH} + 3 x_{tH} - 4 x_{WH}) \log [x_{WH} (x_{tH} - 1)/(x_{tH} - x_{WH})]\nonumber \\ 
& \hskip -2cm
     + (3 x_{tH}^2 - 4 x_{tH} - 3 x_{WH}^2 + 1 ) \log[(x_{tH} - 1)/(x_{tH} - x_{WH})]
     - 5/2 \nonumber \\
& \hskip -2cm
     + (1 - x_{WH})/x_{tH}^2 (3 x_{tH}^3 - kt x_{WH} - 2 x_{tH} x_{WH}^2 + 4 x_{WH}^2)
     + x_{WH} (4 - 3/2 x_{WH})
\Big] \; ,
\end{align}
where $x_{W H} = m_W^2/m_{H^\pm}^2$ and $x_{t H} = [m_t^{\overline{MS}}(\mu_H)]^2/m_{H^\pm}^2$. 

Finally, we turn to the off--shell contributions to the $H^+ \to W^+ A$ mode presented in Eq.~(\ref{WA}). In the parameter space region that we consider ($m_{H^\pm} < m_t$ and $m_A < m_\Upsilon$) the $H^+ \to W^{(*)} A$ decay can proceed through a real or virtual $W$, and we need the exact integration over the Dalitz plot phase space in order to capture both effects. The doubly differential rate is~\cite{Djouadi:1995gv}~\footnote{Note that there is a typo in Eq.~(59) 
of Ref.~\cite{Djouadi:1995gv}.}:
\begin{align}
\frac{d\Gamma (H^+ \to  W^{(*)} A)}{dx_1 dx_2} &=
\cos^2\theta_A \frac{9 G_F^2 m_W^4}{16 \pi^3} m_{H^\pm} \; F_{AW} (x_1,x_2) \; .
\label{WstarA}
\end{align}
The function $F_{AW}(x_1,x_2)$ is given by
\begin{align}
F_{AW}(x_1,x_2) &= \frac{(1 - x_1) (1 - x_2) - x_{AH}}{ 
   \frac{\Gamma_W^2}{m_H^2}\; x_{WH} + (1 - x_{AH} + x_{WH} - x_1 - x_2)^2} \; ,
\end{align}
where $x_{AH} = m_A^2/m^2_{H^\pm}$, $\Gamma_W$ is the total width of the $W$, $x_i = 2 E_i/m_{H^\pm}$ are the normalized energies of decay products of the $W$, and the integration range is:
\begin{align}
0             &\leq x_2 \leq 1-x_{WH} \; , \\
1-x_2 -x_{WH} &\leq x_1 \leq 1- \frac{x_{WH}}{1-x_2} \; .
\end{align}
Note that for an on-shell $W$, the integral over Eq.~(\ref{WstarA}) reproduces the standard two body decay given in Eq.~(\ref{WA}).
 

\end{document}